\documentclass[11pt]{article}

\usepackage[a4paper,margin=1in]{geometry}
\usepackage{amsmath,amssymb,amsthm}
\usepackage{mathrsfs}
\usepackage{bm}
\usepackage{enumitem}
\usepackage{graphicx}
\usepackage{booktabs}
\usepackage{longtable}
\usepackage{array}
\usepackage{hyperref}
\usepackage[most]{tcolorbox}
\usepackage[numbers]{natbib}

\hypersetup{
    colorlinks=true,
    linkcolor=blue,
    citecolor=blue,
    urlcolor=blue
}

\newcommand{\T}{\mathbb T}
\newcommand{\R}{\mathbb R}
\newcommand{\E}{\mathbb E}
\newcommand{\Prob}{\mathbb P}
\newcommand{\F}{\mathcal F}
\newcommand{\Lie}{\mathcal L}
\newcommand{\A}{\mathcal A}
\newcommand{\B}{\mathcal B}

\usepackage{amsthm}

\newtheorem{theorem}{Theorem}[section]

\newtheorem{proposition}[theorem]{Proposition}

\theoremstyle{definition}
\newtheorem{definition}[theorem]{Definition}

\theoremstyle{remark}

\newtcolorbox{roadmap}{
  colback=blue!5,
  colframe=blue!60!black,
  title=Roadmap,
  fonttitle=\bfseries,
  breakable
}

\title{\bf
From Stochastic Transport Models to High-Dimensional Bayesian Inference
}

\author{Dan Crisan\thanks{Imperial College London, d.crisan@imperial.ac.uk} \and Oana Lang\thanks{Babes-Bolyai University, oana.lang@ubbcluj.ro}}
\date{}

\begin{document}

\maketitle

\begin{abstract}
The Stochastic Transport in Upper Ocean Dynamics project\footnote{See \url{https://www.imperial.ac.uk/ocean-dynamics-synergy/} for details.} introduced a new paradigm for representing uncertainty in geophysical fluid dynamics through stochastic transport models calibrated from data. This paper provides a synthesis of the principal theoretical and data assimilation developments achieved as part of the project. 

The central theme of the paper is the transformation of stochastic transport models from a modelling paradigm into a mathematically rigorous framework equipped with analytical, variational, calibration and Bayesian data assimilation methodologies suitable for geophysical fluid dynamics. We describe the development of the theory of stochastic transport equations, including well-posedness, regularity and approximation results, together with their geometric foundations based on stochastic variational principles, rough paths and fractional transport models.

We further review methodologies for calibrating stochastic transport structures from observational and high-resolution simulation data, ranging from reduced-order stochastic models to modern generative approaches. Finally, we discuss Bayesian inference and high-dimensional data assimilation techniques developed for stochastic transport models, including particle filter methodologies and machine learning uncertainty quantification.

Taken together, these developments establish a coherent framework linking stochastic transport modelling, rigorous analysis, calibration and Bayesian inference, thereby providing a foundation for uncertainty quantification and forecasting in geophysical fluid dynamics.\\[0.1mm]


\end{abstract}

\tableofcontents

\section{Introduction}
\label{sec:01_introduction}

Understanding and quantifying uncertainty in geophysical fluid dynamics is one of the central scientific challenges of contemporary climate and environmental science. Oceanic and atmospheric flows evolve across a vast range of interacting spatial and temporal scales, many of which remain unresolved in both numerical simulations and observational datasets. These unresolved processes influence transport, mixing, predictability and forecasting skill, and therefore play a fundamental role in climate modelling, weather prediction and environmental forecasting.

Traditional deterministic models represent the resolved scales of motion with increasing fidelity, but the uncertainty associated with unresolved scales remains difficult to characterise. This challenge has motivated the development of stochastic parameterisation methodologies, which seek to represent the impact of unresolved dynamics through probabilistic rather than deterministic descriptions.

The ERC Synergy project \emph{Stochastic Transport in Upper Ocean Dynamics} (STUOD) was initiated to address this challenge from a fundamentally new perspective. The central idea of the project is that uncertainty should enter fluid models through the transport mechanism itself. Rather than adding stochastic forcing terms to deterministic equations, the project investigated stochastic perturbations of the Lagrangian flow and the resulting stochastic transport equations. This viewpoint gave rise to a coherent mathematical programme connecting stochastic fluid dynamics, geometric mechanics, uncertainty quantification, calibration and Bayesian inference.

The present paper surveys the principal theoretical and data assimilation developments achieved within the STUOD research programme\footnote{Further details of the STUOD research programme can be found at \url{https://www.imperial.ac.uk/ocean-dynamics-synergy/our-mission/}}. These work packages focused on the mathematical foundations of stochastic transport models, their calibration from data and their use in Bayesian inference and data assimilation. 

Over the course of the
STUOD programme, we developed a mathematical framework
that takes stochastic transport models from their physical and geometric origins to
practical methodologies for uncertainty quantification and forecasting. The programme
begins with the introduction of stochastic transport as a new paradigm for modelling
unresolved fluid dynamics, continues with the development of a rigorous analytical theory
for transport-noise SPDEs, extends this theory through stochastic variational principles,
rough paths and fractional transport models, and culminates in methodologies for
calibrating stochastic transport structures from data and incorporating them into
high-dimensional Bayesian data assimilation algorithms.

From a mathematical perspective, these developments make a number of contributions summarised below.
First, they establish a well-posedness theory for a large class of stochastic transport
equations whose first-order noise operators fall outside the scope of classical semilinear
SPDE theory. Secondly, they demonstrate that these equations possess a rich geometric
structure inherited from stochastic variational principles and Lie transport. Thirdly, they develop techniques for
learning stochastic transport structures directly from observational and numerical data,
transforming stochastic transport models from abstract mathematical objects into
data-informed models. Finally, they provide rigorous Bayesian methodologies for
combining these calibrated models with observations through particle filtering and other
high-dimensional data assimilation techniques.

The unifying theme throughout these developments is that uncertainty should be
represented through the transport mechanism itself rather than by externally imposed
random forcing. This principle provides a common mathematical language that connects
all subsequent sections of this survey. Although the individual papers address a variety
of fluid models and mathematical questions, they should be viewed collectively as
successive stages in the development of a single research programme whose objective is
to provide mathematically rigorous and practically useful methodologies for uncertainty
quantification in geophysical fluid dynamics.

\subsection{The STUOD Paradigm}
\label{subsec:01_01_the_stuod_vision}

The scientific vision of STUOD was motivated by two complementary observations.

First, many uncertainties in geophysical fluid dynamics arise from unresolved transport processes. Oceanic and atmospheric observations reveal the presence of coherent structures, eddies, fronts and multiscale interactions that cannot be represented explicitly in operational models.

Second, transport processes possess a strong physical  character. Fluid motion is naturally described through the evolution of Lagrangian trajectories and flow maps. Consequently, uncertainty associated with unresolved scales should also be incorporated at the level of transport.

This perspective led to the development of stochastic transport models in which the Lagrangian flow satisfies an equation of the form
\[
dX_t
=
u_t(X_t)\,dt
+
\sum_{i=1}^{\infty}
\xi_i(X_t)\circ dW_t^i,
\]
where $u$ denotes the resolved velocity field and the vector fields $\xi_i$ represent unresolved transport structures. The resulting stochastic transport equations preserve many of the properties of the underlying deterministic models while providing a mathematically rigorous representation of uncertainty.

\subsection{Scope of the paper}
\label{subsec:01_02_scope_of_the_present_chapter}

The STUOD project generated a broad range of scientific developments spanning stochastic fluid mechanics, oceanography, climate science, numerical analysis and data assimilation.

The paper does not attempt to provide a comprehensive review of all scientific outputs of STUOD. Instead, its objective is to explain how a unified mathematical framework emerged from the interaction between stochastic transport modelling, rigorous analysis and statistical inference.

The results described in this paper were obtained over a period of six years and across a variety of geophysical fluid models. A chronological presentation would therefore obscure the conceptual connections between the different developments. For this reason, we organise the exposition according to the mathematical structure of the theory rather than the chronology of the project.

The central organising principle is the stochastic transport framework developed in \cite{CrisanLangEuler2023} on stochastic Euler equations. This framework provides the analytical and conceptual backbone for many of the subsequent developments. The individual fluid models discussed throughout the paper should therefore be viewed primarily as illustrations of a common mathematical programme rather than as isolated case studies.

\subsection{From Stochastic Transport Models to Bayesian Inference}
\label{subsec:01_03_from_stochastic_transport_models_to_bayesian_inference}

The developments surveyed in this paper may be interpreted as an answer to the following questions:\\

\noindent$\bullet$ Can stochastic transport equations arising from the SALT and Location-Uncertainty paradigms be analysed rigorously?

\noindent$\bullet$ Can stochastic transport equations be derived from variational principles and extended beyond Brownian semimartingale forcing?

\noindent$\bullet$ Can stochastic transport structures be inferred from observational or high-resolution data?

\noindent$\bullet$ How should calibrated stochastic transport models be combined with observations to perform forecasting and uncertainty quantification?\\

\noindent Taken together, these questions define a mathematical programme extending from stochastic transport modelling to high-dimensional Bayesian inference. The developments described in this paper may be grouped into five principal scientific contributions:

{\bf 1. A Unified Analytical Framework.} The first contribution is the development of a rigorous analytical framework for stochastic transport equations.
The framework accommodates transport-noise SPDEs whose stochastic perturbations are generated by first-order differential operators and therefore lie outside the scope of many classical SPDE theories.
The resulting methodology provides well-posedness, stability and regularity results for a broad class of stochastic fluid models.

{\bf 2. Regularity, Stability and Approximation Theory.}
The second contribution is the development of a systematic theory for global behaviour, regularity and approximation. These results demonstrate that stochastic transport models possess sufficient analytical robustness to support numerical approximation, calibration and data assimilation methodologies.

{\bf 3. Semimartingale Variational Foundations.}
The third contribution is the development of semimartingale variational principles, rough-path formulations and fractional transport models, demonstrating that stochastic transport equations arise naturally from variational principles and can be extended beyond the Brownian setting.

{\bf 4. Calibration Methodologies.} The fourth contribution is the development of calibration methodologies capable of inferring stochastic transport structures from data. The resulting approaches range from reduced-order stochastic models and principal-component methods to modern generative-model techniques.

{\bf 5. Bayesian Inference and Data Assimilation}. The fifth contribution is the development of Bayesian inference methodologies for stochastic transport models.
These developments include particle filtering, adaptive assimilation algorithms and machine-learning-based approaches for high-dimensional uncertainty quantification.

Although the individual developments reviewed in this paper span several distinct mathematical disciplines, they are connected by a common principle: Uncertainty should enter geophysical fluid models through the transport mechanism itself. This principle provides a common language linking stochastic fluid dynamics, geometric mechanics, calibration and Bayesian inference. From this perspective, the various results described in the following sections should not be viewed as independent contributions. Rather, they constitute different aspects of a single mathematical programme aimed at understanding, calibrating and assimilating stochastic transport models.

\subsection{Structure of the paper}
\label{subsec:01_06_structure_of_the_chapter}

The remainder of this paper demonstrates how this change in modelling philosophy - from adding stochastic forcing to modelling uncertainty in transport - leads naturally to a unified mathematical framework encompassing well-posedness theory, stochastic variational principles, calibration from data and Bayesian data assimilation. We have organised it as follows.

Section 2 introduces the stochastic transport framework and establishes the notation used throughout the paper. Section 3 develops the analytical foundations of stochastic transport equations, including local and global behaviour, uniqueness of the solution, regularity and approximation theory. Section 4 describes the variational foundations of stochastic transport dynamics and their extensions to rough-path and fractional settings. Section 5 addresses the calibration of stochastic transport models from data.
Section 6 reviews Bayesian inference and high-dimensional data assimilation methodologies.
Finally, Section 7 summarises the principal achievements of the programme and discusses future research directions.

The present paper forms one component of the final STUOD monograph. Its purpose is to provide a synthesis of the mathematical developments associated with stochastic transport theory, calibration and data assimilation.  The present contribution focuses on the development of a coherent mathematical pathway linking stochastic transport modelling, rigorous analysis, calibration and Bayesian inference.


\section{The Framework for Stochastic Transport Models}
\label{sec:02_unified_framework_for_stochastic_transport_models}

The purpose of this section is to introduce a common mathematical framework and notation for the developments surveyed in this paper. The results obtained concern a range of stochastic geophysical fluid models, variational formulations, calibration procedures and data assimilation algorithms. The individual papers often use notation adapted to their particular model. In this synthesis, however, we adopt a single notation wherever possible, with the stochastic transport formulation as the unifying object.

The starting point is the observation that the stochasticity in the STUOD models does not enter as an additive external forcing, nor as a standard multiplicative perturbation. Rather, it enters through the transport mechanism itself. Two closely related paradigms have been used:

\paragraph{The Stochastic Advection by Lie Transport (SALT) paradigm.}

The SALT paradigm, first introduced in \cite{Holm2015Variational}, provides a physically and geometrically consistent
framework for introducing stochasticity into fluid dynamics. Rather than
perturbing the governing equations by adding stochastic forcing terms, the
central idea is to perturb the \emph{Lagrangian trajectories} of the fluid
particles. In its simplest form, the Lagrangian motion of a fluid parcel is represented by
\begin{equation}
	dX_t = u_t(X_t)\,dt
	+ \sum_{i=1}^{\infty} \xi_i(X_t)\circ dW_t^i ,
	\label{eq:lagrangian-salt}
\end{equation}
where $u$ is the resolved velocity field and the vector fields $\xi_i$ encode unresolved transport directions. The stochastic integral in \eqref{eq:lagrangian-salt} is understood in the Stratonovich sense. This form is natural from the point of view of stochastic fluid mechanics because it preserves the ordinary chain rule and is therefore compatible with geometric and variational formulations. Since
all advected quantities are transported by the stochastic flow generated by
$d\chi_t$, the resulting stochastic equations preserve many of the
geometric properties of the underlying deterministic models, including
their transport structure, Kelvin circulation laws and, where appropriate,
their Hamiltonian formulation.

\paragraph{The Location Uncertainty (LU) paradigm.}

The Location Uncertainty (LU) paradigm, first introduced in \cite{Memin2014LocationUncertainty}, provides an alternative stochastic
framework for modelling unresolved transport processes in geophysical fluid
dynamics. As in the SALT paradigm, stochasticity enters through the
Lagrangian motion of fluid particles rather than through additive random
forcing in the governing equations. However, the underlying modelling
philosophy is fundamentally different. The Lagrangian velocity is first
decomposed into a smooth resolved component and a rapidly fluctuating
unresolved component,
\begin{equation}
	dX_t
	=
	u_t(X_t)dt
	+
	\sigma(X_t,t)\,d B_t,
	\label{eq:LU-decomposition}
\end{equation}
where $u$ denotes the resolved velocity field and
$\sigma\dot B_t$ represents unresolved turbulent fluctuations whose
statistics are characterised by the covariance tensor
$
a(x,t)=\sigma(x,t)\sigma(x,t)^T.
$
This procedure leads systematically to stochastic transport equations
containing both multiplicative transport noise and deterministic
correction terms.

A distinctive feature of the LU formulation is that the covariance tensor
appears explicitly in the resolved equations. In particular, the
unresolved fluctuations generate the \emph{It\^o--Stokes drift}
$
u^S=\frac12\nabla\!\cdot a,
$
together with covariance-induced diffusion operators of the form
$
\frac12\nabla\!\cdot(a\nabla\cdot),
$
which represent the mean transport induced by unresolved turbulent
motions. Unlike conventional eddy-viscosity parameterisations, these additional terms are derived systematically from the stochastic Reynolds transport theorem underlying the Location Uncertainty framework, and therefore arise as a consequence of the stochastic transport formulation rather than being introduced empirically. Consequently, the LU framework preserves the energetic
consistency of the coarse-grained dynamics while providing a direct link
between the statistical properties of the unresolved velocity field and
the resulting transport model. 
\vspace{4mm}

Classical stochastic fluid models typically incorporate uncertainty by adding either
additive or multiplicative noise directly to the governing equations. While such
perturbations are often mathematically convenient, they do not generally preserve the
transport mechanisms that underpin fluid motion and may alter fundamental geometric
properties of the deterministic equations. In contrast, the central idea of stochastic
transport models is that uncertainty arises primarily through uncertainty in the
\emph{Lagrangian trajectories} of fluid parcels rather than through external random
forcing. The unresolved velocity fluctuations are therefore introduced by perturbing the
transport velocity itself. This leads naturally to first-order stochastic transport operators,
expressed in terms of Lie derivatives, which preserve the intrinsic transport structure of
the governing equations.

Although the SALT and LU formulations
coincide for certain models, such as the two-dimensional Euler equations,
they generally lead to different stochastic partial differential
equations, as illustrated later for the rotating shallow water equations. For the purposes of this synthesis, the phrase \emph{stochastic transport model} refers to any SPDE in which the stochastic perturbation is generated by first-order transport operators associated with prescribed, inferred, or calibrated vector fields.

\subsection{Probability space and stochastic drivers}
\label{subsubsec:02_00_01_probability_space_and_stochastic_drivers}

All stochastic processes are defined on a filtered probability space
$
(\Omega,\F,(\F_t)_{t\geq0},\Prob),
$
where the filtration satisfies the usual conditions. Unless explicitly stated otherwise, the stochastic drivers are a sequence
$
(W^i)_{i\geq1}
$
of independent real-valued Brownian motions adapted to $(\F_t)_{t\geq0}$. We write stochastic integrals in Stratonovich form when emphasising the transport structure and in It\^o form when deriving estimates.

When rough or fractional extensions are considered, the Brownian family $(W^i)_{i\geq1}$ is replaced by a rough path, a semimartingale family, or a fractional Brownian family. The notation introduced below is chosen so that these extensions can be described without changing the underlying transport philosophy.

\subsection{Spatial domains and function spaces}
\label{subsubsec:02_00_02_spatial_domains_and_function_spaces}

The principal spatial domain will be the $d$-dimensional torus
$
\T^d = \R^d/\mathbb Z^d,
$
with $d=2$ in the vorticity formulation of the Euler equation and
$d=2,3$ for the remaining models.  For $k\in\mathbb N_0$ and $1\le p<\infty$, the Sobolev space
$W^{k,p}(\T^d)$ consists of all functions whose weak derivatives up to
order $k$ belong to $L^p(\T^d)$. It is equipped with the norm
\[
\|u\|_{W^{k,p}}
=
\left(
\sum_{|\alpha|\le k}
\|D^\alpha u\|_{L^p}^p
\right)^{1/p},
\]
where $\alpha=(\alpha_1,\ldots,\alpha_d)$ is a multi-index,
$|\alpha|=\alpha_1+\cdots+\alpha_d$, and
\[
D^\alpha
=
\frac{\partial^{|\alpha|}}
{\partial x_1^{\alpha_1}\cdots\partial x_d^{\alpha_d}}.
\]
For $p=\infty$, the norm is defined by
\[
\|u\|_{W^{k,\infty}}
=
\max_{|\alpha|\le k}
\|D^\alpha u\|_{L^\infty}.
\]
The Hilbert Sobolev spaces are denoted by
$
H^k(\T^d):=W^{k,2}(\T^d),
$
with norm
\[
\|u\|_{H^k}
=
\left(
\sum_{|\alpha|\le k}
\|D^\alpha u\|_{L^2}^2
\right)^{1/2}.
\]
For vector-valued functions the same notation is used componentwise.

For incompressible flow models we write
\[
L^2_\sigma(\T^d)
=
\left\{
u\in L^2(\T^d;\R^d):
\nabla\cdot u=0,\;
\int_{\T^d}u(x)\,dx=0
\right\},
\]
and denote by
$H^k_\sigma(\T^d)$
the corresponding divergence-free Sobolev space.

\subsection{Transport operators}
\label{subsubsec:02_00_03_transport_operators}

Transport by vector fields plays a central role throughout this paper.
Given a sufficiently smooth vector field
$a:\T^d\rightarrow\R^d$,
the associated Lie derivative acting on scalar functions is defined by
\[
\Lie_a f:=a\cdot\nabla f.
\]
When acting on vector fields $u:\T^d\rightarrow\R^d$, the Lie derivative is
given by
\[
\Lie_a u
:=
[a,u]
=
a\cdot\nabla u-u\cdot\nabla a,
\]
where $[\cdot,\cdot]$ denotes the Lie bracket of vector fields. In the
special case in which $a$ is constant, $\Lie_a$ reduces to the directional
derivative in the direction of $a$.

Throughout the paper the stochastic transport vector fields are denoted by
$\{\xi_i\}_{i\ge1}$, and we write
\[
\Lie_i
:=
\Lie_{\xi_i}.
\]
These operators appear as the noise coefficients in the stochastic
transport equations arising from both the SALT and the Location
Uncertainty (LU) formulations. 

In most of the analytical results, the vector fields $\xi_i$ are assumed
to be divergence-free,
\[
\nabla\cdot\xi_i=0,
\]
or to satisfy the appropriate weighted divergence-free condition required
by the particular model. They are also assumed to possess sufficient
regularity, typically
\[
\sum_{i=1}^{\infty}
\|\xi_i\|_{W^{k+1,\infty}}^2
<\infty,
\]
or analogous summability conditions adapted to the functional framework of
the corresponding well-posedness theorem.

The operators $\Lie_i$ are first-order differential operators and are
therefore unbounded on the natural Banach spaces such as
$L^2(\T^d)$ and $H^k(\T^d)$. Consequently, the noise coefficients are not Lipschitz maps in the spaces in which one seeks solutions. This is one of the central analytical obstacles addressed by the well-posedness  theory.

\subsection{Canonical stochastic transport equation}
\label{subsubsec:02_00_04_canonical_stochastic_transport_equation}

The abstract stochastic transport equation used throughout this synthesis has the form
\begin{equation}
dq_t + \A(q_t)\,dt
     + \sum_{i=1}^{\infty} \B_i(q_t)\circ dW_t^i = 0.
\label{eq:abstract-stratonovich}
\end{equation}
Here $q_t$ denotes the state variable. Depending on the model, $q_t$ may be vorticity, buoyancy, layer depth, potential vorticity, velocity, temperature, or a collection of coupled fluid variables. The operator $\A$ denotes the deterministic nonlinear drift, while the operators $\B_i$ are first-order stochastic transport operators.

Although the stochastic fluid models considered throughout this paper arise from
different physical principles and describe different geophysical phenomena, they share a
common mathematical structure. After identifying the state variable, collecting the
deterministic transport and forcing terms into a nonlinear operator, and expressing the
stochastic transport through first-order Lie derivative operators, each model can be
written in the abstract form of equation (\ref{eq:abstract-stratonovich}). This equation therefore serves as a
\emph{master equation} for the analytical developments of the STUOD programme. 
The two-dimensional and three-dimensional Euler equations, the Great Lake equations,
the rotating shallow water equations, the thermal quasi-geostrophic equations, and the
Location Uncertainty models discussed later in this paper all fit naturally within this
framework. Consequently, the master equation provides the unifying mathematical
language that connects the diverse stochastic transport models developed within STUOD
and reveals the common analytical principles underlying seemingly different geophysical
fluid equations.

For scalar transported quantities, the typical structure is
\[
\B_i(q) = \Lie_i q = \xi_i\cdot\nabla q.
\]
For vector-valued quantities or differential forms, $\B_i$ may instead be the appropriate Lie derivative operator. The notation $\B_i$ is used in the abstract framework to cover all such cases.

Provided the operators $\B_i$ are linear, the corresponding It\^o form is
\begin{equation}
dq_t + \A(q_t)\,dt
     + \sum_{i=1}^{\infty} \B_i(q_t)\,dW_t^i
 =
\frac12\sum_{i=1}^{\infty} \B_i(\B_i(q_t))\,dt .
\label{eq:abstract-ito}
\end{equation}
A key point is that the second-order correction in \eqref{eq:abstract-ito} should not in general be regarded as a Laplacian. In particular, the well-posedness analysis does not assume that
\[
\sum_{i=1}^{\infty} \xi_i(x)\xi_i(x)^T
\]
is a spatially constant multiple of the identity. 
The stochastic transport vector fields $
\{\xi_i\}_{i\geq1}$
represent the unresolved component of the fluid motion. In applications, these vector fields may be calibrated from high-resolution simulations or observational data. The resulting fields therefore
capture the dominant coherent transport structures that are not resolved by
the coarse model. This data-driven construction is essential,
since the purpose of the stochastic transport model is to represent the
statistics of the unresolved flow. Unlike conventional 
parameterisations based on isotropic eddy diffusion, the calibrated fields
$
\xi_i
$
encode anisotropic, flow-dependent transport structures that reflect the
underlying physics of the unresolved dynamics.

\subsection{The stochastic Euler vorticity equation as anchor model}
\label{subsubsec:02_00_05_the_stochastic_euler_vorticity_equation_as_anchor_model}

The foundational example for the analytical developments presented in this
paper is the two-dimensional stochastic Euler equation in vorticity
formulation. It is the simplest nonlinear stochastic transport equation in
which the principal analytical difficulties already arise, while retaining
the structure that underlies more complex models such as the
stochastic quasi-geostrophic and rotating shallow water equations. On $\T^2$, let
\(
u:\T^2\times[0,T]\rightarrow\R^2
\)
denote the incompressible velocity field satisfying
\(
\nabla\cdot u=0
\).
The corresponding scalar vorticity is
\[
\omega=\operatorname{curl}u
=\partial_1u_2-\partial_2u_1.
\]
Conversely, the velocity is recovered from the vorticity through the
Biot--Savart law
\[
u
=
K*\omega
=
\nabla^\perp\Delta^{-1}\omega,
\]
where
\(
\nabla^\perp=(-\partial_2,\partial_1)
\),
\(
\Delta^{-1}
\)
denotes the inverse Laplacian acting on mean-zero functions, and
\(K\) is the Biot--Savart kernel on the torus. Throughout, the vorticity is
assumed to have zero spatial mean, ensuring that the inverse Laplacian is
well defined. The stochastic Euler vorticity equation with transport noise is
\begin{equation}
d\omega_t + u_t\cdot\nabla\omega_t\,dt
+ \sum_{i=1}^{\infty} \xi_i\cdot\nabla\omega_t\circ dW_t^i = 0.
\label{eq:stoch-euler-vorticity}
\end{equation}
Its It\^o form is
\begin{equation}
d\omega_t + u_t\cdot\nabla\omega_t\,dt
+ \sum_{i=1}^{\infty} \xi_i\cdot\nabla\omega_t\,dW_t^i
=
\frac12\sum_{i=1}^{\infty}
\xi_i\cdot\nabla\bigl(\xi_i\cdot\nabla\omega_t\bigr)\,dt.
\label{eq:stoch-euler-ito}
\end{equation}
This equation is the prototype for the analytical framework used in later STUOD work. It displays the main difficulties: the noise is transport-type, the It\^o correction is not assumed to be elliptic, and the nonlinear drift is the usual Euler transport term.

In the language of \eqref{eq:abstract-stratonovich},
\[
q=\omega,
\qquad
\A(\omega)=u\cdot\nabla\omega,
\qquad
\B_i(\omega)=\xi_i\cdot\nabla\omega.
\]
\begin{figure}[htbp]
	\centering
	\includegraphics[width=0.82\textwidth]{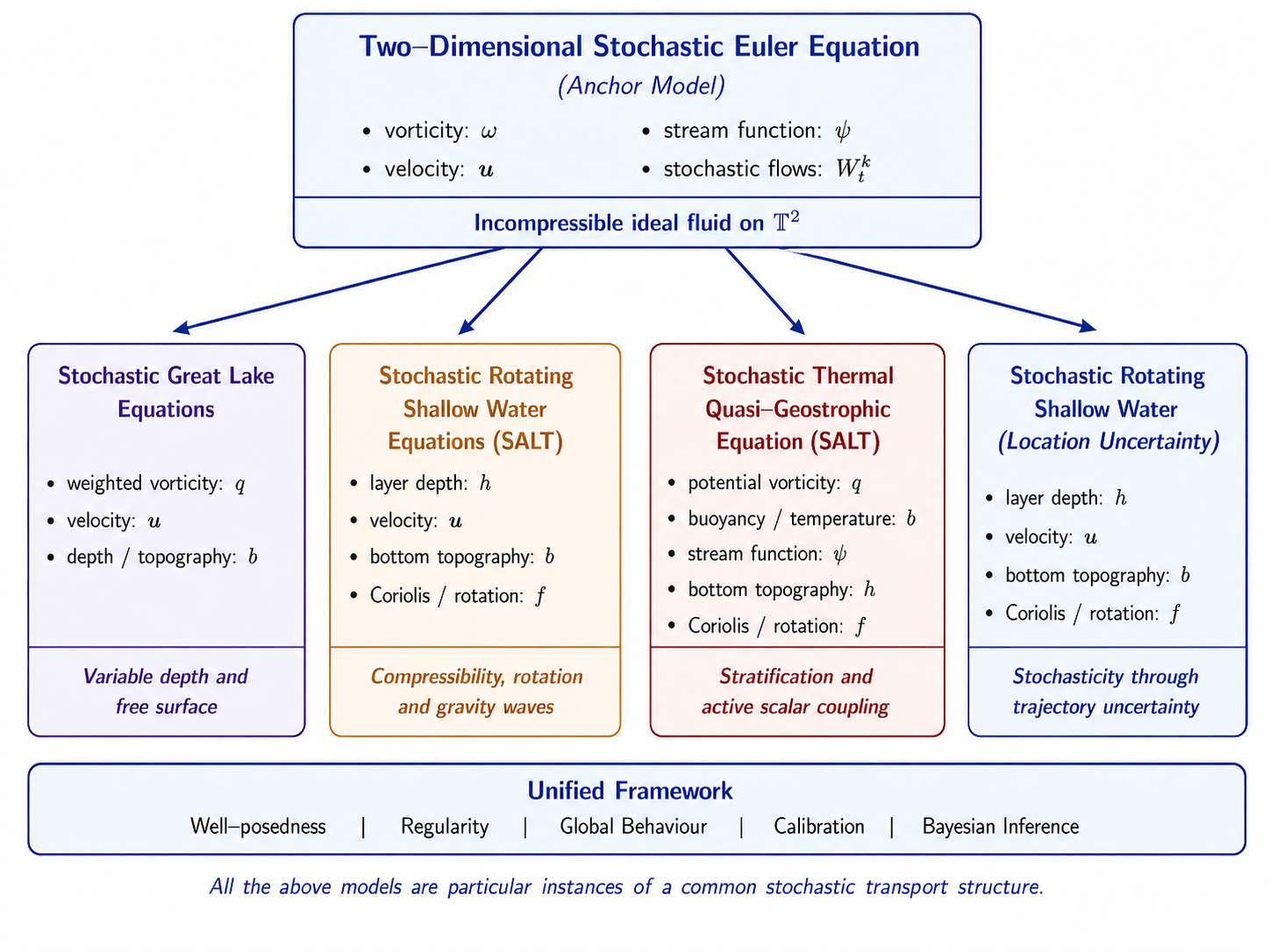}
	\caption{The two-dimensional stochastic Euler equation is a prototype for the analytical framework. Other stochastic transport models considered here are obtained by enriching the physical model while preserving the common transport structure.}
	\label{fig:stuod-unified-framework}
\end{figure}

The stochastic Euler equation serves as the natural starting point for the
analytical developments reviewed in this paper. Although we consider progressively more sophisticated geophysical fluid
models, including the Great Lake equations, the rotating shallow water
equations and the stochastic thermal quasi-geostrophic equations, the
Euler equation already contains the principal mathematical difficulties
associated with stochastic transport. In particular, the stochastic
perturbations are first-order Lie derivative operators acting at the same
differential order as the nonlinear advection term. Consequently, the
equation lies outside the scope of classical semilinear SPDE theory,
and the analytical framework must be developed from first principles.

From the mathematical perspective, the two-dimensional stochastic Euler
equation provides the simplest setting in which the essential ingredients
of the analytical programme can be developed. These include
localisation through truncated equations, compactness arguments at the
level of probability laws, Skorohod representation, pathwise uniqueness,
the Yamada--Watanabe principle, logarithmic Sobolev estimates and
continuation arguments leading to global solutions. The geometric structure
of the SALT formulation also gives rise to fundamental transport
cancellations that underpin the energy estimates used throughout the
subsequent analysis.

Figure~\ref{fig:stuod-unified-framework} illustrates the progression of the analytical developments within the STUOD programme. The two-dimensional stochastic Euler equation occupies a central position, serving as the prototype stochastic transport equation from which the principal analytical techniques were developed. These include the localisation strategy, compactness arguments, pathwise uniqueness, continuation principles and global well-posedness results that underpin the unified analytical framework presented in this paper. The subsequent models—the stochastic Great Lake equations, the stochastic rotating shallow water equations with SALT transport noise, the stochastic thermal quasi-geostrophic equations, and the rotating shallow water equations under the Location Uncertainty paradigm—introduce progressively richer physical mechanisms, including variable bathymetry, compressibility, rotation and buoyancy, while preserving the common stochastic transport structure. 

\subsection{Solution concepts}
\label{subsubsec:02_00_06_solution_concepts}

Different parts of the analytical developments require different notions of solution. We use the following terminology.

\paragraph{Strong probabilistic solutions.}

A process $q$ is a strong probabilistic solution of \eqref{eq:abstract-stratonovich} if it is defined on a prescribed stochastic basis, is adapted to $(\F_t)_{t\geq0}$, has the required spatial regularity, and satisfies the equation in the appropriate strong,  weak or distributional sense.

\paragraph{Martingale solutions.}

A martingale solution consists of a stochastic basis, stochastic drivers, and an adapted process satisfying the equation in law. Such solutions often arise from compactness arguments and are then upgraded to stronger notions when pathwise uniqueness is available.

\paragraph{Classical solutions.}

When the initial data and coefficients are sufficiently smooth, the solution may satisfy the equation pointwise in space, or in an equivalent classical sense. Such results are usually obtained after proving strong Sobolev well-posedness.

\paragraph{Mild solutions.}

For semigroup-based formulations, one writes the equation in integral form. If $S(t)$ denotes the semigroup generated by the linear part, then a mild solution satisfies a formula of the type
\[
q_t=S(t)q_0
-\int_0^t S(t-s)\A(q_s)\,ds
-\sum_i\int_0^t S(t-s)\B_i(q_s)\circ dW_s^i.
\]
Mild formulations are especially useful in extensions involving analytic semigroups, Young integration or sewing methods.

\paragraph{Pathwise strong solutions.}

When the stochastic basis and driving noises are fixed and pathwise uniqueness holds, a solution is called pathwise strong. This terminology will be used when emphasising uniqueness and continuous dependence on the initial condition.

\subsection{Variational and geometric notation}
\label{subsubsec:02_00_07_variational_and_geometric_notation}

Several STUOD developments show that stochastic transport models arise naturally from variational principles. In this setting the basic object is a stochastic flow map $g_t$, and the Lagrangian trajectory satisfies
\[
dg_t \circ g_t^{-1}
=
u_t\,dt+\sum_{i=1}^{\infty}\xi_i\circ dW_t^i.
\]
The vector field
\[
d\chi_t := u_t\,dt+\sum_{i=1}^{\infty}\xi_i\circ dW_t^i
\]
will be used as a shorthand for the stochastic transport velocity.

For advected quantities $a_t$, the geometric transport law is expressed as
\[
da_t + \Lie_{d\chi_t}a_t = 0.
\]
This notation is intentionally compatible with both scalar transport and transport of tensorial quantities or differential forms. It also provides the bridge to the semimartingale, rough-path and fractional formulations discussed later in the paper.

\subsection{Calibration notation}
\label{subsubsec:02_00_08_calibration_notation}

The stochastic transport fields $\xi_i$ are not merely analytical devices. In the STUOD programme they are intended to represent unresolved transport structures inferred from data. We therefore distinguish between:\\[2mm]
$\bullet$ the resolved velocity $u$;\\
$\bullet$ the stochastic transport fields $(\xi_i)_{i\geq1}$;\\
$\bullet$ the stochastic amplitudes or driving paths;\\
$\bullet$ the observed or high-resolution reference data used for calibration.\\

\indent A generic calibration problem may be written schematically as:
\[
\text{given data } \mathcal D,\qquad
\text{infer } \{\xi_i\}_{i=1}^r
\text{ and possibly their amplitudes}.
\]
In finite-dimensional or reduced-order models, the stochastic dynamics may take the form
\[
dX_t = F(X_t)\,dt+\sum_{i=1}^r G_i(X_t)\circ dW_t^i,
\]
where the vector fields $G_i$ are obtained from a projection of the stochastic transport structure or learned from data.

\subsection{Observation and filtering notation}
\label{subsubsec:02_00_09_observation_and_filtering_notation}

For the data assimilation sections, we use $q_t$ for the true or modelled state and $Y_t$ for the observations. In continuous time a standard observation model is
\[
dY_t = h(q_t)\,dt + dV_t,
\]
where $h$ is the observation function and $V_t$ is an observation noise process. In discrete time, observations are denoted by
\[
Y_n = H(q_{t_n})+\eta_n,
\]
where $H$ is an observation operator and $\eta_n$ is observational noise.

The filtering distribution is denoted by
\[
\pi_t(\varphi)
=
\E\bigl[\varphi(q_t)\mid \mathcal Y_t\bigr],
\]
where $\mathcal Y_t$ is the observation sigma-algebra and $\varphi$ is a test function. In particle methods, $\pi_t$ is approximated by an empirical measure
\[
\pi_t^N = \sum_{j=1}^N w_t^j\,\delta_{q_t^j},
\]
where $q_t^j$ are particles and $w_t^j$ are weights. When all particles have equal weight, this reduces to
\[
\pi_t^N = \frac1N\sum_{j=1}^N \delta_{q_t^j}.
\]
Further details for the filtering framework can be found in \cite{BainCrisan2009}, \cite{CrisanRozovsky2011}.
This notation will be used in the discussion of high-dimensional particle filters, localisation, tempering, nudging and jittering.

\subsection{Summary of notation and framework}
\label{subsubsec:02_00_10_summary_of_notation}

\begin{center}
\begin{tabular}{ll}
\hline
Symbol & Meaning \\
\hline
$(\Omega,\F,(\F_t),\Prob)$ & filtered probability space \\
$W^i$ & independent Brownian drivers \\
$\T^d$ & $d$-dimensional torus \\
$H^k$ & Sobolev space $W^{k,2}$ \\
$\xi_i$ & stochastic transport vector fields \\
$\Lie_a$ & transport operator $a\cdot\nabla$ \\
$q_t$ & generic transported state variable \\
$\A(q)$ & deterministic nonlinear drift \\
$\B_i(q)$ & stochastic transport operator \\
$d\chi_t$ & stochastic transport velocity \\
$\omega$ & vorticity \\
$u$ & velocity field \\
$Y_t$ & observation process \\
$\pi_t$ & filtering distribution \\
$\pi_t^N$ & particle approximation of $\pi_t$ \\
\hline
\end{tabular}
\end{center}

\noindent The framework introduced in this section plays three roles:
First, it provides the notation for the analytical developments in Section 3. The stochastic Euler vorticity equation is the anchor example, but the same structure appears in the Great Lake equation, thermal quasi-geostrophic dynamics, rotating shallow water models and other stochastic geophysical systems.
Second, it provides the geometric language for Section 4. The expression
\[
da_t+\Lie_{d\chi_t}a_t=0
\]
is the common starting point for semimartingale variational principles, rough-path formulations and fractional transport extensions.
Third, it provides the state-space notation for Sections 5 and 6. Calibration concerns the estimation of the transport fields $\xi_i$ or their finite-dimensional analogues, while data assimilation concerns the conditional distribution of $q_t$ given observations.

\section{Analytical Foundations of Stochastic Transport Equations}
\label{sec:03_analytical_foundations_of_stochastic_transport_equations}

This section develops the analytical backbone of the paper. It is centred on the stochastic Euler equation with transport noise, which provides the prototype for which provides the prototype for the subsequent analytical developments. The emphasis is on identifying the common analytical ideas that underpin further extensions to the Great Lake equations, thermal quasi-geostrophic dynamics and stochastic rotating shallow water models.

\subsection{The analytical challenge}
\label{subsec:03_01_the_analytical_challenge}

From an analytical perspective, stochastic transport equations differ fundamentally from the semilinear parabolic SPDEs that form the basis of much of the classical theory. In the latter, the stochastic perturbation is typically a bounded or Lipschitz operator acting on a parabolic equation whose diffusion generates regularisation and smoothing. In contrast, the stochastic perturbations arising from SALT and Location Uncertainty are first-order differential operators corresponding to Lie derivatives. As already stated, these operators are unbounded, are of the same differential order as the nonlinear transport terms, and provide no additional regularisation. Moreover, the It\^o correction is itself a second-order operator constructed from compositions of Lie derivatives, but it is generally not uniformly elliptic and hence does not generate the smoothing properties associated with classical diffusion operators. 

The mathematical framework traditionally used for nonlinear stochastic
evolution equations is the \emph{variational approach}, based on a
\emph{Gelfand triple}
\[
V\hookrightarrow H\hookrightarrow V^{*},
\]
where \(V\) is a reflexive Banach space densely and continuously embedded
into the Hilbert space \(H\), and \(V^{*}\) denotes the dual of \(V\).  Under suitable conditions, this framework provides a powerful existence and uniqueness theory for a broad class of nonlinear stochastic partial differential equations. Standard references include \cite{PrevotRockner2007}, \cite{LiuRockner2015}, and \cite{KrylovRozovskii1981}. However, the stochastic transport equations arising in stochastic fluid
dynamics and throughout the STUOD programme fall largely outside this variational setting. The principal difficulty is that the transport operator is a first-order differential operator which 
does not satisfy the coercivity or monotonicity assumptions required by
the Gelfand triple theory.  Consequently, the analytical techniques used throughout these proceedings
are necessarily different from those employed in the classical theory of SPDEs, see, e.g., \cite{DaPratoZabczyk1992} for details. Instead, one must exploit the geometric structure of the transport operators, derive delicate commutator and energy estimates, establish compactness at the level of probability laws, and combine localisation arguments with pathwise uniqueness to obtain strong probabilistic solutions. These analytical challenges motivate the development of the well-posedness theory presented in the remainder of this paper.

The principal objective of the analytical programme is therefore to develop a well-posedness theory adapted specifically to transport-noise SPDEs while preserving the stochastic transport structure inherited from the deterministic equations.

\subsection{The incompressible inviscid fluid model}
\label{subsec:03_02_crisan_lang_framework}

The starting point is the stochastic two-dimensional Euler equation in vorticity form
\[
d\omega
+u\cdot\nabla\omega\,dt
+\sum_{i=1}^{\infty}\xi_i\cdot\nabla\omega\circ dW_t^i
=0,
\]
where the velocity is recovered from the vorticity through the Biot--Savart law. This equation is regarded throughout the paper as the prototype stochastic transport equation.

The key contribution of the paper \cite{CrisanLangEuler2023} is the construction of a complete analytical framework for this equation without imposing artificial isotropy assumptions on the transport fields. The analysis combines transport cancellations, Sobolev energy estimates, compactness arguments and pathwise uniqueness to obtain well-posedness results in natural function spaces.

\subsection{Energy estimates and transport cancellations}
\label{subsec:03_03_energy_estimates}

The proof of well-posedness for stochastic transport equations differs substantially from the classical theory for deterministic evolution equations and semilinear SPDEs. The stochastic perturbations are
first-order differential operators acting at the same order as the
nonlinear transport term, so no parabolic regularisation is available.
Instead, the analysis relies on a combination of geometric transport
identities, a priori estimates, localisation, compactness and probabilistic
representation theorems. Although the details vary from one model to
another, the analytical strategy developed throughout the analytical programme
follows the same general pattern, which is summarised below.

The first cancellation is the familiar skew-adjointness of Lie transport.
If the transport vector field $\xi$ is divergence free, then the Lie
derivative satisfies
\[
\langle \mathcal L_\xi f,g\rangle_{L^2}
=
-
\langle
f,\mathcal L_\xi g
\rangle_{L^2},
\]
and consequently
\[
\langle
\mathcal L_\xi f,f
\rangle_{L^2}
=
0.
\]
Thus, exactly as in the deterministic Euler equation, the stochastic
transport operator does not contribute directly to the basic
$L^2$-energy estimate.

An important observation concerns the interaction between the It\^o
correction and the quadratic-variation term. Since $\Lie_\xi$ is
skew-adjoint when $\xi$ is divergence-free, one has the exact identity
\[
\langle \Lie_\xi^2 f,f\rangle_{L^2}
+
\|\Lie_\xi f\|_{L^2}^2
=0.
\]
Thus, the second-order contribution generated by the It\^o correction
is cancelled by the corresponding quadratic-variation term.
After differentiation, the same cancellation applies to the
highest-order contributions, while the remaining commutator terms
can be controlled by the Sobolev norm of $f$.

Applying second-order derivatives to the equation and
commuting them through the Lie derivative yields the higher-order
estimate
\[
\langle
\partial^\alpha\mathcal L_\xi^2f,
\partial^\alpha f
\rangle
+
\|
\partial^\alpha\mathcal L_\xi f
\|_{L^2}^2
\le
C\|f\|_{W^{2,2}}^2,
\qquad |\alpha|=2,
\]
which again involves no loss of derivatives. This estimate provides the
Sobolev control required for the existence theory and ultimately leads to
the construction of strong solutions.

A further ingredient is provided by commutator estimates. Differentiating
the stochastic transport equation generates terms involving
$
[\partial^\alpha,\mathcal L_\xi]f,
$
which can be estimated entirely in terms of the Sobolev norm of $f$,
without introducing derivatives of higher order than those already
present. Consequently, differentiation of the equation does not destroy
the transport structure or lead to a loss of regularity.

Taken together, these identities generate a set of
Sobolev energy estimates that control the nonlinear terms uniformly
throughout the approximation procedure. They constitute the cornerstone of the well-posedness theory developed for the stochastic Euler
equation and reappear, in various forms, throughout the analytical developments,
including the analyses of the stochastic Great Lake equations, the
stochastic thermal quasi-geostrophic equation, the stochastic rotating
shallow water equations under both the SALT and Location Uncertainty
paradigms, and the general abstract theory of transport-noise SPDEs.

\subsection{Approximation and compactness}
\label{subsec:03_04_approximation_and_compactness}

Relative compactness arguments for SPDEs are substantially more delicate than their deterministic PDE counterparts. In deterministic compactness methods, one typically obtains uniform bounds for a sequence of approximate solutions in suitable function spaces and then extracts a subsequence converging in a topology strong enough to identify the nonlinear terms. For SPDEs, the approximate solutions are random variables taking values in path spaces, and compactness is usually first obtained at the level of their laws. Thus one proves tightness of the sequence of probability measures induced by the approximate processes, rather than direct relative compactness of the sample paths themselves. This is the stochastic analogue of compactness, and it is one of the reasons why stochastic Aubin--Lions type arguments play a central role in the analysis of transport-noise SPDEs. 

Once tightness has been established, one typically uses a Skorokhod representation theorem, or one of its variants, to realise a subsequence on a new probability space on which almost-sure convergence holds. This step converts convergence in law into pathwise convergence, but at a price: the limiting process is initially only a martingale solution, defined on a newly constructed stochastic basis. Additional work is then needed to identify the limit as a solution of the original SPDE and to recover a solution driven by the prescribed noise on the original probability space. In many well-posedness arguments this is achieved by combining pathwise uniqueness with a Yamada--Watanabe type principle: compactness and Skorokhod representation yield existence of a martingale solution, while pathwise uniqueness upgrades this to strong probabilistic well-posedness. This mechanism is particularly important for stochastic transport equations, where the noise coefficients are unbounded first-order differential operators and where compactness must be combined with the cancellation structure of the transport terms, see \cite{CrisanLangEuler2023}, 
\cite{GoodairCrisanLang2024}.

We need the following definition in order to state the main well-posedness result for the Euler equation.

\begin{definition}[Local, maximal and global solutions]
	\label{def:local-maximal}
	
	Let $k\ge2$ and let
	$\omega_0\in W^{k,2}(\mathbb T^2)$
	be $\mathcal F_0$-measurable.
	
	A pair $(\omega,\tau)$ is called a \emph{local solution} of the
	stochastic Euler equation if $\tau$ is an $\{\mathcal F_t\}$-stopping time with
		$\tau>0$ almost surely;
		$
		\omega(\cdot\wedge\tau)
		\in
		C\!\left([0,\infty);W^{k,2}(\mathbb T^2)\right)
		\quad\text{a.s.},
		$
		and for every $t\ge0$, the stopped process
		$\omega(\cdot\wedge\tau)$ satisfies the integral formulation of the
		stochastic Euler equation on $[0,t\wedge\tau]$ almost surely.
	
	A pair $(\omega,\tau_{\max})$ is called a \emph{maximal solution} of the
	stochastic Euler equation if $\tau_{\max}$ is an $\{\mathcal F_t\}$-stopping time for which there exists an increasing sequence of
	stopping times $(\tau_n)_{n\ge1}$ such that
	$
	\tau_n\uparrow\tau_{\max}$ almost surely,
	and each pair $(\omega,\tau_n)$ is a local solution, and
	\[
	\limsup_{t\uparrow\tau_{\max}}
	\|\omega(t)\|_{W^{k,2}}
	=
	\infty
	\qquad
	\text{almost surely on }
	\{\tau_{\max}<\infty\}.
	\]
	
	A maximal solution is called a \emph{global solution} if $
	\tau_{\max}=\infty$
	almost surely.
\end{definition}

The objective of the well-posedness theory is therefore to construct a maximal solution and then, when possible,  establish a priori estimates showing that the maximal lifetime satisfies $\tau_{\max}=\infty$ almost surely. The latter implies the existence of a global solution. Combined with the pathwise-uniqueness result established below, this yields a unique global solution.

For the stochastic Euler equation, this programme can be carried out
completely, yielding the existence of a unique global solution. 
For more complex stochastic fluid models, however, such global a priori
estimates are generally unavailable. This is the case, for example, for
the stochastic rotating shallow water equations, the Great Lake equations,
and several classes of stochastic compressible fluid models. In these
situations one typically establishes the existence and uniqueness of a
maximal solution together with an associated blow-up criterion, while the
question of whether the maximal lifetime is almost surely infinite remains
open. Developing techniques capable of ruling out finite-time blow-up for
these nonlinear stochastic transport equations remains one of the major
analytical challenges in stochastic fluid dynamics.

\begin{theorem}[Fundamental Existence Theorem]
	\label{thm:EulerExistence}
	
	Let $k\ge2$ and let
	$
	\omega_0\in W^{k,2}(\mathbb T^2).
	$
	Assume that the stochastic transport vector fields
	$\{\xi_i\}_{i\ge1}$ are divergence-free and satisfy
	\[
	\sum_{i=1}^{\infty}
	\|\xi_i\|_{W^{k+1,\infty}}^2<\infty .
	\]
	Then the stochastic Euler equation
	\begin{equation}
		d\omega_t
		+
		u_t\cdot\nabla\omega_t\,dt
		+
		\sum_{i=1}^{\infty}
		(\xi_i\cdot\nabla\omega_t)\circ dW_t^i
		=
		0
		\label{eq:Euler}
	\end{equation}
	admits a global
	$\{\mathcal F_t\}$-adapted solution
	$
	\omega
	\in
	C\!\left(
	[0,\infty);
	W^{k,2}(\mathbb T^2)
	\right),
	$
	almost surely.
	Moreover, if $k\ge4$, then the solution is classical.
\end{theorem}

\paragraph{Outline of the proof of the Fundamental Existence Theorem.}

The proof proceeds by localisation. The first step is not to solve the
full nonlinear stochastic Euler equation directly, but to introduce a
truncated equation in which the nonlinear drift is cut off at large
Sobolev norm. More precisely, one replaces the Euler transport term by a
globally controlled version, for example of the form
\[
u\cdot\nabla\omega
\quad\leadsto\quad
\theta_n(\|\omega\|_{W^{k-1,2}})\,u\cdot\nabla\omega,
\]
where $\theta_n$ is a smooth cut-off function equal to one below level
$n$ and zero above a larger level. The truncated equation has coefficients
with better global bounds, and one can construct a global solution to this
modified problem.

The next step is to obtain estimates for the truncated equation that are
uniform up to the stopping time
\[
\tau_n
=
\inf\{t\ge0:\|\omega_t^n\|_{W^{k-1,2}}\ge n\}.
\]
On the interval $[0,\tau_n)$ the cut-off is inactive, and therefore the
solution of the truncated equation is also a solution of the original
stochastic Euler equation. This gives a local solution of the original
equation up to the stopping time $\tau_n$.

The construction of the truncated solution uses compactness methods. One
first solves regularised or finite-dimensional approximations of the
truncated equation and derives uniform Sobolev estimates. These estimates
imply tightness of the laws of the approximate processes in appropriate
path spaces. A Skorokhod representation theorem is then used to realise a
convergent subsequence on a new probability space and to identify the
limit as a martingale solution of the truncated equation.

Finally, pathwise uniqueness for the truncated equation allows one to
upgrade the martingale solution to a pathwise solution on the original
stochastic basis by a Yamada--Watanabe type argument. Since the cut-off is
inactive before $\tau_n$, this produces a pathwise solution of the
original stochastic Euler equation on $[0,\tau_n)$. The local solutions
obtained for increasing values of $n$ are then patched together. The following 
global a priori estimate shows that the corresponding explosion time is
infinite, and hence the local solution extends to a global solution.

\begin{proposition}[Logarithmic bound and global solution]
	\label{prop:log-bound-globalisation}
	
	Assume the hypotheses of Theorem~\ref{thm:EulerExistence}.
	For each $n\in\mathbb N$, let $\omega^n$ denote the global solution
	of the truncated stochastic Euler equation, and define
	\[
	\tau_n
	:=
	\inf\left\{
	t\ge0:
	\|\omega_t^n\|_{W^{k-1,2}}\ge n
	\right\},
	\]
	with the convention $\inf\varnothing=\infty$.
	Assume that the truncated solutions are chosen consistently, so that
	the stopped solutions agree on their common intervals of existence and
	\[
	\tau_n\uparrow\tau_{\max}
	\qquad\text{almost surely}.
	\]
	Then, for every $T>0$, there exists a constant there exists a constant $C_T>0$, depending on $T$, the initial
	condition and the regularity and summability bounds on the transport
	fields, but independent of $n$, such that
	\[
	\mathbb E
	\left[
	\sup_{0\le t\le T}
	\log\left(
	e+\|\omega_t^n\|_{W^{k,2}}^2
	\right)
	\right]
	\le C_T.
	\]
	Consequently,
	\[
	\mathbb P(\tau_n\le T)
	\le
	\frac{C_T}{\log(e+n^2)}
	\longrightarrow 0
	\qquad
	\text{as }n\to\infty.
	\]
	It follows that
	\[
	\mathbb P(\tau_{\max}\le T)
	=
	\lim_{n\to\infty}\mathbb P(\tau_n\le T)
	=
	0
	\qquad
	\text{for every }T>0.
	\]
	Hence
	\[
	\tau_{\max}=\infty
	\qquad
	\text{almost surely},
	\]
	and therefore the maximal solution of the stochastic Euler equation is
	global.
\end{proposition}

\subsection{Pathwise uniqueness and stability}
\label{subsec:03_05_pathwise_uniqueness}

The compactness argument described in the previous subsection yields the existence of a martingale solution to the stochastic Euler equation. To identify this solution with the desired pathwise solution, one first establishes pathwise uniqueness. The proof relies on stability estimates that exploit the first-order transport structure of the noise and the associated cancellation properties. Once pathwise uniqueness has been obtained, the Yamada--Watanabe principle upgrades the martingale solution produced by the compactness argument to a unique strong probabilistic solution on the prescribed stochastic basis. We state the result in the following theorem.

\begin{theorem}[Stability and pathwise uniqueness]
	\label{thm:EulerUniquenessStability}
	
	Let $k\geq2$ and assume that the transport vector fields
	$\{\xi_i\}_{i\geq1}$ satisfy the hypotheses of
	Theorem~\ref{thm:EulerExistence}. Let
	$\omega$ and $\widetilde{\omega}$ be two global pathwise strong solutions of the
	stochastic Euler equation corresponding to initial data
	\[
	\omega_0,\widetilde{\omega}_0\in W^{k,2}(\mathbb T^2),
	\]
	driven by the same Brownian motions on the same stochastic basis.
	
	Define
	\[
	A_t
	:=
	\int_0^t
	\left(
	1+\|\omega_s\|_{W^{k,2}}
	\right)\,ds,
	\qquad t\geq0.
	\]
	Then there exists a constant $C>0$, independent of the two solutions,
	such that, for every $t\geq0$,
	\[
	\mathbb E
	\left[
	e^{-CA_t}
	\|\omega_t-\widetilde{\omega}_t\|_{W^{k-1,2}}^2
	\right]
	\leq
	\|\omega_0-\widetilde{\omega}_0\|_{W^{k-1,2}}^2.
	\]
	In particular, if $\omega_0=\widetilde{\omega}_0$, then
	$\omega$ and $\widetilde{\omega}$ are indistinguishable. Hence pathwise
	uniqueness holds on $[0,\infty)$.
	Moreover, the estimate gives weighted Lipschitz stability of the solution
	with respect to the initial condition in the
	$W^{k-1,2}(\mathbb T^2)$ norm.
\end{theorem}

The above stability result also provides continuous dependence on the initial condition, an essential property for later calibration and data assimilation.

\vspace{3mm}

The two stochastic transport formulations (SALT and LU) are equivalent for the two-dimensional incompressible Euler equations: after conversion from Stratonovich to Itô form they lead to the same stochastic vorticity equation and therefore have identical analytical properties. This coincidence relies on the special transport structure of the two-dimensional vorticity equation, where the nonlinear dynamics consist entirely of advection of a scalar by a divergence-free velocity field. Beyond this setting, however, the two approaches are genuinely different. In particular, for three-dimensional Euler equations the vortex stretching term causes the SALT and LU formulations to diverge, leading to different stochastic models. Similar differences arise for more complex geophysical fluid models, including the rotating shallow water and thermal quasi-geostrophic equations discussed below.

\subsection{Other well-posedness results}
\label{subsec:03_06_local_and_maximal_solutions}

The techniques developed for the stochastic Euler equation naturally extend to local and maximal solution theories for more general nonlinear stochastic fluid equations. These developments form the basis for the later work on stochastic rotating shallow water and related models.

\subsubsection{The three-dimensional stochastic Euler equation}
\label{subsec:3d-stochastic-euler}

The three-dimensional stochastic Euler equation introduced in
\cite{CrisanFlandoliHolm2019} is written in vorticity form. The Lie derivative of a vector field is defined by
\[
\mathcal L_\xi \omega
=
[\xi,\omega]
=
(\xi\cdot\nabla)\omega
-
(\omega\cdot\nabla)\xi .
\]
Then the Euler equation in 3D is given by 
\begin{equation}
	d\omega_t
	+
	\mathcal L_{u}\omega_t\,dt
	+
	\sum_{i=1}^{\infty}
	\mathcal L_{\xi_i}\omega_t\circ dW_t^i
	=
	0 .
	\label{eq:3d-euler-equation2-format}
\end{equation}
Thus, in the notation of the master equation \eqref{eq:abstract-stratonovich},
\[
q=\omega,
\qquad
\mathcal A(\omega)=\mathcal L_u\omega,
\qquad
\mathcal B_i(\omega)=\mathcal L_{\xi_i}\omega.
\]
This is the three-dimensional analogue of the two-dimensional stochastic Euler
vorticity equation, but the analytical theory is fundamentally different: in three
dimensions one obtains a maximal solution and a blow-up criterion, rather than
global well-posedness. Under the regularity and summability assumptions specified in \cite{CrisanFlandoliHolm2019}, the following result holds.

\begin{theorem}[Existence and uniqueness of a maximal solution]
	\label{thm:3d-euler-maximal}
	Let
	$
	\omega_0\in W^{2,2}_\sigma(\mathbb T^3;\mathbb R^3)
	$
	where
	\[
	W^{2,2}_\sigma(\mathbb T^3;\mathbb R^3)
	=
	\left\{
	\omega\in W^{2,2}(\mathbb T^3;\mathbb R^3):
	\nabla\cdot\omega=0,\;
	\int_{\mathbb T^3}\omega(x)\,dx=0
	\right\}.
	\]
Then there exists a maximal solution
	$
	(\omega,\tau_{\max})
	$
	of the three-dimensional stochastic Euler equation
	\eqref{eq:3d-euler-equation2-format}, where
	\[
	\omega\in
	C\!\left([0,\tau_{\max});W^{2,2}_\sigma(\mathbb T^3;\mathbb R^3)\right)
	\quad\text{almost surely}.
	\]
	Moreover, the maximal solution is pathwise unique. If
	$(\omega',\tau'_{\max})$
	is another maximal solution with the same initial condition and driven by the same Brownian motions, then
	\[
	\tau_{\max}=\tau'_{\max},
	\qquad
	\omega=\omega'
	\quad
	\text{on }[0,\tau_{\max})
	\]
	almost surely.	Finally, either
	$
	\tau_{\max}=\infty
	$
	or
	$
	\limsup_{t\uparrow\tau_{\max}}
	\|\omega_t\|_{W^{2,2}}
	=
	\infty.
	$
\end{theorem}

\begin{theorem}[Beale--Kato--Majda criterion]
	\label{thm:3d-euler-bkm}
	Let
	$
	\omega_0\in W^{2,2}_\sigma(\mathbb T^3;\mathbb R^3),
	$
	and let \((\omega, \tau_{\max})\) be the maximal solution of
	\eqref{eq:3d-euler-equation2-format}. Then
	$
	\tau_{\max}<\infty
	$
	if and only if 
	\[
	\int_0^{\tau_{\max}}
	\|\omega_t\|_{L^\infty}\,dt
	=
	\infty .
	\]
	In particular,
	$
	\limsup_{t\uparrow\tau_{\max}}
	\|\omega_t\|_{L^\infty}
	=
	\infty 
	$ if $\tau_{\max}<\infty.
	$
\end{theorem}

\paragraph{Discussion.}
The results presented in this subsection do not formally constitute one of the scientific outputs of the STUOD programme. They are included here for two closely related
reasons. First, 
the three-dimensional stochastic Euler equation illustrates the geometric
structure of SALT noise in its full vector-valued form, where stochastic
transport is expressed through Lie derivatives of vector fields rather than
through the scalar transport operator that appears in the two-dimensional
vorticity equation. Secondly, the comparison between the two- and
three-dimensional theories highlights a fundamental analytical difference.
While the two-dimensional stochastic Euler equation admits a unique global
Sobolev solution, the three-dimensional theory currently establishes the
existence and uniqueness of a maximal solution together with a
Beale--Kato--Majda continuation criterion. This contrast illustrates both
the additional mathematical challenges posed by three-dimensional fluid
dynamics and the level of analytical results that can presently be obtained
for nonlinear stochastic transport equations.

\subsubsection{The stochastic Great Lake equations}
\label{subsec:great-lake}

One of the first extensions of the stochastic transport framework developed
for the Euler equation concerns the stochastic Great Lake equations. These
equations model incompressible flows over a basin with spatially varying
depth and therefore provide a more realistic description of large-scale
geophysical flows than the constant-depth Euler equations.

Let $b(x)$ denote the bottom topography, assumed to satisfy
\[
0<b_*\le b(x)\le b^*,
\]
and let $u$ denote the horizontal velocity field. The weighted
incompressibility condition is
\[
\nabla\cdot(bu)=0,
\]
which replaces the standard divergence-free constraint
$\nabla\cdot u=0$ of the Euler equations.

Introducing the weighted vorticity
\[
q=\frac{1}{b}\,\operatorname{curl}u,
\]
the stochastic Great Lake equation may be written in the transport form
\begin{equation}
	dq
	+
	u\cdot\nabla q\,dt
	+
	\sum_{i=1}^{\infty}
	\xi_i\cdot\nabla q
	\circ dW_t^i
	=0,
	\label{eq:great-lake}
\end{equation}
where the velocity is recovered from $q$ through the corresponding weighted
Biot--Savart operator.

Although equation \eqref{eq:great-lake} has the same transport structure as
the stochastic Euler equation, the presence of the depth function $b$
modifies the elliptic operator determining the velocity and introduces
additional analytical difficulties. In particular, the Sobolev estimates
must be adapted to weighted function spaces, while the transport
cancellations have to be reformulated to account for the weighted
divergence constraint.

\begin{theorem}[Existence and uniqueness of maximal solutions]
	\label{thm:great-lake}
		Let $k\geq 2$. Assume that the bottom topography
		$b:\mathbb T^2\to(0,\infty)$ is sufficiently smooth and satisfies
		\[
		0<b_*\leq b(x)\leq b^*,
		\qquad x\in\mathbb T^2,
		\]
		and let the transport vector fields
		$\{\xi_i\}_{i\geq1}$ satisfy the regularity, summability and weighted
		divergence conditions imposed in \cite{CrisanLangGreatLake2021}.
		For every initial vorticity
		$
		q_0\in W^{k,2}_b(\mathbb T^2),
		$
		the stochastic Great Lake equation
		\eqref{eq:great-lake}
		admits a unique maximal solution
		$
		(q,\tau_{\max})
		$
		of the stochastic Great Lake equation
		\eqref{eq:great-lake}
		such that
		$
		q
		\in
		C\!\left(
		[0,\tau_{\max});
		W^{k,2}
		\right)
		$ almost surely.
		\end{theorem}
	
	\paragraph{Discussion.}
	
	The significance of this result is not merely that it extends the
	well-posedness theory from the Euler equation to a more realistic
	geophysical model. Rather, it demonstrates that the analytical framework
	developed for stochastic transport equations is sufficiently robust to
	accommodate weighted transport structures arising from variable bathymetry.
	The compactness arguments, localisation procedure and pathwise uniqueness
	estimates developed for the stochastic Euler equation continue to apply,
	although substantial modifications are required to deal with the weighted
	elliptic operator and the associated Sobolev estimates. This paper
	therefore provides the first indication that the analytical framework
	introduced in Section 2 extends beyond idealised incompressible flows to
	the more general stochastic transport models encountered in geophysical
	fluid dynamics.

\subsubsection{Stochastic rotating shallow water equations with SALT transport noise}
\label{enough3}

The rotating shallow water (RSW) equations constitute one of the fundamental reduced
models of geophysical fluid dynamics. They retain the essential balance between nonlinear
advection, gravity waves and the Coriolis force while remaining considerably simpler than
the primitive equations. Within STUOD, they served as the first compressible geophysical
model for which a complete analytical theory was developed in the presence of (SALT) noise \cite{CrisanLangRSW2024}.
Following the SALT framework,  unresolved transport is represented
through stochastic perturbations of the Lagrangian flow map. The resulting stochastic rotating shallow water system
takes the Stratonovich form
\begin{align}
	dv_t
	&+
	\Big(
	L_{u_t}v_t
	+
	f\hat z\times u_t
	+
	\nabla p_t
	\Big)\,dt
	+
	\sum_{i=1}^{\infty}(L_i+A_i)v_t
	\circ dW_t^i
	=
	\nu\Delta v_t\,dt,
	\\
	dh_t
	&+
	\nabla\!\cdot(h_tu_t)\,dt
	+
	\sum_{i=1}^{\infty}
	\nabla\!\cdot(\xi_i h_t)
	\circ dW_t^i
	=
	\eta\Delta h_t\,dt,
\end{align}
where
\[
L_i v=\xi_i\cdot\nabla v,
\qquad
A_i v
=
v_1\nabla\xi_i^1
+
v_2\nabla\xi_i^2,
\]
and $v=u+R$ with $\operatorname{curl}R=f\hat z$.  The operators
$L_i$ and $A_i$ arise naturally from the SALT transport formulation and preserve the
underlying geometric transport structure of the deterministic equations. 

The well-posedness analysis is considerably more difficult than for the stochastic Euler equations.
Unlike the incompressible case, the pressure cannot be eliminated through a vorticity
formulation because it depends on the free-surface height. Consequently, the velocity and
height equations remain strongly coupled, and there is no transport estimate for the
vorticity alone. The proof therefore relies on a viscous regularisation together with a
carefully constructed truncation argument, stochastic energy estimates and compactness
methods. 

The principal analytical result establishes the local well-posedness of the SALT rotating
shallow water equations.

\begin{theorem}[Existence and uniqueness of maximal solutions]
	Suppose that
	$
	a_0=(v_0,h_0)\in (W^{1,2}(\mathbb T^2))^3.
	$
	Then the stochastic rotating shallow water system admits a unique pathwise maximal
	solution
	\[
	a=(v,h)
	\in
	L^2\!\left(
	\Omega;
	C([0,T);W^{1,2})
	\right)
	\cap
	L^2\!\left(
	\Omega;
	L^2(0,T;W^{2,2})
	\right),
	\]
	which is unique and depends continuously on the initial data. Moreover, the maximal
	solution satisfies the usual blow-up alternative: if the maximal existence time is finite,
	then
	\[
	\limsup_{t\uparrow T}
	\|a(t)\|_{W^{1,2}}
	=
	\infty.
	\]
\end{theorem} 

The second principal result concerns global existence for sufficiently small initial data.
Although the stochastic perturbation prevents proving almost sure global existence, the
authors establish that the stochastic dynamics remain globally well posed with strictly
positive probability.

\begin{proposition}[Global existence with positive probability]
	There exists a constant $C>0$ such that whenever
	$
	\|a_0\|_{W^{1,2}}<C,
	$
	the maximal solution satisfies
	$
	\mathbb P(T=\infty)>0.
	$
	In other words, sufficiently small initial conditions generate global solutions with positive
	probability.
\end{proposition} 

This work represented the first complete well-posedness theory within STUOD for a
compressible geophysical fluid model driven by SALT transport noise. Beyond its intrinsic
analytical interest, it provided the mathematical foundation for the subsequent Bayesian
data assimilation framework developed for the stochastic rotating shallow water equations.
The well-posedness theory guarantees that the signal process defining the Bayesian inverse
problem is itself well defined, thereby allowing the associated nonlinear filtering and
parameter estimation problems to be formulated rigorously. This work
provides the well-posed stochastic dynamics on which the probability
measure evolution and the corresponding master equation are built.
	
\subsubsection{The Stochastic Thermal Quasi-Geostrophic Equation}

The thermal quasi-geostrophic (TQG) equations arise as an asymptotic approximation of the thermal rotating shallow-water
equations in the regime of small Rossby number and incorporate the interaction
between buoyancy gradients, potential vorticity and bottom topography.
Unlike the classical quasi-geostrophic equations, the TQG model retains the
coupling between the buoyancy field and the potential vorticity and therefore
captures important submesoscale transport phenomena. The analysis for the TQG is covered in \cite{CrisanHolmLuesinkMensahPanTQGComputational2023}. Its stochastic counterpart is analysed in 
\cite{CrisanHolmLangMensahPanTQG2023}, where   
the stochasticity is introduced through the
 SALT framework. 

Let $b(x,t)$ denote the buoyancy,
$q(x,t)$ the potential vorticity and
$\psi(x,t)$ the stream function.
The velocity field is

\[
u=\nabla^\perp\psi,
\qquad
u_h=\frac12\nabla^\perp h,
\qquad
q=(\Delta-1)\psi+f,
\]
where $h$ denotes the bathymetry and $f$ the Coriolis parameter. The SALT stochastic thermal quasi-geostrophic equations are
\begin{align}
	db
	+(u\cdot\nabla)b\,dt
	+(\xi_i\cdot\nabla)b\circ dW_t^i
	&=0,
	\\
	dq
	+(u\cdot\nabla)(q-b)\,dt
	+(\xi_i\cdot\nabla)(q-b)\circ dW_t^i
	&=
	-(u_h\cdot\nabla)b\,dt,
\end{align}
where $(\xi_i)_{i\ge1}$ are prescribed divergence-free transport
vector fields obtained from observational data.

The principal analytical result establishes that the stochastic system is
locally well posed in Sobolev spaces.

\begin{theorem}[Existence and uniqueness of maximal solutions]
	\label{thm:stqg-wellposedness}
	
	Assume
	\[
	b_0\in W^{3,2}(\mathbb T^2),
	\qquad
	q_0\in W^{2,2}(\mathbb T^2),
	\]
	and let the stochastic transport vector fields
	$\{\xi_i\}_{i\ge1}$ satisfy the regularity and summability assumptions
	specified in \cite{CrisanHolmLangMensahPanTQG2023}.
	Then the stochastic thermal quasi-geostrophic equations admit a unique
	maximal pathwise strong solution
	\[
	(b,q,\tau_{\max}),
	\]
	where $\tau_{\max}$ denotes the maximal lifetime of the solution.
	The solution is unique pathwise and is strong both in the probabilistic
	sense and as a solution of the stochastic partial differential equation.
\end{theorem}

The solution map is stable with respect to perturbations of the initial
condition.

\begin{theorem}[Continuous dependence]
	\label{thm:stqg-stability}
	
	The maximal solution depends continuously on the initial data in the
	Sobolev spaces
	$
	W^{3,2}(\mathbb T^2)
	\times
	W^{2,2}(\mathbb T^2).
	$
	More precisely, if
	$
	(b_0^{(n)},q_0^{(n)})
	\longrightarrow
	(b_0,q_0)
	$
	in
	$
	W^{3,2}(\mathbb T^2)
	\times
	W^{2,2}(\mathbb T^2),
	$
	then the corresponding maximal solutions converge in probability on every
	compact time interval contained in their common interval of existence.
\end{theorem}

As for the stochastic Euler equation, the local solution satisfies a
continuation criterion identifying the only possible mechanism through
which regularity can be lost.

\begin{theorem}[Beale--Kato--Majda criterion]
	\label{thm:stqg-bkm}
	
	Assume
	$
	b_0\in W^{3,2}(\mathbb T^2),
	\qquad
	q_0\in W^{2,2}(\mathbb T^2),
	$
	and let the stochastic transport vector fields
	$\{\xi_i\}_{i\ge1}$ satisfy the regularity and summability assumptions
	specified in \cite{CrisanHolmLangMensahPanTQG2023}
	and let $
	(b,q,\tau_{\max}),
	$ be a maximal solution of the 
	 the stochastic thermal quasi-geostrophic equations. 
	Then
	$
	\tau_{\max}<\infty
	$
	if and only if 
	\[
	\int_0^{\tau_{\max}}
	\left(
	\|\nabla b(t)\|_{L^\infty}
	+
	\|q(t)\|_{L^\infty}
	\right)\,dt
	=
	\infty.
	\]
	\end{theorem}

The Beale-Kato-Majda criterion shows that finite-time singularities can occur
only through the loss of control of the Lipschitz norm of the buoyancy
field or the $L^\infty$ norm of the potential vorticity. As in the
stochastic Euler equation, global regularity is therefore reduced to
obtaining suitable a priori estimates for these quantities.

\subsubsection{A general class of rotating shallow water equations under the
	Location Uncertainty paradigm}
\label{subsec:LU-rsw}

The Location Uncertainty (LU) paradigm provides an alternative stochastic
representation of unresolved transport processes. In contrast with the SALT
framework, where stochasticity is introduced through stochastic Lie
transport, the LU formulation starts from a decomposition of the Lagrangian
velocity into a smooth resolved component and a rapidly fluctuating
unresolved component with prescribed covariance. Averaging over the
unresolved fluctuations produces both stochastic transport and additional
deterministic drift terms that depend explicitly on the covariance tensor
of the unresolved velocity field. These drift corrections model the
backscatter of energy from unresolved to resolved scales and give rise to
an inhomogeneous diffusion of the resolved flow.

A convenient general class of viscous LU rotating shallow water models is
given by
\begin{align}
	du_t
	&+
	\Big(
	\nabla\!\cdot(u_t\otimes u_t)
	+
	\alpha u_t(\nabla\!\cdot u_t)
	-
	u_t^S\!\cdot\nabla u_t
	+
	f\times u_t
	+
	g\nabla h_t
	\Big)dt
	\nonumber\\
	&\qquad
	+
	\sigma\!\cdot\nabla u_t\,dB_t
	=
	\nu\Delta u_t\,dt
	+
	\frac12\nabla\!\cdot(a\nabla u_t)\,dt,
	\label{eq:LU-general-u}
	\\
	dh_t
	&+
	\Big(
	\nabla\!\cdot(h_tu_t)
	+
	\beta h_t(\nabla\!\cdot u_t)
	-
	u_t^S\!\cdot\nabla h_t
	\Big)dt
	\nonumber\\
	&\qquad
	+
	\sigma\!\cdot\nabla h_t\,dB_t
	=
	\eta\Delta h_t\,dt
	+
	\frac12\nabla\!\cdot(a\nabla h_t)\,dt,
	\label{eq:LU-general-h}
\end{align}
where
\[
u^S=\frac12\nabla\!\cdot a,
\qquad
a=\sigma\sigma^T,
\]
denotes the It\^o--Stokes drift generated by the covariance tensor of the
unresolved velocity field. Here $\sigma$ is the correlation operator,
$a$ is the covariance tensor, while $\nu$ and $\eta$ are viscosity
coefficients. The constants $\alpha$ and $\beta$ define a family of
transport models, the particular LU rotating shallow water model analysed
in \cite{CrisanLangMemin2023} corresponding to the choice
\[
\alpha=-1,
\qquad
\beta=0.
\]
Introducing
\[
X=(u,h),
\]
system
\eqref{eq:LU-general-u}--\eqref{eq:LU-general-h}
can be written compactly as
\begin{equation}
	dX_t
	+
	\mathcal A(X_t)\,dt
	+
	\mathcal G(X_t)\,dB_t
	=
	\frac12\nabla\!\cdot(a\nabla X_t)\,dt
	+
	\gamma\Delta X_t\,dt,
	\label{eq:LU-master}
\end{equation}
where
\[
\mathcal A(X)
=
\begin{pmatrix}
	\nabla\!\cdot(u\otimes u)
	+\alpha u(\nabla\!\cdot u)
	-u^S\!\cdot\nabla u
	+f\times u
	+g\nabla h
	\\[2mm]
	\nabla\!\cdot(hu)
	+\beta h(\nabla\!\cdot u)
	-u^S\!\cdot\nabla h
\end{pmatrix},
\]
and
\[
\mathcal G(X)
=
\begin{pmatrix}
	\sigma\!\cdot\nabla u\\
	\sigma\!\cdot\nabla h
\end{pmatrix}.
\]
Consequently, the LU rotating shallow water equations constitute another
instance of the master equation introduced in
Section~\ref{subsubsec:02_00_04_canonical_stochastic_transport_equation}. Compared with the SALT formulation, the
principal difference is the appearance of the covariance-induced drift and
diffusion terms determined by the covariance tensor \(a=\sigma\sigma^T\).

The first principal analytical result establishes the local well-posedness
of the entire class of models (see Theorem 5 in \cite{CrisanLangMemin2023} for details).

\begin{theorem}[Existence and Uniqueness of of a Maximal Strong Solutions]
	Let $
	X_0\in W^{k,2}(\mathbb T^2)^3,$
	$k\ge1. $
	Then equation~\eqref{eq:LU-master} admits a unique pathwise maximal strong
	solution $
	(X,  \tau_{\max}).
	$
	Moreover, for every sequence of stopping times
	\((\tau_n)_{n\geq1}\) such that each
	\((X,\tau_n)\) is a local solution and
	$
	\tau_n\uparrow\tau_{\max}$
	almost surely,
	one has, for every \(T>0\),
	\[
	X_{\cdot\wedge\tau_n}
	\in
	L^2\!\left(
	\Omega;
	C\!\left([0,T];W^{k,2}(\mathbb T^2)^3\right)
	\right),
	\]
	and
	\[
	X\,\mathbf 1_{\{t\leq\tau_n\}}
	\in
	L^2\!\left(
	\Omega;
	L^2\!\left(0,T;W^{k+1,2}(\mathbb T^2)^3\right)
	\right).
	\]
\end{theorem}

The second result identifies a distinguished member of the family for
which global weak solutions exist (see Theorem 6 in \cite{CrisanLangMemin2023} for details).

\begin{theorem}[Global weak solution]
	Suppose
	$
	X_0\in L^2(\mathbb T^2)^3,
	$
	and choose
	$
	\alpha=\beta=-\frac12.
	$
	Then equation~\eqref{eq:LU-master} admits an
	\(L^2\)-valued continuous global weak solution satisfying
	\[
	\sup_{0\le s\le T}\|X_s\|_{L^2}^2
	+
	\int_0^T
	\|X_s\|_{W^{1,2}}^2\,ds
	\le
	\|X_0\|_{L^2}^2,
	\qquad
	\mathbb P\text{-a.s.}
	\]
\end{theorem}

\paragraph{Discussion.}

The proof of the well posedness result follows a different strategy from the compactness
arguments employed for the stochastic Euler, Great Lake and thermal
quasi-geostrophic equations. Rather than establishing relative compactness
of the laws of approximate solutions and invoking Prokhorov compactness,
the Skorohod representation theorem and the Yamada--Watanabe principle,
the analysis uses an approximating sequence of finite-dimensional stochastic differential
equations taking values in Littlewood--Paley spaces. Uniform Sobolev estimates are then established independently of
the approximation level, and the essential step consists in proving that
the approximate solutions form a Cauchy sequence in suitable stochastic
Sobolev spaces. This allows the strong solution to be constructed directly
as the limit of the approximations, without passing through martingale
solutions. Although conceptually close to deterministic Galerkin methods,
the stochastic Cauchy argument is substantially more delicate because one
must control not only the nonlinear drift but also stochastic integrals,
quadratic variation terms and It\^o corrections. The proof
combines Littlewood--Paley estimates, Burkholder--Davis--Gundy
inequalities and stochastic Gronwall arguments to obtain convergence of
the approximations.

\subsubsection{Other SALT Models}
\label{subsec:03_07_connections}

An important extension of the SALT framework to climate science is presented in \cite{CrisanHolmKorn2023}, where the authors develop a stochastic atmosphere--ocean model that provides a geometric implementation of Hasselmann's paradigm for stochastic climate modelling. Instead of introducing random forcing directly into the governing equations, stochasticity is incorporated into the fast atmospheric dynamics through stochastic advection by Lie transport (SALT), while the slow ocean component remains deterministic. This construction preserves Kelvin's circulation theorem and therefore retains the geometric structure of the underlying deterministic climate model. The paper also introduces a new Lagrangian-Averaged SALT (LA-SALT) formulation, in which the drift velocity of the stochastic transport is replaced by its expectation. A remarkable consequence of this averaging procedure is that the statistical moments of the climate variables satisfy a closed hierarchy of deterministic evolution equations, providing a mathematically consistent framework for studying climate statistics while retaining the physical interpretation of unresolved atmospheric variability. The authors establish local well-posedness of the deterministic, SALT and LA-SALT models, thereby placing this implementation of Hasselmann's programme on a rigorous analytical foundation.

A complementary development is the abstract well-posedness theory developed in
 \cite{GoodairCrisanLang2024}. Rather than studying a
particular stochastic fluid equation, the authors formulate a general framework for
semilinear stochastic evolution equations driven by first-order transport operators.
The abstract assumptions are sufficiently general to include a broad family of SALT
viscous models. Here the 
construction of maximal pathwise strong solutions is based on finite-dimensional
Galerkin approximations and a stochastic Cauchy convergence argument. 

\subsection{Global solutions of nonlinear evolution equations by stochastic control}

One of the most recent developments (see \cite{CrisanLang2025} for a more detailed account) arising from the analytical programme is the
observation that stochasticity can be used not only to model uncertainty but
also to improve the analytical properties of nonlinear evolution equations.
This philosophy differs substantially from both the SALT and the Location
Uncertainty paradigms introduced earlier. There, the stochastic perturbation is
introduced for physical reasons, namely to represent unresolved transport
processes while preserving the geometric structure of the underlying fluid
equations. Here, instead, the stochastic perturbation is chosen
analytically so that its quadratic variation counteracts the nonlinear growth
responsible for finite-time blow-up.

The resulting framework applies to a broad class of nonlinear
evolution equations, including the three-dimensional Navier--Stokes equation,
the three-dimensional Euler equation in vorticity form, viscous and inviscid
rotating shallow water equations, Burgers' equation and many other nonlinear
fluid models. Although the theory is formulated for deterministic partial
differential equations, the analytical arguments extend naturally to stochastic
transport equations and therefore provide a general stabilization mechanism for
transport-type SPDEs.

The starting point is the abstract evolution equation

\begin{equation}
	dX_t=A_t(X_t)\,dt,
	\label{eq:abstractPDE}
\end{equation}
where
$
A:[0,T]\times F_i\times\Omega\longrightarrow G
$
is a nonlinear operator acting between Hilbert spaces
$
D\hookrightarrow F_1\hookrightarrow F_0\hookrightarrow G,
$
with compact embeddings. Depending on the particular model, the space $F_0$, $F_1$ or $D$ becomes the
natural space of existence of the solution.
The central idea is to replace
\eqref{eq:abstractPDE} by the stochastic equation
\begin{equation}
	dX_t
	=
	A_t(X_t)\,dt
	+
	B_t(X_t)\,dW_t,
	\label{eq:abstractSPDE}
\end{equation}
where the diffusion coefficient is chosen as
\begin{equation}
	B_t(X_t)
	=
	\theta
	\|X_t\|_{F_i}^{\alpha}
	X_t,
	\label{eq:controlnoise}
\end{equation}
for suitable constants $\theta>0$ and $\alpha>0$.

The key observation is that if one applies It\^o's formula to a logarithmic
Lyapunov functional and control all the terms appearing in the resulting evolution equation. Consequently the logarithmic Lyapunov functional remains bounded,
which prevents blow-up of the solution.

This mechanism is fundamentally different from deterministic stabilization.
The stochastic perturbation does not act directly on the nonlinear drift.
Instead, stabilization is achieved through the second-order It\^o correction,
that is, through the quadratic variation of the Brownian motion.

The principal result of the work \cite{CrisanLang2025} is the following.

\begin{theorem}\label{gsdo}
	Under certain assumptions the stochastic evolution equation
	\[
	dX_t
	=
	A_t(X_t)\,dt
	+
	\theta
	\|X_t\|_{F_i}^{\alpha}
	X_t\,dW_t
	\]
	admits a global strong solution
	$
	X\in C([0,\infty);G),
	$
	and, for every $T>0$,
	\[
	\mathbb P
	\left(
	\sup_{0\le t\le T}
	\|X_t\|_G
	<
	\infty
	\right)
	=
	1.
	\]
	
\end{theorem}

These estimates imply that every local solution generated by the corresponding
deterministic equation becomes global after the addition of the stochastic
control.

The proof of Theorem \ref{gsdo} follows a similar strategy from earlier works. Galerkin approximations are first constructed in finite dimensions,
where global existence follows from classical finite-dimensional stochastic
analysis.
The principal analytical contribution is the derivation of logarithmic
Lyapunov estimates which remain uniform with respect to the Galerkin
dimension. Unlike the compactness arguments used in earlier work on the
rotating shallow water equations, the proof relies on pathwise logarithmic
estimates, tightness of the approximating sequence, and convergence of the
Galerkin solutions in suitable Hilbert spaces. The resulting bounds control
both the deterministic drift and the stochastic integral despite the latter
being only a local martingale.

The stabilization result fits naturally into the general framework introduced
earlier through the master equation \[
dX_t
=
A(X_t)\,dt
+
B(X_t)\,dW_t.
\]

Earlier sections considered diffusion operators $B$ arising from physical
modelling principles, such as SALT and Location Uncertainty. In \cite{CrisanLang2025}, the diffusion operator is instead constructed analytically in order to
improve the qualitative behaviour of the equation. Consequently the master
equation provides a unified framework in which physically motivated transport
noise, uncertainty quantification, and analytically designed stabilizing noise
appear simply as different choices of the diffusion operator.

\paragraph{Extension to stochastic transport equations.}

Although the results in \cite{CrisanLang2025} are formulated for deterministic evolution equations, the
arguments depend only on the availability of suitable pathwise energy
estimates, logarithmic Lyapunov functions, and compactness properties of the
Galerkin approximations. These ingredients are precisely those established in
the earlier sections of this review for stochastic transport equations driven
by SALT and Location Uncertainty noise.

Consequently, the stabilization mechanism developed here extends naturally to
a broad class of transport-type SPDEs. In that setting, the original transport
noise becomes part of the drift-diffusion operator defining the stochastic
fluid model, while the additional multiplicative control acts independently to
prevent blow-up. This suggests a new research direction in which physically
motivated stochastic transport models are combined with analytically designed
stabilizing noises to obtain globally well-posed stochastic fluid equations.

\section{Variational, Rough and Fractional Extensions}
\label{sec:05_variational_rough_and_fractional_extensions}

The previous section focused on the analytical foundations of stochastic transport equations. The purpose of the present section is to explain the variational origins of these equations and the subsequent extensions of the stochastic transport paradigm beyond semimartingale forcing.

Together these works extend the geometric foundations of stochastic fluid dynamics far beyond the original Brownian setting.

\subsection{Semi-martingale driven variational principles}
\label{subsec:semimartingale_variational}

The SALT framework discussed throughout the previous sections is rooted in
Hamilton's principle of stationary action. A rigorous mathematical
foundation for stochastic variational principles was developed in  \cite{StreetCrisan2021}, who introduced the notion of
\emph{semi-martingale driven variational principles}. Rather than deriving a
particular stochastic fluid model, their objective is to formulate a general
variational framework in which stochasticity enters directly at the level of
the action functional. This provides a common mathematical setting for a wide
class of stochastic continuum models, including the SALT equations
introduced by Holm.

In classical mechanics, the evolution of a system is determined by seeking
stationary points of the action
\[
\mathcal A(q)
=
\int_{t_0}^{t_1}
L(t,q,\dot q)\,dt,
\]
where \(q(t)\) denotes the configuration of the system, \(\dot q(t)\)
its time derivative, and
\(L:TQ\rightarrow\mathbb R\) is the Lagrangian defined on the tangent
bundle of the configuration manifold \(Q\).
For continuum mechanics it is more convenient to collect all physical
variables, together with any Lagrange multipliers used to impose
constraints, into a single variable
\[
v=v(t,x),
\]
where \(x\) denotes the spatial variable. The corresponding action then
takes the form
\[
\mathcal A(v)
=
\int_{t_0}^{t_1}
\widetilde L(t,v(t))\,dt,
\]
where \(\widetilde L\) denotes the spatially integrated Lagrangian.

In \cite{StreetCrisan2021} the deterministic time integration is replaced by integration 
with respect to a general continuous semi-martingale \(S_t\), leading to the
stochastic action
\[
\mathcal A(v)
=
\int_{t_0}^{t_1}
\widetilde L(t,v(t))
\circ dS_t,
\]
where the integration is understood in the Stratonovich sense. The
deterministic theory is recovered by choosing the driving semi-martingale to
be simply \(S_t=t\). This observation shows that deterministic and
stochastic variational principles belong to a single unified mathematical
framework.

A central concept introduced in \cite{StreetCrisan2021} is that of
\emph{compatibility}. Every dynamical variable, every Lagrangian and every
constraint is required to evolve with respect to the same driving
semi-martingale. Rather than viewing stochasticity as an external forcing
added to an already existing deterministic model, the randomness becomes an
intrinsic part of the variational structure itself. Compatibility guarantees
that all stochastic integrals appearing in the action are well defined and
that the various constraints remain consistent with the stochastic
evolution.

One of the principal mathematical contributions of the paper is the proof of
a stochastic analogue of the fundamental lemma of the calculus of
variations. This result establishes that variations of stochastic action
integrals may be computed in essentially the same manner as in the classical
deterministic theory. Consequently, the Euler--Lagrange and
Euler--Poincar\'e equations continue to arise as stationarity conditions of
the stochastic action functional, thereby placing previous stochastic
variational derivations on a rigorous mathematical foundation.

As an application of the general framework, the paper derives the stochastic
Euler--Poincar\'e equations with stochastic Lie transport constraints and
recovers the SALT equations introduced by Holm. The theory is further
illustrated by deriving stochastic incompressible Euler equations with the
correct stochastic pressure formulation, stochastic rotating shallow-water
equations, and stochastic wave--current interaction models. These examples
demonstrate that a wide variety of stochastic fluid equations can be
obtained from a common geometric variational principle.

\paragraph{Pressure Reconstruction}
\label{subsec:05_03_pressure_reconstruction}

One of the additional results of the semimartingale variational work in \cite{StreetCrisan2021} is the treatment of pressure. The paper shows that, in the stochastic
setting, the Lagrange multipliers themselves must be compatible with the
driving semi-martingale. As a consequence, the pressure enforcing
incompressibility is no longer purely deterministic but becomes a
semi-martingale containing both drift and martingale components. The paper
derives the corresponding stochastic pressure equation directly from the
variational principle, explaining why stochastic pressure terms necessarily
appear in incompressible SALT fluid models. This result is significant because it shows that the pressure is not merely a technical artifact of the incompressibility constraint. It is an intrinsic component of the stochastic variational structure.

\subsection{Variational Principles for Fluid Dynamics on Rough Paths}
\label{sec:rough_variational}

The semimartingale variational framework described in the previous section
provides a geometric derivation of stochastic fluid equations driven by
Brownian motion. Although Brownian semimartingales form a natural class of
driving signals, they represent only one example of the irregular
trajectories encountered in applications. Many physical systems exhibit
temporal correlations, long-range dependence or memory effects that cannot
be represented adequately within the semimartingale framework. Examples
include turbulent transport, climate variability and unresolved oceanic
motions. These considerations motivate the search for a geometric theory of
fluid dynamics driven by more general irregular signals while preserving
the variational structure of the underlying deterministic equations.

This problem is addressed in
\cite{CrisanHolmLeahyNilssen2022}, where the semimartingale theory is
extended to the framework of \emph{geometric rough paths}\footnote{Details of rough paths theory can be found, for example,  in \cite{FrizHairer2014}.}. The principal
idea is to replace Brownian motion by a geometric rough path
\[
\mathbf{Z}=(Z,\mathbb Z),
\]
where $Z$ denotes the driving signal and $\mathbb Z$ represents its second
order iterated integrals (the L\'evy area). The additional second-order
information is essential for defining differential equations driven by
highly irregular paths. Working within Lyons' rough path theory allows the
authors to formulate stochastic transport equations entirely in a pathwise
fashion, eliminating the dependence on stochastic calculus while
preserving the geometric structure of the dynamics. In particular, the
framework naturally includes fractional Brownian motion and more general
Gaussian rough paths as admissible drivers. Consequently, deterministic,
stochastic and rough transport models are all incorporated within a single
variational theory. 

The principal innovation of \cite{CrisanHolmLeahyNilssen2022} is the introduction of
\emph{Geometric Rough Path Parametrisation Schemes} (GRPPS), in which the
Lagrangian trajectories of fluid particles are described by geometric rough
paths on the infinite-dimensional manifold of diffeomorphisms.
This extends the SALT paradigm from Brownian transport noise to arbitrary
rough transport signals while retaining the geometric interpretation of
Lagrangian transport. In the special case in which the rough path is
generated by a semimartingale, the theory recovers the SALT models
developed previously. However, the rough-path formulation has the important
advantage that it remains entirely pathwise and therefore accommodates
non-Markovian transport models possessing memory effects.

An important aspect of the work is the development of three
equivalent variational formulations, each extending a classical principle
from geometric mechanics to the rough-path setting:
The first is the \emph{Clebsch variational principle}, in which the
constraints are imposed through rough advection equations.
The second is the \emph{Hamilton--Pontryagin principle}, where the
constraints are incorporated through right-invariant rough vector fields on
the diffeomorphism group.
The third is the \emph{Euler--Poincar\'e variational principle}, obtained
after reduction by symmetry and expressed in terms of constrained
variations of the Eulerian velocity field.
The equivalence of these three formulations demonstrates that the familiar
hierarchy of deterministic geometric mechanics survives unchanged in the
rough setting.

These constrained variational principles lead directly to the corresponding
Lie--Poisson Hamiltonian formulation of fluid dynamics on geometric rough
paths. Consequently, the geometric structures that characterise
deterministic ideal fluid mechanics are preserved. In particular, the
paper derives a rough version of Kelvin's circulation theorem, showing that
circulation remains conserved along rough Lagrangian trajectories.
This result provides one of the strongest indications that rough transport
should be regarded as a genuine geometric extension of deterministic fluid
mechanics rather than merely a class of irregular perturbations.

The paper illustrates the general theory by deriving several important
rough partial differential equations arising in fluid dynamics. These
include the rough incompressible Euler equations, the rough
Camassa--Holm equations together with their limiting rough Burgers
equation, and the rough equations governing compressible adiabatic fluid
flow. Each of these models is obtained systematically from the
corresponding rough variational principle, thereby demonstrating the
generality of the framework and its applicability across a wide range of
continuum mechanical systems.

From the perspective of the STUOD programme, \cite{CrisanHolmLeahyNilssen2022} represents a
conceptual advance. The semimartingale framework of the previous section
shows that stochastic transport equations arise naturally from Hamilton's
principle. The rough variational theory extends this observation far beyond
the Brownian setting by replacing semimartingales with arbitrary geometric
rough paths. The resulting framework provides a common geometric language
for deterministic fluid dynamics, SALT models, rough transport equations
and, ultimately, the fractional transport models discussed in the
following section.

\subsection{Fractional Brownian Motion Driven Transport Equations}
\label{sec:fbm}

As explained above, most of the stochastic transport models considered throughout the previous sections
are driven by Brownian motion and are therefore formulated within the
classical framework of semimartingale stochastic calculus. Although this
setting is mathematically natural and encompasses the SALT models discussed
earlier, Brownian motion is not always an adequate model for unresolved
transport processes in geophysical fluid dynamics. Oceanic and atmospheric
flows frequently exhibit temporal correlations over a wide range of scales,
and tracer trajectories often display persistent or anti-persistent
behaviour that cannot be captured by a Markov process. These observations
motivate the replacement of Brownian motion by more general stochastic
processes possessing memory.

A natural candidate is fractional Brownian motion. Fractional Brownian motion is a centred Gaussian process whose
regularity is controlled by the Hurst parameter
$
H\in(0,1).
$
When
$
H=\frac12,
$
fractional Brownian motion reduces to classical Brownian motion.
For
$
H>\frac12,
$
the increments are positively correlated, leading to persistent sample
paths, whereas
$
H<\frac12
$
corresponds to negatively correlated increments. Consequently, fractional
Brownian motion provides a natural framework for modelling transport
processes with long-range temporal dependence.

Replacing Brownian motion by fractional Brownian motion introduces
substantial analytical challenges. Fractional Brownian motion is no longer
a semimartingale unless $H=\frac12,$
and therefore It\^o calculus is no longer available. Instead, stochastic integration must be replaced by deterministic pathwise integration. This permits the evolution equations to be analysed entirely pathwise using a
generalised sewing lemma, which plays the role occupied by stochastic
integration in the Brownian setting. The results presented in this section are based on the work in
\cite{BlessingCrisanLang2026}.

A one-dimensional fractional Brownian motion with Hurst parameter
\(H\in(0,1)\)
is the centred Gaussian process
$
W^H=\{W_t^H:t\ge0\},
$
whose covariance function is
\[
\mathbb E
\left[
W_t^HW_s^H
\right]
=
\frac12
\left(
t^{2H}
+s^{2H}
-|t-s|^{2H}
\right).
\]
Fractional Brownian motion possesses almost surely
\(\gamma\)-H\"older continuous sample paths for every
$
\gamma<H.
$
We restrict attention to the regime
$
H>\frac12,
$
for which Young integration is available.

The analysis is carried out in a scale of Banach spaces
$
(\mathcal B_\alpha)_{\alpha\in\mathbb R},
$
where the index \(\alpha\) measures spatial regularity. In the vorticity
application considered below, one takes
$
\mathcal B_\alpha
=
W^{2\alpha,2}(\mathbb T^2)
=
H^{2\alpha}(\mathbb T^2),
$
with norm denoted by
$
\|x\|_\alpha:=\|x\|_{H^{2\alpha}(\mathbb T^2)}.
$
The space
$
\mathcal B_\infty
:=
\bigcap_{\alpha\ge0}W^{2\alpha,2}(\mathbb T^2)
=
C^\infty(\mathbb T^2)
$
is used as the space of smooth test functions.
The use of a Banach scale is one of the key analytical features of the theory.
Since the transport operator generally does not preserve the regularity of the
solution, the nonlinear drift and the stochastic transport naturally act between
different levels of regularity. 

Let
\[
A:D(A)\subset\mathcal B_\alpha
\rightarrow
\mathcal B_\alpha
\]
be the infinitesimal generator of a strongly continuous analytic semigroup
$
S(t)=e^{tA}.
$
The smoothing properties of this semigroup play a central role throughout
the analysis. 

The abstract stochastic evolution equation considered in
\cite{BlessingCrisanLang2026}
is
\begin{equation}
	\label{eq:abstract-fbm}
	dX_t
	=
	\left(
	AX_t
	+
	F(X_t)
	\right)\,dt
	+
	G(X_t)\,dW_t^H.
\end{equation}
The nonlinear drift is described by a mapping
$
F:\mathcal B_\alpha
\longrightarrow
\mathcal B_{\alpha-\delta},
$
while the transport operator
$
G:\mathcal B_\alpha
\longrightarrow
\mathcal B_{\alpha-1/2}
$
determines the action of the fractional Brownian motion.

Equation
\eqref{eq:abstract-fbm}
is understood through its mild formulation
\begin{equation}
	\label{eq:mild-fbm}
	X_t
	=
	S(t)X_0
	+
	\int_0^t
	S(t-s)F(X_s)\,ds
	+
	\int_0^t
	S(t-s)G(X_s)\,dW_s^H.
\end{equation}

The first integral is the usual Bochner integral,
whereas the second is a Young integral.
The principal analytical difficulty is that the semigroup
\(S(t-s)\)
becomes singular as
\(s\uparrow t\),
so classical Young integration is insufficient.
To overcome this difficulty, a
generalised sewing lemma, which combines the smoothing properties of the
analytic semigroup with the H\"older regularity of the driving fractional
Brownian motion.
This yields a robust pathwise integration theory adapted to semilinear
stochastic evolution equations.
Equation
\eqref{eq:abstract-fbm} forms the master equation for the remainder of this section.

In the following,  a local well-posedness
theorem for \eqref{eq:abstract-fbm} is stated, for the  two-dimensional incompressible vorticity
equation on the torus perturbed by transport-type fractional Brownian noise:
\begin{equation}
	\label{eq:fbm-vorticity}
	d\omega_t
	+
	u_t\cdot\nabla\omega_t\,dt
	+
	\mathcal L_\xi\omega_t\,dW_t^H
	=
	\Delta\omega_t\,dt,
\end{equation}
where
\[
\omega_t=\operatorname{curl}u_t,
\qquad
\nabla\cdot u_t=0,
\qquad
\mathcal L_\xi\omega_t:=\xi\cdot\nabla\omega_t.
\]
Here \(\xi\) is a time-independent divergence-free vector field and
\(W^H\) is a fractional Brownian motion with Hurst parameter
\(H>\frac12\). The velocity is recovered from the vorticity by the
Biot--Savart law,
\[
u_t=K*\omega_t.
\]
The mild formulation of \eqref{eq:fbm-vorticity} is
\begin{equation}
	\label{eq:fbm-vorticity-mild}
	\omega_t
	=
	S(t)\omega_0
	-
	\int_0^t S(t-s)(u_s\cdot\nabla\omega_s)\,ds
	-
	\int_0^t S(t-s)(\xi\cdot\nabla\omega_s)\,dW_s^H.
\end{equation}
In the vorticity equation, \(A=\Delta\) and \(S(t)=e^{t\Delta}\) is the heat
semigroup. The analytic semigroup provides the parabolic regularisation that
compensates for the loss of derivatives caused by the transport operator. Its
smoothing estimates form the basis of the existence theory developed below. For \(\sigma\in[0,1]\), the semigroup satisfies estimates of the form
\begin{align}
	\|S(t)x\|_{\mathcal B_\alpha}
	&\lesssim
	\|x\|_{\mathcal B_\alpha},
	\label{eq:semigroup-bound-1}
	\\
	\|S(t)x\|_{\mathcal B_{\alpha+\sigma}}
	&\lesssim
	t^{-\sigma}
	\|x\|_{\mathcal B_\alpha},
	\label{eq:semigroup-bound-2}
	\\
	\|(S(t)-I)x\|_{\mathcal B_\alpha}
	&\lesssim
	t^\sigma
	\|x\|_{\mathcal B_{\alpha+\sigma}}.
	\label{eq:semigroup-bound-3}
\end{align}
For \(T>0\), define
\[
\mathcal V^T
:=
C([0,T];\mathcal B_\alpha)
\cap
C^\gamma([0,T];\mathcal B_{\alpha-\gamma}),
\]
with norm
\[
\|y\|_{\mathcal V^T}
:=
\max\left\{
\sup_{0\le t\le T}\|y_t\|_\alpha,
\sup_{0\le s<t\le T}
\frac{\|y_t-y_s\|_{\alpha-\gamma}}{|t-s|^\gamma}
\right\}.
\]
The key technical result in this work is a generalized sewing lemma adapted to analytic semigroups. This
result extends the sewing approach to operator-valued kernels possessing
singularities at the diagonal. It provides a pathwise construction of the
stochastic integral that simultaneously exploits the H\"older regularity of the
fractional Brownian motion and the smoothing properties of the semigroup.

The construction of the stochastic convolution is the first main result. Let
\[
Y\in
C([0,T];\mathcal B_{\alpha-1/2})
\cap
C^\gamma([0,T];\mathcal B_{\alpha-\gamma-1/2}).
\]
The object to be constructed is
\[
I_t
:=
\int_0^t S(t-r)Y_r\,dW_r^H.
\]
This includes the transport term \(Y_r=\xi\cdot\nabla\omega_r\), since
\(\xi\in\mathcal B_\infty\) and \(\omega\in\mathcal V^T\) imply the required
regularity. For dyadic partitions, define
\[
I_t^k
:=
\sum_{[n2^{-k},(n+1)2^{-k}]\subset[0,t]}
S\left(t-n2^{-k}\right)Y_{n2^{-k}}
\left(W^H_{(n+1)2^{-k}}-W^H_{n2^{-k}}\right).
\]

\begin{theorem}[Sewing Lemma]
	\label{thm:young-semigroup}
	Let
	\[
	Y\in
	C([0,T];\mathcal B_{\alpha-1/2})
	\cap
	C^\gamma([0,T];\mathcal B_{\alpha-\gamma-1/2}),
	\qquad
	\frac12<\gamma<H.
	\]
	Then the sequence \((I_t^k)_k\) is Cauchy in \(\mathcal B_\alpha\), and one defines
	\[
	I_t
	=
	\int_0^t S(t-r)Y_r\,dW_r^H
	:=
	\lim_{k\to\infty} I_t^k.
	\]
	Moreover,
	\[
	I\in C([0,T];\mathcal B_\alpha)
	\cap C^\gamma([0,T];\mathcal B_{\alpha-\gamma}).
	\]
	In particular, the map
	\[
	Y
	\longmapsto
	\left(t\mapsto\int_0^tS(t-r)Y_r\,dW_r^H\right)
	\]
	is well defined from
	\[
	C([0,T];\mathcal B_{\alpha-1/2})
	\cap C^\gamma([0,T];\mathcal B_{\alpha-\gamma-1/2})
	\]
	into
	\[
	C([0,T];\mathcal B_\alpha)
	\cap C^\gamma([0,T];\mathcal B_{\alpha-\gamma}).
	\]
\end{theorem}

As for \(Y=\xi\cdot\nabla\omega\) one obtains a well-defined stochastic
convolution
\begin{equation}
	\label{eq:transport-young-convolution}
	\mathcal I_t(\omega)
	:=
	\int_0^t S(t-r)(\xi\cdot\nabla\omega_r)\,dW_r^H.
\end{equation}
Moreover, the map \(\omega\mapsto\mathcal I(\omega)\) is locally Lipschitz on
\(\mathcal V^T\), and for sufficiently small \(T\) its Lipschitz constant can be made
small. This is the crucial ingredient in the fixed-point construction of the
solution.

Define next the map
\[
\Lambda:\mathcal V^T\longrightarrow\mathcal V^T
\]
by
\begin{equation}
	\label{eq:fbm-fixed-point-map}
	\Lambda(\omega)_t
	=
	S(t)\omega_0
	-
	\int_0^tS(t-s)(u_s\cdot\nabla\omega_s)\,ds
	-
	\int_0^tS(t-s)(\xi\cdot\nabla\omega_s)\,dW_s^H,
\end{equation}
where \(u_s=K*\omega_s\). A fixed point of \(\Lambda\) is precisely a mild solution
of \eqref{eq:fbm-vorticity}.

\begin{theorem}[Local mild well-posedness for the fractional vorticity equation]
	\label{thm:fbm-vorticity-local}
	Let
	\[
	\xi\in\mathcal B_\infty,
	\qquad
	\omega_0\in\mathcal B_\alpha,
	\qquad
	\frac12<\gamma<H.
	\]
	Then, for \(T>0\) sufficiently small, the stochastic vorticity equation
	\eqref{eq:fbm-vorticity} admits a unique mild solution
	\[
	\omega\in\mathcal V^T
	=
	C([0,T];\mathcal B_\alpha)
	\cap
	C^\gamma([0,T];\mathcal B_{\alpha-\gamma}).
	\]
	Equivalently, \(\omega\) is the unique fixed point of the map
	\eqref{eq:fbm-fixed-point-map} in \(\mathcal V^T\).
\end{theorem}

The theorem is obtained by applying Banach's fixed-point theorem. The nonlinear
drift term is controlled using the Biot--Savart law, Sobolev product estimates and
semigroup smoothing, while the fractional transport term is controlled by the
semigroup-adapted Young integral described above. The local nature of the result
comes from the fact that the contraction constants depend on powers of \(T\), so
choosing \(T\) sufficiently small makes the fixed-point map contractive.

The mild formulation is the natural one for constructing solutions, because it
uses the smoothing generated by the heat semigroup. For applications, however, it
is important to know that the mild solution also satisfies the weak formulation of
the original SPDE. 

\begin{theorem}
	\label{thm:fbm-mild-weak}
	Let the assumptions of Theorem~\ref{thm:fbm-vorticity-local} hold. Then the mild
	solution
	\[
	\omega_t
	=
	S(t)\omega_0
	-
	\int_0^t S(t-r)(u_r\cdot\nabla\omega_r)\,dr
	-
	\int_0^t S(t-r)(\xi\cdot\nabla\omega_r)\,dW_r^H
	\]
	is equivalent to the weak solution satisfying, for every test function
	\(\varphi\in\mathcal B_\infty\),
	\begin{equation}
		\langle \omega_t,\varphi\rangle
		=
		\langle \omega_0,\varphi\rangle
		+
		\int_0^t
		\langle \omega_r,u_r\cdot\nabla\varphi+\Delta\varphi\rangle\,dr		+
		\int_0^t
		\langle \omega_r,\xi\cdot\nabla\varphi\rangle\,dW_r^H.
		\label{eq:fbm-weak-final}
	\end{equation}
\end{theorem}

This result confirms that the solution constructed by semigroup methods is a
solution of the original fractional stochastic vorticity equation in the usual
weak PDE sense. 

\paragraph{Estimation of the Hurst parameter.}

An important feature of the framework developed in \cite{BlessingCrisanLang2026} is that it provides not only a well-posedness theory for
fractional transport equations but also a statistical methodology for
estimating the Hurst parameter from observations of the solution. The key
observation is that, despite the nonlinear dynamics, suitable projections
of the solution inherit the local regularity of the driving fractional
Brownian motion. By analysing the asymptotic behaviour of appropriately
normalised quadratic variations of these projected processes, a strongly consistent estimator of the Hurst parameter can be constructed.
Consequently, the memory parameter governing the transport noise can be inferred directly from observations of the evolving fluid. This 
provides an important connection between the analytical theory of fractional SPDEs and
statistical calibration from data.

\paragraph{Extension to a general class of transport equations.}

Although the stochastic vorticity equation provides the principal
application studied in the paper, the analytical framework is far more
general. The use of Banach scales, analytic semigroups and the generalized
sewing lemma leads to an abstract local well-posedness theory for a broad
class of semilinear evolution equations driven by fractional Brownian
motion governed by equation \eqref{eq:abstract-fbm}. The assumptions are formulated entirely in terms of the analytic
properties of the semigroup and the nonlinear operators, making the
results applicable beyond the vorticity equation considered here.

\section{Calibration of Stochastic Transport Models}
\label{sec:06_calibration_of_stochastic_transport_models}

The previous sections established the stochastic transport framework and its analytical foundations. In this section we address the question: 
How should the stochastic transport structure be determined from data?

The stochastic transport models considered throughout this paper contain transport vector fields
$
(\xi_i)_{i\ge1}
$
that represent the effects of unresolved scales. While the analytical theory treats these fields as given, practical applications require them to be estimated from observations or high-resolution simulations. To use stochastic transport models operationally one must determine:  how many transport modes should be retained (how many vector fields $\xi_i$), how these modes should be estimated, and how stochastic amplitudes should be calibrated. 

Let $u^{\mathrm{ref}}$ denote a high-resolution reference velocity field obtained either from observations or from a trusted numerical simulation.
The objective is to construct a stochastic transport model
\[
dX_t
=
u(X_t,t)\,dt
+
\sum_{i=1}^{r}
\xi_i(X_t)\circ dW_t^i
\]
whose statistical behaviour reproduces the unresolved variability present in the high-resolution reference velocity field.

Schematically, the calibration problem can be formulated as

\[
\text{Reference data}
\quad\longrightarrow\quad
\text{Transport fields } \{\xi_i\}
\quad\longrightarrow\quad
\text{Stochastic transport model}.
\]

The challenge is that the transport fields are not directly observable. They must be inferred from information contained in the unresolved fluctuations.
This section describes three different  calibration paradigms: Lagrangian calibration, Eulerian calibration and calibration based on generative models.

\subsection{Lagrangian Calibration}
\label{subsec:06_03_lagrangian_calibration}

Lagrangian callibration has been developed in 
\cite{CotterCrisanHolmPanShevchenko2019}, \cite{CotterCrisanHolmPanShevchenko2020FODS},  prior 
to the STUOD programme itself. We include it here for two reasons.
First, it provided much of the conceptual motivation for the later
calibration developments undertaken within STUOD. Second, among all the
approaches discussed in this paper, it is arguably the one that remains
closest to the modelling philosophy underlying both the SALT and the LU 
frameworks. In both paradigms, uncertainty is introduced through the
Lagrangian motion of fluid parcels, making it natural to calibrate the
stochastic model directly at the level of trajectories.

The central idea is to compare particle trajectories generated by a
coarse-resolution flow with those obtained from a trusted high-resolution
reference simulation. The resolved velocity field determines the large-scale
transport, while the discrepancy between the two sets of trajectories
contains information about the cumulative effect of unresolved scales. Rather
than modelling this discrepancy deterministically, it is interpreted as the
signature of unresolved stochastic transport.
More precisely, let
\[
\dot X^{\,r}(t)=u^{\,r}(X^{\,r}(t),t)
\]
denote the Lagrangian trajectory associated with the resolved velocity field $u^{\,r}$,
and let
\[
\dot X^{\,f}(t)=u^{\,f}(X^{\,f}(t),t)
\]
be the corresponding trajectory generated by a fine-resolution reference
flow. The trajectory difference
\[
\Delta X(t)=X^{\,f}(t)-X^{\,r}(t)
\]
provides a direct measure of the cumulative transport error arising from
unresolved dynamics. Since this discrepancy is generated by the transport
mechanism itself, it is natural to use its statistical properties to infer
the stochastic transport vector fields appearing in the SALT or LU
formulation.

The calibration proceeds by constructing an ensemble of trajectory
differences from many particle releases. Their empirical covariance
structure is then analysed to identify the dominant modes of unresolved
transport. These modes define a collection of spatial vector fields which
constitute the stochastic transport directions in the resulting model. In
practice, this is achieved through a principal-component decomposition of the
Lagrangian displacement field, yielding a finite number of transport modes
that capture most of the unresolved variability.

A distinctive feature of this methodology is that it calibrates the
stochastic model at the level of particle motion rather than at the level of
Eulerian velocity fields. Consequently, the calibrated model reproduces the
dispersion and mixing properties of the reference flow, which are precisely
the quantities that stochastic transport models are designed to represent.

\subsection{Eulerian Calibration}
\label{subsec:06_04_eulerian_calibration}

The Lagrangian calibration procedure described in the previous subsection is
conceptually attractive because it follows directly from the transport
interpretation of stochastic fluid models. 
An Eulerian calibration methodology was introduced in 
\cite{LCLcalibration2023}\footnote{See also \cite{LangPanPathwise2023} for a different Eulerian approach.}. Although the paper
uses the stochastic rotating shallow water equations as its principal
application, the methodology itself is completely generic. 
Its objective is to estimate the stochastic transport operator directly
from either observational data or high-resolution numerical simulations.

Let \(m^f\) denote the state of a high-resolution (fine-scale) model.
Its evolution is assumed to satisfy the deterministic equation
\begin{equation}
	\frac{d m^f}{dt}
	=
	A(m^f),
	\qquad t\ge0,
	\label{eq:fine_model}
\end{equation}
where \(A\) denotes the deterministic model operator.

The corresponding coarse-scale model is obtained by introducing a stochastic
parameterisation that represents the effect of unresolved scales. We will denote by $m^c$ the coarse scale model. The effect of the unresolved scales
is modelled by a stochastic integral of the form
\begin{equation}
	\int_0^t
	M(m_s^c)\,dW_s,
	\label{eq:generic_noise}
\end{equation}
where \(M\) is a suitably chosen operator and
\(W=\{W_t\}_{t\ge0}\) is a space-time Brownian motion.
The coarse-scale model therefore satisfies the stochastic evolution equation
\begin{equation}
	d m_t^c
	=
	A(m_t^c)\,dt
	+
	M(m_t^c)\,dW_t,
	\qquad t\ge0.
	\label{eq:coarse_model}
\end{equation}

Although the deterministic and stochastic models evolve on the same
physical domain, they are generally discretised on different spatial
grids. The fine-scale solution is computed on a sufficiently refined mesh,
whereas the stochastic model is intended to reproduce its large-scale
behaviour on a much coarser discretisation.

For the calibration procedure it is convenient to represent the driving
Brownian field by a representation of the form 
\begin{equation}
	W_t
	=
	\sum_{k=1}^{\infty}
	\xi_k\,W_t^k,
	\label{eq:KL_expansion}
\end{equation}
where the vector fields
\(\{\xi_k\}_{k\ge1}\)
depend only on space and
\(\{W^k\}_{k\ge1}\)
are independent one-dimensional Brownian motions.

For practical purposes, we need to approximate the infinite-dimensional
noise by a finite-dimensional representation. More precisely, the unresolved forcing is approximated by
\begin{equation}
	\int_0^t
	M(m_s^c)\,dW_s^N,
	\label{eq:finite_noise}
\end{equation}
where
\begin{equation}
	W_t^N
	=
	\sum_{k=1}^{N}
	\xi_k\,W_t^k.
	\label{eq:finite_KL}
\end{equation}

The calibration procedure therefore consists of estimating both the number
of stochastic modes \(N\) and the associated spatial vector fields
\(\{\xi_k\}_{k=1}^N\) directly from data. Once these quantities have been
identified, the calibrated stochastic model can be evolved on the coarse
grid instead of the much more expensive deterministic model on the fine
grid, thereby providing an efficient stochastic representation of the
effects of unresolved scales.

To do so we compare a high-resolution reference solution
\(m^f\) with a suitably coarsened version \(C(m^f)\) obtained by applying a
mollification procedure, such as a low-pass filter or Gaussian smoothing.
The difference
\[
\widehat m = m^f-C(m^f)
\]
is interpreted as the contribution of the unresolved scales. Assuming that
these unresolved fluctuations admit a stochastic representation, the
increments of \(\widehat m\) over suitably decorrelated time intervals may
be viewed as samples of the stochastic forcing. The calibration problem then
reduces to estimating the spatial covariance structure of these samples.
Unlike the earlier Lagrangian methodology, no particle trajectories are
required; all computations are performed directly on Eulerian fields. 

The methodology was applied in \cite{LCLcalibration2023} for the stochastic rotating shallow water
equations by calibrating the transport noise using the free-surface
elevation, one of the most readily available observational variables in
oceanography. Numerical experiments showed that the resulting calibrated
stochastic model accurately reproduces the uncertainty generated by model
reduction. In particular, the ensemble spread closely matches the
root-mean-square error, the truth remains well contained within the
stochastic ensemble, and standard uncertainty quantification diagnostics,
including rank histograms, indicate that the calibrated model provides a
realistic representation of forecast uncertainty. 

An important feature of this approach is that the calibration procedure is
independent of the particular stochastic parameterisation used. Once a
relationship between the unresolved fluctuations and the stochastic forcing
operator has been established, exactly the same methodology can be applied
to SALT, Location Uncertainty, additive-noise or multiplicative-noise
models.

\subsection{Generative Modelling of Stochastic Transport Noise}
\label{subsec:06_05_generative}

The Eulerian calibration procedure described in the previous section
constructs stochastic parametrizations by assuming that the unresolved
increments are Gaussian. This assumption leads naturally to a Principal
Component Analysis (PCA) representation of the stochastic forcing through a
finite Karhunen--Lo\`eve expansion. Although this approach proved highly
successful, there is no physical reason why the unresolved transport should
be Gaussian. 
To overcome this limitation, 
replaced the PCA representation was replaced in \cite{CrisanLangLobbeGenerativeRSW2024} by a data-driven generative model.
The central idea is to learn directly the probability distribution of the
stochastic transport increments from calibration data, without imposing any
parametric assumptions on their distribution.
Consequently, the stochastic forcing is no longer represented by Gaussian
random variables but by samples generated from a learned probability
distribution.

As in the Eulerian calibration methodology, let
$
m^f
$
denote the fine-scale deterministic solution and let
$
C(m^f)
$
be its coarse-scale projection obtained through an appropriate
mollification procedure.
The unresolved component is again defined by
\[
\widehat m
=
m^f-C(m^f).
\]
Instead of assuming that the increments satisfy
\[
\widehat m_{n+1}-\widehat m_n
=
M(m_n^c)
\sum_{i=1}^{N}
\xi_i
\Delta W_n^i,
\]
the paper assumes the more general stochastic representation
\begin{equation}
	\widehat m_{n+1}-\widehat m_n
	=
	M(m_n^c)N_n,
	\label{eq:general_noise_model}
\end{equation}
where
\(
N_n
\)
are independent identically distributed random fields whose distribution is
unknown and is learned directly from data.
Equation \eqref{eq:general_noise_model} therefore contains the Gaussian
model as a special case, while allowing substantially richer stochastic
structures. 

The unknown distribution of the random fields \(N_n\) is approximated
using a score-based diffusion generative model based on the Diffusion
Schr\"odinger Bridge (DSB) algorithm of De Bortoli et al.
Rather than estimating only the covariance of the unresolved transport,
the DSB algorithm learns the entire probability distribution by repeatedly
transforming the training data into a simple Gaussian reference
distribution and then learning the reverse transformation.
After training, independent samples from the learned distribution can be
generated efficiently and used as stochastic parametrizations in the
coarse-resolution model. 

The methodology was applied in \cite{LCLcalibration2023} for the stochastic rotating shallow water
equations.
As in the previous section, the calibration data are obtained from the
difference between fine- and coarse-resolution simulations.
The calibration equation is solved to recover stream functions
corresponding to the stochastic velocity perturbations.
These stream functions constitute the training data for the generative
model.
After training, new samples of stochastic transport are generated,
converted into divergence-free velocity perturbations, and inserted into
the stochastic rotating shallow water equations. 

The numerical experiments demonstrate that the learned transport noise is
genuinely non-Gaussian.
Kolmogorov--Smirnov tests reject the hypothesis of Gaussianity at all
tested spatial locations.
Forecast experiments further show that the generative model produces lower
root-mean-square errors (RMSE) and Continuous Ranked Probability Scores
(CRPS) than Gaussian models in regimes of small initial uncertainty.
The improvement is particularly pronounced for the free-surface elevation,
while the generated ensembles also exhibit more realistic rank histograms.
These results indicate that learning the full probability distribution of
the unresolved transport yields a more accurate representation of model
uncertainty than covariance-based Gaussian approximations. 

Conceptually, this work extends the Eulerian calibration framework rather
than replacing it.
Both approaches employ the same calibration philosophy and the same
transport operator \(M\).
The difference lies solely in the statistical representation of the
unresolved scales: the earlier methodology estimates only second-order
statistics through PCA, whereas the generative approach learns the complete
distribution of the stochastic transport directly from data.

\subsection{Calibration of stochastic parametrisations for transport SPDEs}
\label{subsec:calibration_transport_spdes}

A central objective of the STUOD programme has been to turn stochastic
transport models into data-informed reduced models. In this setting, the
fine-scale deterministic model is assumed to provide a reference dynamics,
while the stochastic coarse-scale model is designed to reproduce the
statistical effect of the unresolved scales. The calibration problem is
therefore to infer the stochastic transport fields
$\{\xi_i\}_{i=1}^{N}$
from either high-resolution
simulations or observational data. The common philosophy is that the
difference between the fine-scale dynamics and its coarse representation
should be interpreted as unresolved transport noise. In the following we give example of several models where the stochastic parametrization was calibrated using the methodologies presented above.

\subsubsection{The damped and forced two-dimensional Euler equation.}
\label{sec:05_04_calibration}

In \cite{CotterCrisanHolmPanShevchenko2019}, a stochastic parame\-tri\-zation was calibrated for the damped
and forced incompressible two-dimensional Euler equation. 
The fluid occupies the unit square
$
D=[0,1]^2\subset\mathbb{R}^2,
$
and the deterministic model is written in vorticity--streamfunction form as
\begin{align}
	\partial_t\omega+u\cdot\nabla\omega
	&=Q-r\omega,
	\label{eq:damped-forced-euler-deterministic}\\
	u&=\nabla^\perp\psi,
	\label{eq:euler-velocity-streamfunction}\\
	\Delta\psi&=\omega.
	\label{eq:euler-vorticity-streamfunction}
\end{align}
Here,
\(\omega=\widehat z\cdot\operatorname{curl}u\) is the scalar vorticity,
\(u=(u_1,u_2)\) is the incompressible velocity field,
\(\psi\) is the streamfunction,
$
\nabla^\perp\psi:=(-\partial_y\psi,\partial_x\psi),
$
and
\(\Delta=\partial_x^2+\partial_y^2\) is the two-dimensional Laplacian.
The forcing is prescribed by
\[
Q(x,y)=\alpha\sin(\beta\pi x),
\]
where \(\alpha\) determines the strength of the forcing and \(\beta\)
determines its spatial oscillation, or equivalently the number of forced
gyres. The constant \(r>0\) is the linear damping rate.
The model is supplemented by the (impermeability) boundary
condition
$
\left.\psi\right|_{\partial D}=0.
$
Since \(u=\nabla^\perp\psi\), this condition implies
$
u\cdot n=0$ on $\partial D$,
so that fluid particles cannot cross the boundary of the square. Tangential
motion along the boundary is permitted.

\begin{figure}[h]
	\centering
	\includegraphics[width=0.8\textwidth]{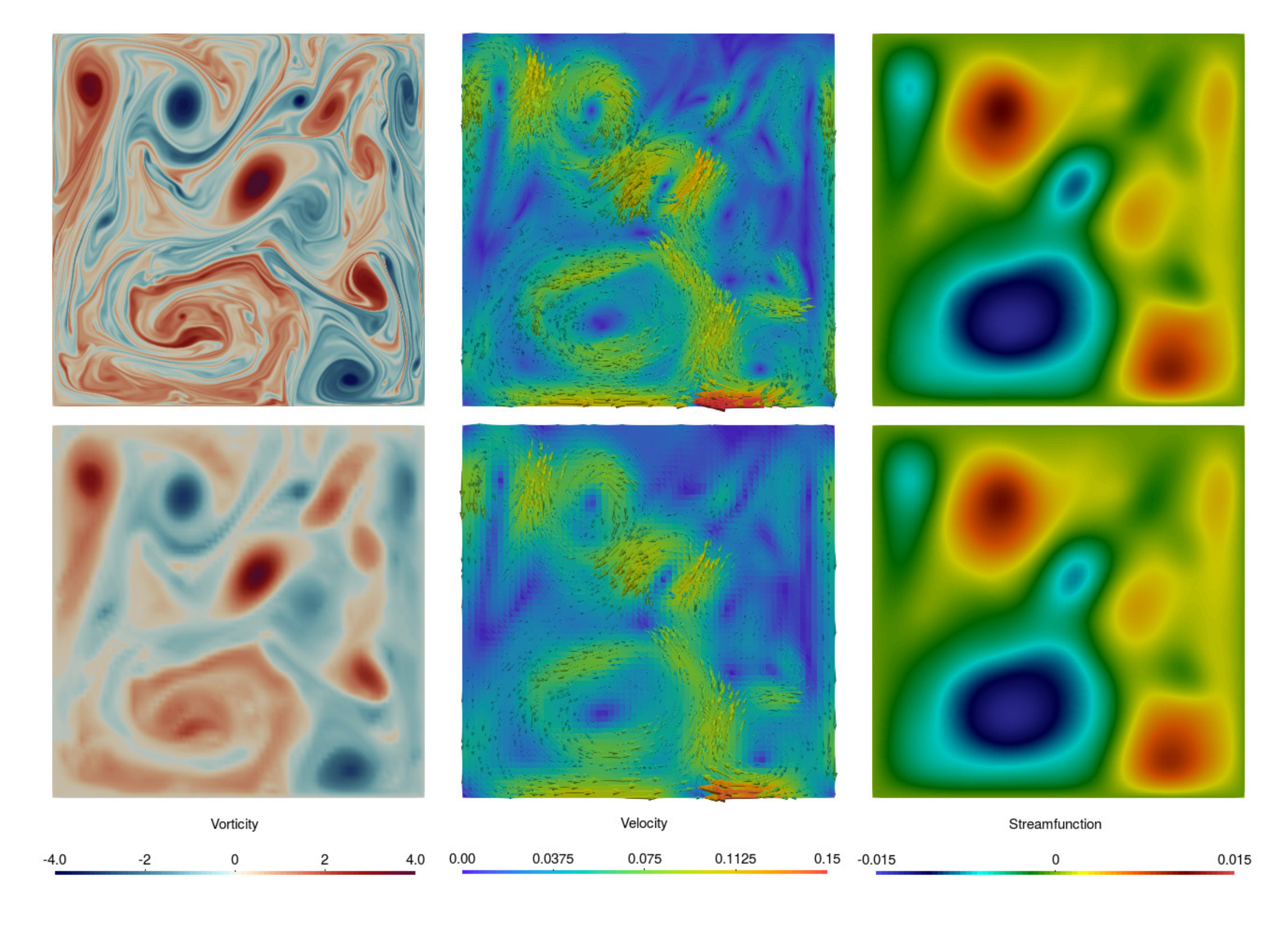}
	\caption{The fine and coarse-grained truth for the damped and forced
		two-dimensional Euler equation at Plot of the numerical PDE solution at time $t=t_{initial} + 146$ large eddy
		turnover times (ett). The coarse-graining is done via spatial averaging and
		projection of the fine grid streamfunction to the coarse grid time. From left to
		right, the panels display the vorticity \(\omega\), the velocity field
		\(u=\nabla^\perp\psi\), and the streamfunction \(\psi\). In the
		vorticity panel, the two colours indicate regions of opposite
		vorticity sign and their intensity indicates the magnitude. In the
		velocity panel, arrows show the direction and relative magnitude of the
		velocity, superimposed on a colour representation of its magnitude.
		In the streamfunction panel, the colour levels represent contours of
		\(\psi\), along which the incompressible velocity is tangent. Adapted
		from \cite{CotterCrisanHolmPanShevchenko2019}.}
	\label{fig:euler-truth-vorticity-velocity-streamfunction}
\end{figure}

The corresponding SALT model is
\begin{equation}
	d\omega
	+
	u\cdot\nabla\omega\,dt
	+
	\sum_{i=1}^{N}
	\xi_i\cdot\nabla\omega\circ dW_t^i
	=
	(Q-r\omega)\,dt,
	\label{eq:damped-forced-euler-salt}
\end{equation}
where
\(\{W^i\}_{i=1}^{N}\) are independent standard Brownian motions and
\(\{\xi_i\}_{i=1}^{N}\) are time-independent, divergence-free stochastic
transport vector fields.

The fine deterministic model was computed on a \(512\times512\) grid and
then coarse-grained to obtain reduced models on the
\(64\times64\), \(128\times128\), and \(256\times256\) grids. The
calibration was performed in a Lagrangian framework. For each calibration
interval, fluid particles were released from the coarse-grid points and
were transported once by the fine-grid velocity and once by its
coarse-grained counterpart. The differences between the endpoints of these
two sets of trajectories were regarded as samples of the unresolved
Lagrangian displacement. A velocity--velocity covariance tensor was
constructed from these displacement errors, and its leading empirical
orthogonal functions were used to define the stochastic transport vector
fields \(\xi_i\).
On the \(64\times64\) grid, \(51\), \(107\), and \(225\) vector fields $\xi_i$ were
required to explain \(50\%\), \(70\%\), and \(90\%\) of the
trajectory-error variance, respectively. For the \(90\%\) variance
threshold, the number of retained modes increased to \(277\) on the
\(128\times128\) grid and \(300\) on the \(256\times256\) grid. The
resulting stochastic ensembles reproduced the large-scale statistics of
the coarse-grained truth over several large-eddy turnover times.

\medskip
\noindent
\textbf{Calibration variable.}
The fields \(\xi_i\) were calibrated from the differences between
fine-resolution and coarse-resolution \emph{Lagrangian particle
	trajectories}; they were therefore inferred from unresolved transport
displacements, rather than directly from the Eulerian vorticity field
\(\omega\).

\subsubsection{The two-layer quasi-geostrophic model.}
\label{enough}

In \cite{CotterCrisanHolmPanShevchenko2020FODS}, a stochastic parame\-tri\-zation was calibrated for the
two-layer quasi-geostrophic (QG) equations on a mid-latitude
$\beta$-plane channel. The model describes the evolution of the potential vorticity $q$ in two
superposed homogeneous fluid layers and provides one of the standard
reduced models for large-scale ocean circulation.
\begin{equation}
	\begin{aligned}
		\frac{\partial q_1}{\partial t}+\mathbf{u}_1\cdot\nabla q_1&=\nu\Delta^2\psi_1-\beta\frac{\partial \psi_1}{\partial x},\\
		\frac{\partial q_2}{\partial t}+\mathbf{u}_2\cdot\nabla q_2&=\nu\Delta^2\psi_2-\mu\Delta\psi_2-\beta\frac{\partial \psi_2}{\partial x},
	\end{aligned}
	\label{eq:pv}
\end{equation}
where $\psi$ is the stream function, $\beta$ is the planetary vorticity gradient, 
$\mu$ is the bottom friction parameter, $\nu$ is the lateral eddy viscosity,
and $\mathbf{u}=(u,v)$ is the velocity vector.
The computational domain $\Omega=[0,L_x]\times[0,L_y]\times[0,H]$ is a horizontally periodic flat-bottom channel of depth
$H=H_1+H_2$ given by two stacked isopycnal fluid layers of depth
$H_1$ and $H_2$. 

Forcing in \eqref{eq:pv} is introduced via a vertically sheared, baroclinically unstable background flow 
\begin{equation}
	\psi_i\rightarrow-U_i\,y+\psi_i,\quad i=1,2, 
	\label{eq:forcing}
\end{equation}
where the parameters $U_i$ are background-flow zonal velocities. The potential vorticity and stream function are related through two elliptic equations:
\begin{subequations}
	\begin{align}
		q_1=\Delta\psi_1+s_1(\psi_2-\psi_1),\\
		q_2=\Delta\psi_2+s_2(\psi_1-\psi_2),
	\end{align}
	\label{eq:q_psi}
\end{subequations}
with stratification parameters $s_1$, $s_2$. The system~(\ref{eq:pv})-(\ref{eq:q_psi}) is augmented by the integral mass conservation constraint
\begin{equation}
	\frac{\partial}{\partial t}\iint\limits_{\Omega}(\psi_1-\psi_2)\ dydx=0,
	\label{eq:masscon}
\end{equation}
by the periodic horizontal boundary conditions,
$
\boldsymbol{\psi}\Big|_{\Gamma_2}=\boldsymbol{\psi}\Big|_{\Gamma_4}\,,\quad \boldsymbol{\psi}=(\psi_1,\psi_2)\,,
$ and no-slip boundary conditions 
$\boldsymbol{u}\Big|_{\Gamma_1}=\boldsymbol{u}\Big|_{\Gamma_3}=0\,
$ set at northern and southern boundaries of the domain.

\begin{figure}[h]
	\centering
	\includegraphics[width=0.95\textwidth]
	{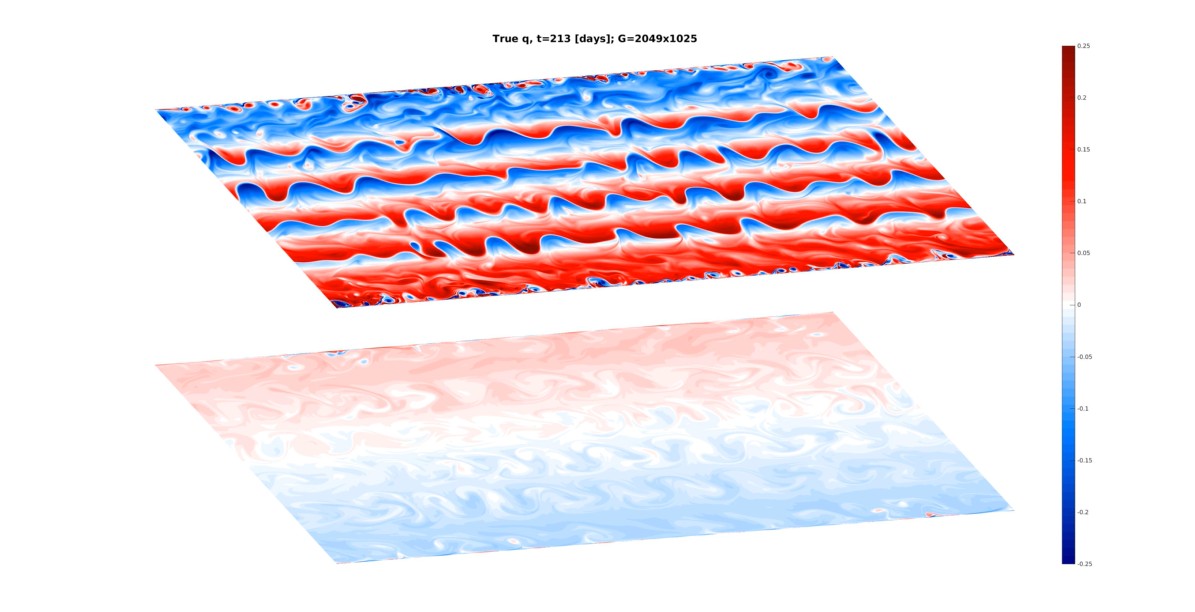}
	\caption{High-resolution reference solution of the two-layer
		quasi-geostrophic model in the heterogeneous, jet-dominated regime at
		\(t=213\) days, computed on the fine grid
		\(G^{f}=2049\times1025\). The upper panel shows the potential-vorticity
		anomaly \(q_1\) in the upper layer, where the dynamics are highly
		energetic and organised into several zonal jets together with numerous
		small-scale vortices. The lower panel shows the corresponding
		potential-vorticity anomaly \(q_2\) in the deeper layer. The lower-layer
		flow is substantially weaker and smoother, reflecting the reduced
		dynamical activity at depth and the coupling between the two layers.
		The common colour scale indicates the signed potential-vorticity
		anomaly, with red and blue corresponding to positive and negative
		values, respectively. This fine-resolution simulation provides the
		reference dynamics from which the coarse-grained truth and the
		Lagrangian calibration data are constructed.}
	\label{fig:two-layer-qg-high-resolution}
\end{figure}

The stochastic parametrization of the QG equations~\eqref{eq:pv} is given by:
\begin{align}
	\label{eq:SLTpv}
	\begin{split}
		d q_1+\left(\mathbf{u}_1\,dt + \sum\limits^K_{k=1} \xi^k_1 \circ dW^k_t\right)\cdot\nabla q_1
		&=\left(\nu\Delta^2\psi_1-\beta\frac{\partial \psi_1}{\partial x}\right)\,dt
		,\\
		d q_2+\left(\mathbf{u}_2\,dt + \sum\limits^K_{k=1} \xi^k_2 \circ dW^k_t\right)\cdot\nabla q_2
		&=\left(\nu\Delta^2\psi_2-\mu\Delta\psi_2-\beta\frac{\partial \psi_2}{\partial x}\right)\,dt.
	\end{split}
\end{align}
where \(\{W_t^k\}\) are independent Brownian motions,
and the divergence-free vector fields
\(\{\xi_k\}\) represent the unresolved transport.

The calibration followed exactly the Lagrangian methodology introduced
for the Euler equations.
A high-resolution deterministic simulation was performed on a
\(2049\times1025\) grid and coarse stochastic models were constructed on
\(257\times129\) and \(129\times65\) grids.
Fluid particles were advected simultaneously by the fine-grid velocity
field and by its coarse-grained approximation.
The differences between the two sets of trajectories were interpreted as
samples of the unresolved transport velocity.
A covariance tensor of these trajectory differences was computed and its
leading empirical orthogonal functions (EOFs) were used to define the
stochastic transport vector fields
\(\{\xi_k\}\).
In the heterogeneous jet-dominated regime, the first \(64\) EOFs captured
approximately \(96\%\) of the variance, while the first \(128\) EOFs
captured approximately \(99\%\).
The resulting stochastic ensembles successfully reproduced the
large-scale uncertainty of the unresolved dynamics and demonstrated that
the calibrated SALT model provides an accurate reduced-order
representation of the coarse-grained quasi-geostrophic flow.

\medskip
\noindent
\textbf{Calibration variable.}
As in the Euler case, the stochastic transport fields
\(\{\xi_k\}\) were calibrated from differences between
fine-resolution and coarse-resolution \emph{Lagrangian particle
	trajectories}. Thus the calibration is based on unresolved transport
velocities rather than directly on the Eulerian potential vorticities
\(q_1\) and \(q_2\).

\subsubsection{The rotating shallow water model}
\label{enough2}

The third calibration study considered the stochastic rotating shallow
water (SRSW) equations
\cite{LangVanLeeuwenCrisanPotthastRSW2022},
which provide one of the fundamental models for large-scale geophysical
fluid dynamics. Unlike the previous examples, where calibration relied on
Lagrangian particle trajectories, this work introduced a fully Eulerian
calibration that is applicable to a much broader class of
transport SPDEs.

The deterministic rotating shallow water equations are
\begin{align}
	\frac{\partial h}{\partial t}
	+
	\nabla\cdot(h u)
	&=0,
	\\
	\frac{\partial u}{\partial t}
	+
	(u\cdot\nabla)u
	+
	f u^\perp
	+
	g\nabla h
	&=
	\nu\Delta u,
\end{align}
where
\(h(x,t)\) denotes the free surface elevation,
\(u=(u_1,u_2)\) is the horizontal velocity field,
\(u^\perp=(-u_2,u_1)\),
\(g\) is the gravitational acceleration,
\(f\) is the Coriolis parameter,
and \(\nu\) is the viscosity coefficient.

The computational domain is a rectangular channel
$
D=[0,L_x]\times[0,L_y],
$
with periodic boundary conditions in the zonal direction and impermeable
walls in the meridional direction. Consequently,
$
u\cdot n=0$,
on $\partial D,$
so that no mass crosses the northern or southern boundaries.

\begin{figure}[h]
	\centering
	\includegraphics[width=0.9\textwidth]
	{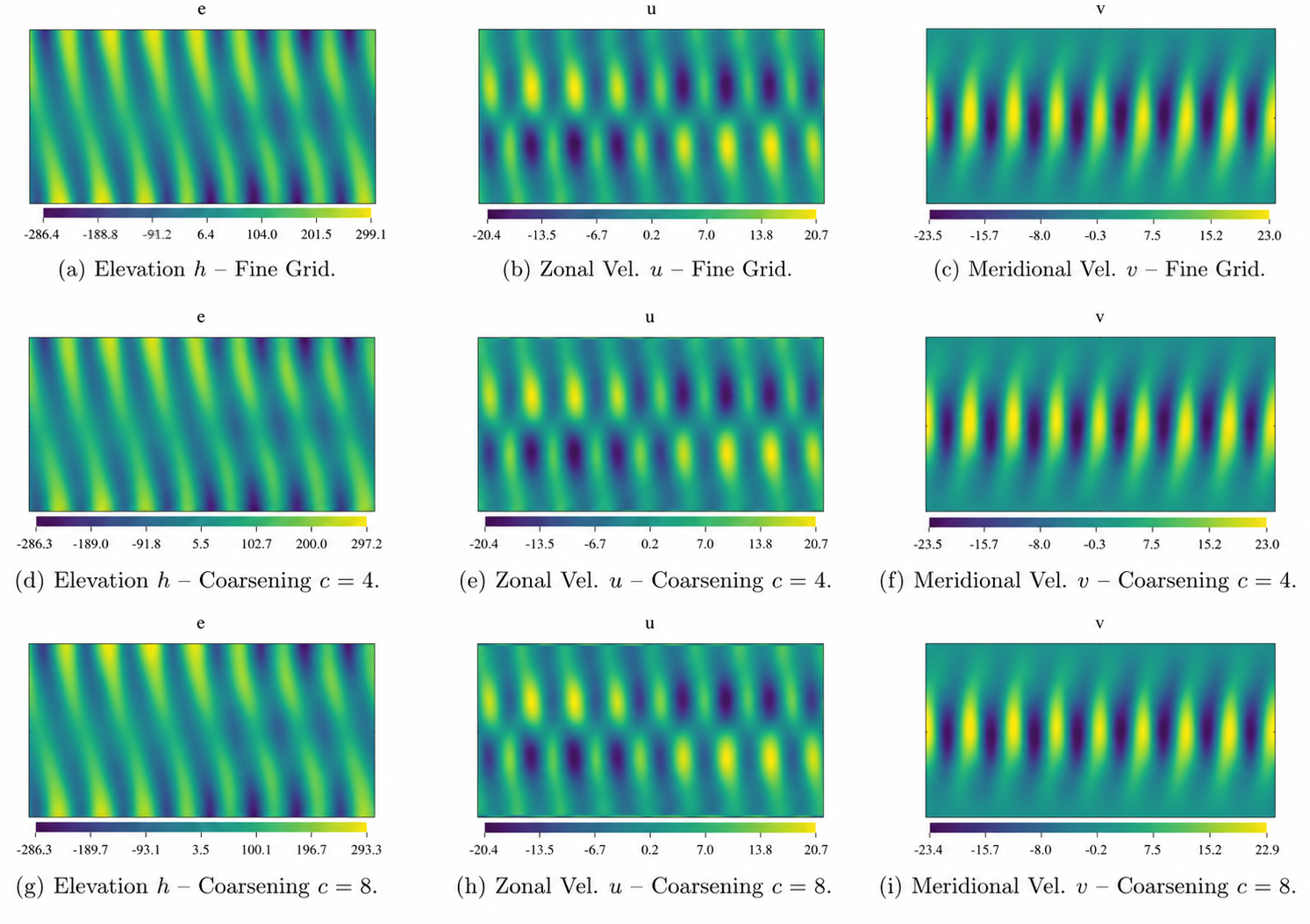}
	\caption{Fine- and coarse-resolution rotating shallow water fields. The
		top row shows the elevation \(\eta\), zonal velocity \(u\), and meridional
		velocity \(v\) on the fine grid. The middle and bottom rows show the same
		fields after coarsening by factors \(c=4\) and \(c=8\), respectively.
		The loss of small-scale structure caused by coarsening motivates the
		stochastic parametrisation. In the calibration procedure, time increments
		of the discrepancy between the fine elevation field and its filtered
		coarse representation are used to infer the stochastic transport fields
		\(\{\xi_i\}\).}	
	\label{fig:rsw-fine-coarse-comparison}
\end{figure}

The stochastic rotating shallow water model is obtained by replacing the
deterministic transport velocity by a stochastic transport velocity.
In Stratonovich form the equations become
\begin{align}
	dh
	+
	\mathcal{L}_{dX_t}h
	&=0,
	\\
	du
	+
	\mathcal{L}_{dX_t}u
	+
	f u^\perp\,dt
	+
	g\nabla h\,dt
	&=
	\nu\Delta u\,dt,
\end{align}
where
\[
dX_t
=
u\,dt
+
\sum_{i=1}^{N}
\xi_i\circ dW_t^i.
\]
The divergence-free vector fields
\(\{\xi_i\}\)
represent unresolved transport processes and constitute the stochastic
parameterisation to be calibrated.

The calibration differs fundamentally from the previous Lagrangian
approach.
A fine deterministic rotating shallow water simulation was first
computed on a
\(2224\times320\)
grid.
The solution was then coarse-grained using coarsening factors
$c=4$ and $c=8$,
leading to coarse grids of size
\(556\times80\)
and
\(278\times40\),
respectively.

Rather than comparing particle trajectories, the calibration is based on
the unresolved Eulerian component of the free-surface elevation.
The fine elevation field
\(h^f\)
is first low-pass filtered to obtain a coarse representation
\(C(h^f)\).
The discrepancy
between the fine solution and its filtered approximation is interpreted as
the unresolved transport.
From this discrepancy a linear hyperbolic calibration equation is derived,
whose solution is a streamfunction generating the stochastic transport
velocity.
Consequently, the calibration variable is the unresolved increment of the
free-surface elevation,
while the quantities ultimately inferred are the divergence-free transport
fields
\(\{\xi_i\}\).

The resulting streamfunctions are assembled into a data matrix and a
principal component analysis (PCA), or equivalently a singular value
decomposition (SVD), is performed.
The dominant empirical orthogonal functions provide the stochastic
transport modes used in the reduced-order model.
Two variance thresholds were considered,
$
90\%$ and $99\%$, 
corresponding to retaining the smallest number of EOFs explaining these
fractions of the calibration variance.
For the coarsening factor
\(c=8\),
only nine EOFs were required to explain
\(99\%\)
of the variance.
The calibrated stochastic ensembles were subsequently assessed using the
ensemble spread, bias, RMSE, relative \(L^2\)-error and rank histograms,
demonstrating that the stochastic model accurately reproduces the
large-scale statistics of the fine-resolution dynamics.

\begin{figure}[h]
	\centering
	\includegraphics[width=0.9\textwidth]
	{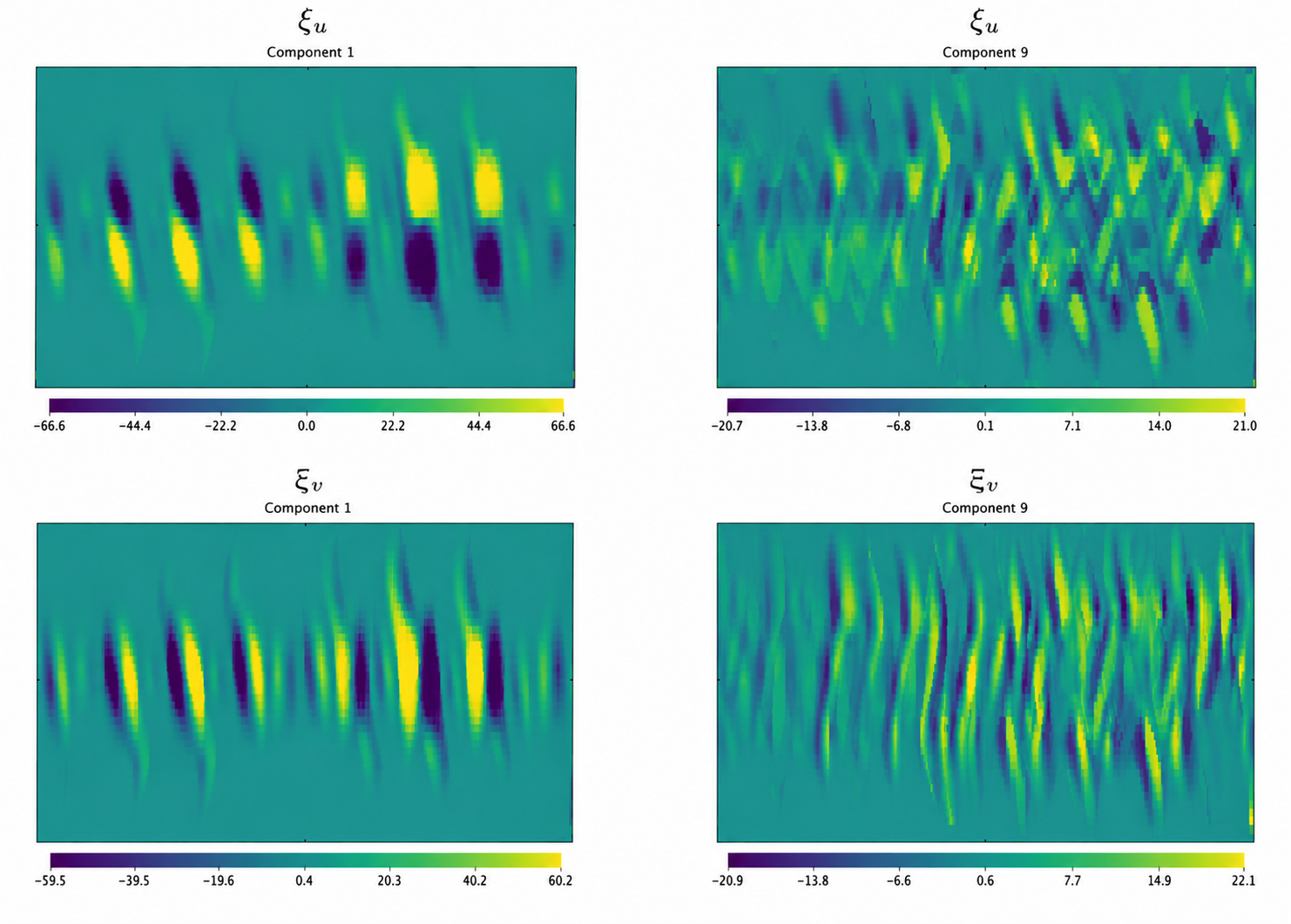}
	\caption{Estimated stochastic transport modes for the strongly coarsened
		rotating shallow water model on the \(256\times40\) grid. The figure shows
		the horizontal components of the divergence-free transport vector fields
		\(\xi=(\xi_u,\xi_v)\) obtained from the Eulerian calibration procedure.
		The left column displays the first principal component, while the right
		column shows the ninth (and final) component retained in the truncated
		Karhunen--Lo\`eve expansion. These nine modes explain approximately
		\(99\%\) of the variance of the calibration data obtained from the
		solutions of the calibration equation. The upper row shows the zonal
		component \(\xi_u\), while the lower row shows the meridional component
		\(\xi_v\). Together, these calibrated transport modes define the
		stochastic velocity perturbation used in the stochastic rotating shallow
		water model.}
	\label{fig:rsw-fine-coarse-comparison2}
\end{figure}

\medskip
\noindent
\textbf{Calibration variable.}
Unlike the Euler and quasi-geostrophic models, where the stochastic
transport fields were inferred from unresolved Lagrangian particle
displacements, the rotating shallow water model is calibrated entirely in
an Eulerian framework.
The calibration is based on the unresolved increments of the free-surface
elevation \(h\), from which the stochastic streamfunctions and hence the
transport vector fields
\(\{\xi_i\}\)
are recovered through the calibration equation.

\paragraph{Generative modelling for the stochastic rotating shallow water model.}

In \cite{LobbeCrisanLang2025}, we introduce 
a new development of the calibration methodology which  replaces the
Gaussian assumptions underlying principal component analysis by a fully
data-driven generative modelling approach
As explained in section \eqref{subsec:06_05_generative},
instead of approximating the unresolved fluctuations
by a finite-dimensional Gaussian random field through a truncated
Karhunen--Lo\`eve expansion, the new methodology seeks to learn the
complete probability distribution of the unresolved transport from data. Proceeding as above, the deterministic reference solution of the rotating shallow water
equations is first computed on a fine grid, while a reduced-order model is
constructed on a much coarser grid. The unresolved transport increments
are obtained from the discrepancy between the fine-resolution elevation
field and its coarse representation, exactly as in the Eulerian
calibration procedure described in the previous subsection. Solving the
calibration equation produces a collection of divergence-free stochastic
velocity perturbations, which serve as training samples for the generative model.

The probability distribution of these samples is learned using a
diffusion Schr\"odinger bridge model. Unlike principal component analysis,
which characterises only the first two moments of the calibration data,
the generative model learns the entire distribution and is therefore able
to represent strongly non-Gaussian stochastic transport. Once trained, the
model generates new realisations of the stochastic transport fields that
are statistically indistinguishable from those obtained from the
high-resolution simulations.

\begin{figure}[h]
	\centering
	\includegraphics[width=0.9\textwidth]{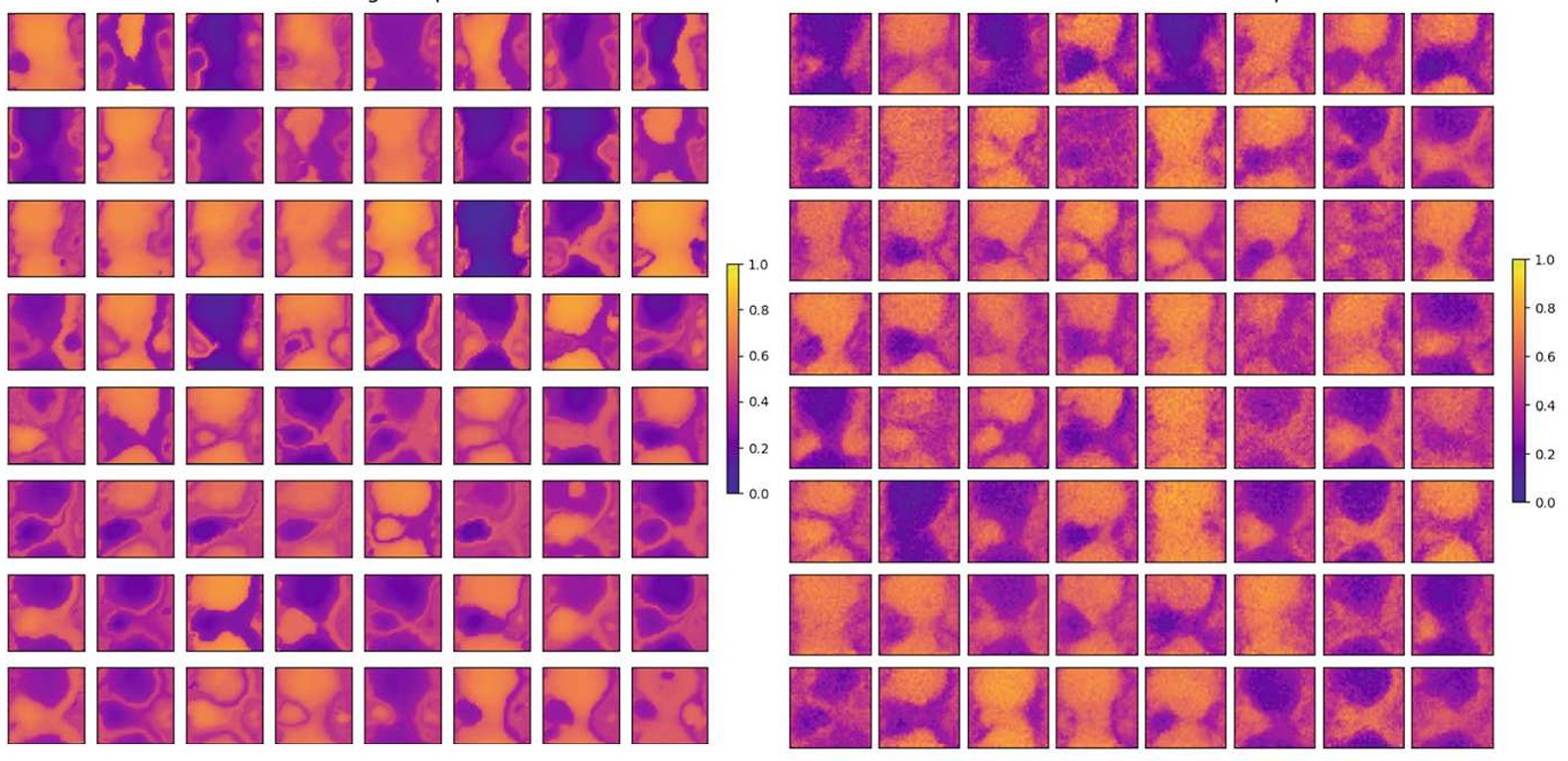}
	\caption{Comparison between calibration data and samples generated by the
		trained diffusion Schr\"odinger bridge model. The left panel shows samples
		of the stochastic transport obtained from the solutions of the Eulerian
		calibration, while the right panel shows synthetic samples
		generated by the learned diffusion model. Although generated entirely from
		the learned probability distribution, the synthetic samples reproduce the
		large-scale coherent structures, spatial variability and statistical
		characteristics of the original calibration data.}
	\label{fig:generative_samples}
\end{figure}

Figure~\ref{fig:generative_samples} illustrates the performance of the
generative model. The left panel shows samples taken directly from the
training data obtained through the calibration procedure, while the right
panel shows independently generated samples produced by the trained
diffusion model. The synthetic realisations reproduce the large-scale
coherent structures, spatial correlations and variability of the original
calibration data remarkably well, despite being generated entirely by the
learned probability model. This demonstrates that the diffusion model has
successfully captured the distribution of the unresolved stochastic
transport.

The learned stochastic transport model is subsequently incorporated into
the stochastic rotating shallow water equations and used within a Bayesian
particle filtering framework. Numerical experiments reported in
\cite{LobbeCrisanLang2025} demonstrate that the generative
approach produces lower forecast errors than Gaussian
Karhunen--Lo\`eve-based parametrisations, particularly in regimes where
the unresolved transport exhibits pronounced non-Gaussian behaviour. The
method therefore provides a natural extension of the calibration
framework, replacing covariance estimation by distribution learning while
remaining fully consistent with the stochastic transport philosophy
developed throughout the STUOD programme.

\paragraph{Generative calibration of stochastic rotating shallow water noise.}

In the
numerical experiments, the fine model was run on a \(128\times128\) grid
and the reduced stochastic model on a \(32\times32\) grid. 

This work therefore replaces covariance-based calibration by a
distribution-learning approach and provides a route towards non-Gaussian
stochastic parametrisation of unresolved transport.

\section{Bayesian Inference and High-Dimensional Data Assimilation}
\label{sec:07_bayesian_inference_and_high_dimensional_data_assimilation}

The previous section addressed the calibration of stochastic transport models. Once a stochastic model has been calibrated, a natural question arises: How should the model state be updated when new observations become available?
This question lies at the heart of data assimilation.
The objective of data assimilation is to combine dynamical models and observational information in order to produce estimates of the evolving state of a physical system together with a quantitative description of uncertainty.

The developments reviewed in this section connect stochastic transport modelling with Bayesian inverse problems, particle filtering, ensemble methods and machine-learning-based inference.

\subsection{Bayesian Formulation of Data Assimilation}
\label{subsec:07_01_bayesian_formulation_of_data_assimilation}

The objective of data assimilation is to estimate the evolving state of a
dynamical system by combining information from a mathematical model with
partial and noisy observations. In geophysical applications the state
typically consists of the velocity, pressure, temperature or sea-surface
height of the ocean or atmosphere, while observations are obtained from
satellites, drifting buoys, Argo floats, radar or other sensing
technologies. Since observations are incomplete and contaminated by
measurement errors, the true state cannot be reconstructed uniquely.
Instead, the problem is naturally formulated in a probabilistic framework.

\begin{figure}[h]
	\centering
	\includegraphics[width=0.9\textwidth]{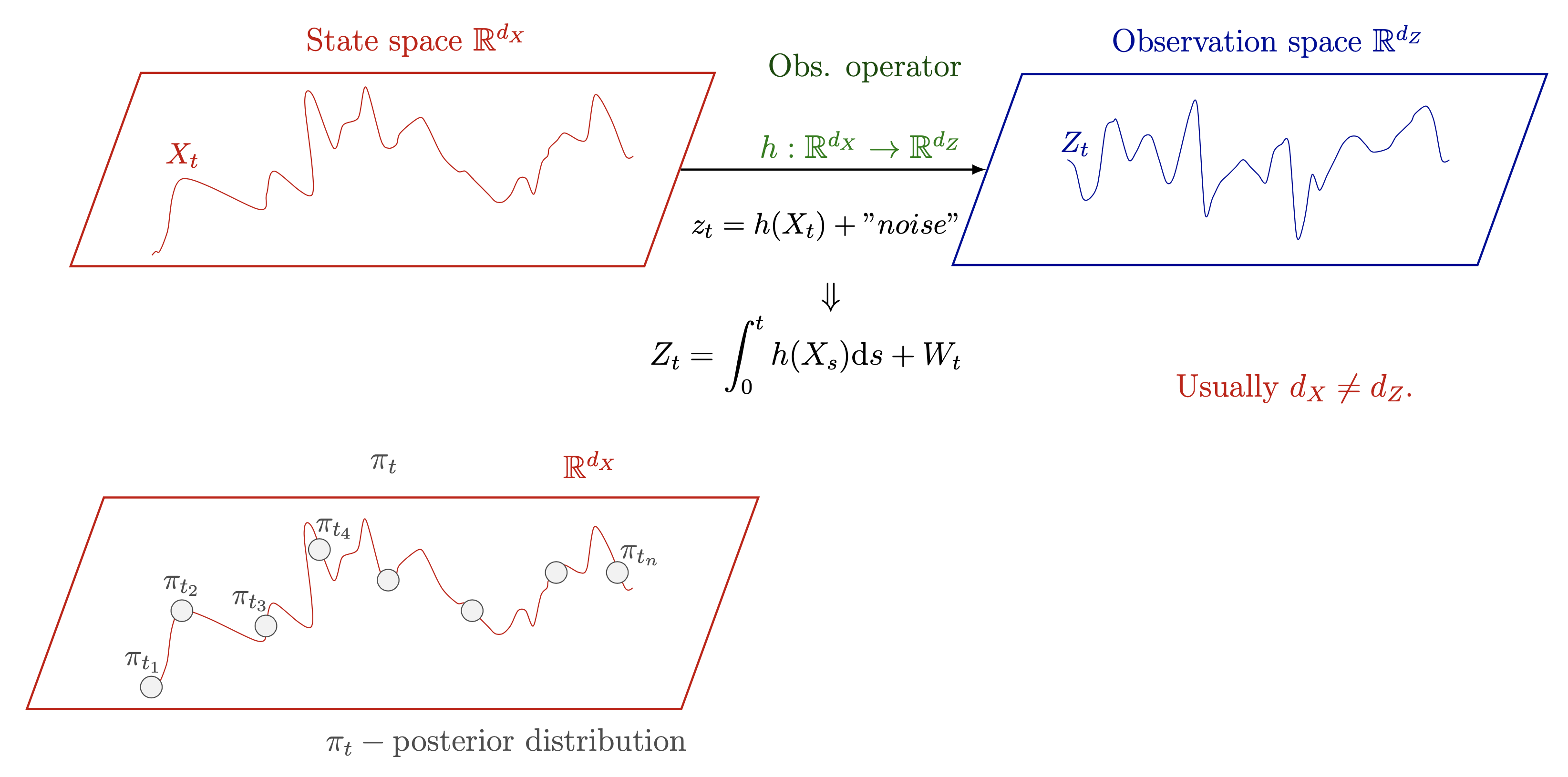}
	\caption{The nonlinear filtering problem. The hidden signal
		\(X_t\) evolves in the high-dimensional state space
		\(\mathbb R^{d_x}\), while the observations
		\(Z_t\) belong to a generally different observation space
		\(\mathbb R^{d_z}\). The observation operator
		\(h:\mathbb R^{d_x}\rightarrow\mathbb R^{d_z}\)
		maps the physical state into the observed variables. The objective of
		Bayesian filtering is not to reconstruct the trajectory itself, but to
		determine the posterior probability distribution
		\(\pi_t\) of the hidden state conditioned on the observations. The lower
		panel illustrates the evolution of the posterior distribution through
		successive observation times.}
	\label{fig:filtering_setup}
\end{figure}

Figure ~\ref{fig:filtering_setup} illustrates probabilistic framework commonly known as the filtering
problem. The unknown state (or signal) $X_t$ evolves in a high-dimensional state
space
$
\mathbb{R}^{d_x},
$
while the observations $Z_t$ belong to a generally different observation space
$
\mathbb{R}^{d_z},
$
where typically
$
d_z \ll d_x.
$

The two spaces are connected by an observation operator
$
h:\mathbb{R}^{d_x}\rightarrow\mathbb{R}^{d_z},
$
which maps the physical state into the quantities that are actually
measured. 

In continuous time, the observations are commonly modelled by
\begin{equation}
	Z_t
	=
	\int_0^t h(X_s)\,ds
	+
	W_t,
	\label{eq:continuous_observations}
\end{equation}
where \(W_t\) denotes the observation noise, usually assumed to be a
Brownian motion independent of the signal process. Equation
\eqref{eq:continuous_observations} expresses the fact that the observations
contain both deterministic information about the underlying state and
random measurement errors.

In many practical applications observations are available only at discrete
times. In this case one writes
\begin{equation}
	Z_n
	=
	h(X_n)
	+
	V_n,
	\label{eq:discrete_observations}
\end{equation}
where \(\{V_n\}\) is a sequence of independent observation errors.
Equations \eqref{eq:continuous_observations} and
\eqref{eq:discrete_observations} represent the continuous- and
discrete-time formulations of the same estimation problem. Although their
mathematical analysis differs, both lead to the same Bayesian formulation
of data assimilation.

The central object of Bayesian filtering is therefore not the trajectory
\(X_t\) itself but the evolving posterior distribution
\[
\pi_t(A)
=
\mathbb P
\left(
X_t\in A
\mid
\mathcal Z_t
\right),
\]
where
\[
\mathcal Z_t
=
\sigma(Z_s,\;0\le s\le t)
\]
denotes the filtration generated by the observations. The posterior
distribution summarises all information available about the hidden state
and therefore provides a complete probabilistic description of the current
state of knowledge.

The lower panel of Figure~\ref{fig:filtering_setup} provides a geometric
interpretation of this viewpoint. Rather than producing a single estimate
of the system state, Bayesian filtering constructs a stochastic process
taking values in the space of probability measures,
\[
t
\longmapsto
\pi_t.
\]
Each posterior distribution incorporates all observations collected up to
the current time and evolves as new information becomes available. The
problem of Bayesian filtering may therefore be viewed as the problem of
determining the evolution of the conditional distribution of the hidden
signal rather than the signal itself.

\subsubsection{The Bayesian Prediction--Analysis Cycle}

The posterior distribution introduced in the previous subsection evolves
recursively as new observations become available. The recursive structure
of Bayesian filtering is one of its most attractive features, since it
allows information to be incorporated sequentially without recomputing the
entire posterior distribution from scratch each time a new observation is
received.

Suppose that observations are available at the discrete times
\[
0=t_0<t_1<\cdots<t_n<\cdots .
\]
Assume that at time \(t_{n-1}\) the posterior distribution
\[
\pi_{n-1}
=
\mathbb P
\left(
X_{t_{n-1}}\in dx
\mid
Z_1,\ldots,Z_{n-1}
\right)
\]
has already been computed. The objective is to determine the updated
posterior distribution after the next observation \(Z_n\) becomes
available.

\begin{figure}[h]
	\centering
	\includegraphics[width=0.7\textwidth]{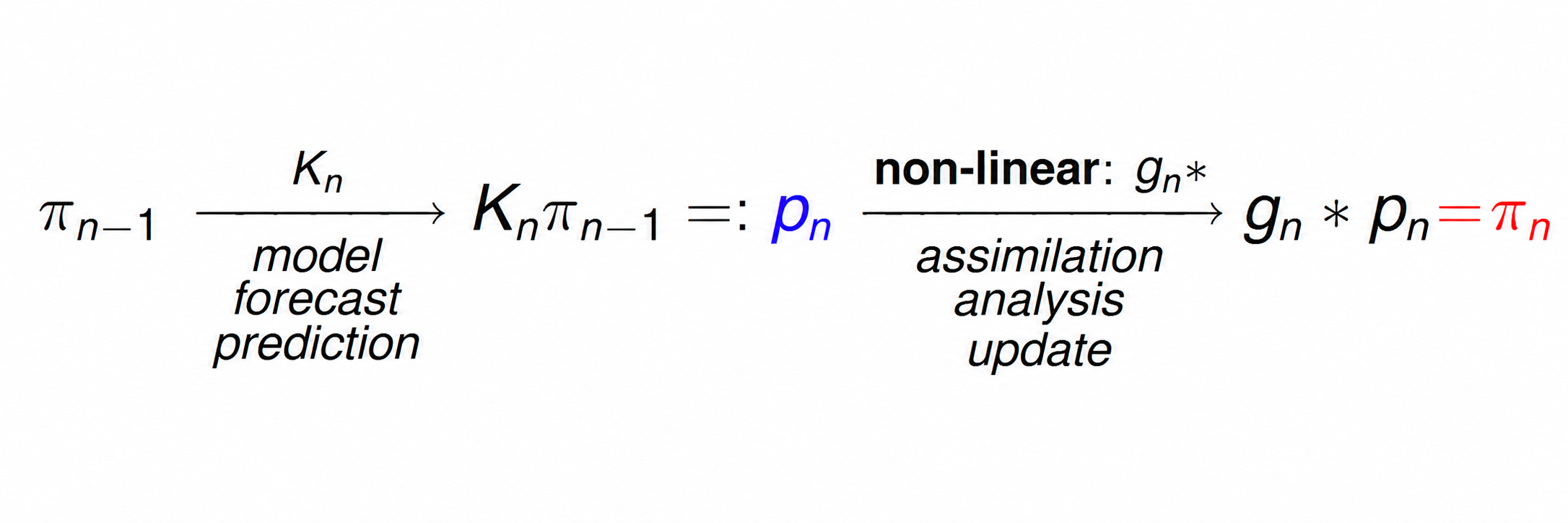}
	\caption{The Bayesian prediction--analysis cycle. The posterior
		distribution \(\pi_{n-1}\) is propagated forward by the model transition
		operator \(K_n\) to obtain the predictive distribution
		\(p_n=K_n\pi_{n-1}\). Once a new observation becomes available, Bayes'
		theorem is applied through the nonlinear analysis operator \(g_n\),
		yielding the updated posterior distribution
		\(\pi_n=g_n*p_n\).}
	\label{fig:prediction_analysis}
\end{figure}

The Bayesian recursion naturally decomposes into two successive stages:
the \emph{forecast} (or prediction) step and the \emph{analysis} (or
update) step. These two stages are illustrated schematically in
Figure~\ref{fig:prediction_analysis}.

\paragraph{Prediction.}

Between two successive observations the state evolves according to the
forecast model. If the signal is Markovian, its evolution is completely
described by the transition kernel
\[
K_n(x,A)
=
\mathbb P
\left(
X_{t_n}\in A
\mid
X_{t_{n-1}}=x
\right),
\]
which propagates probability distributions forward in time.

Applying this transition kernel to the posterior distribution
\(\pi_{n-1}\) produces the \emph{predictive} (or background)
distribution
\[
p_n
=
K_n\pi_{n-1},
\]
which represents our knowledge of the system immediately before the new
observation is assimilated. Equivalently,
\[
p_n(A)
=
\int
K_n(x,A)
\,\pi_{n-1}(dx),
\]
showing that the predictive distribution is simply the push-forward of the
previous posterior through the stochastic dynamics.

Within the STUOD programme the transition operator \(K_n\) is generated by
the stochastic transport models described in the previous sections. The
forecast uncertainty therefore reflects not only uncertainty in the
initial condition but also uncertainty arising from unresolved transport
processes represented through the calibrated stochastic parameterisations.

\begin{figure}[h]
	\centering
	\includegraphics[width=0.5\textwidth]{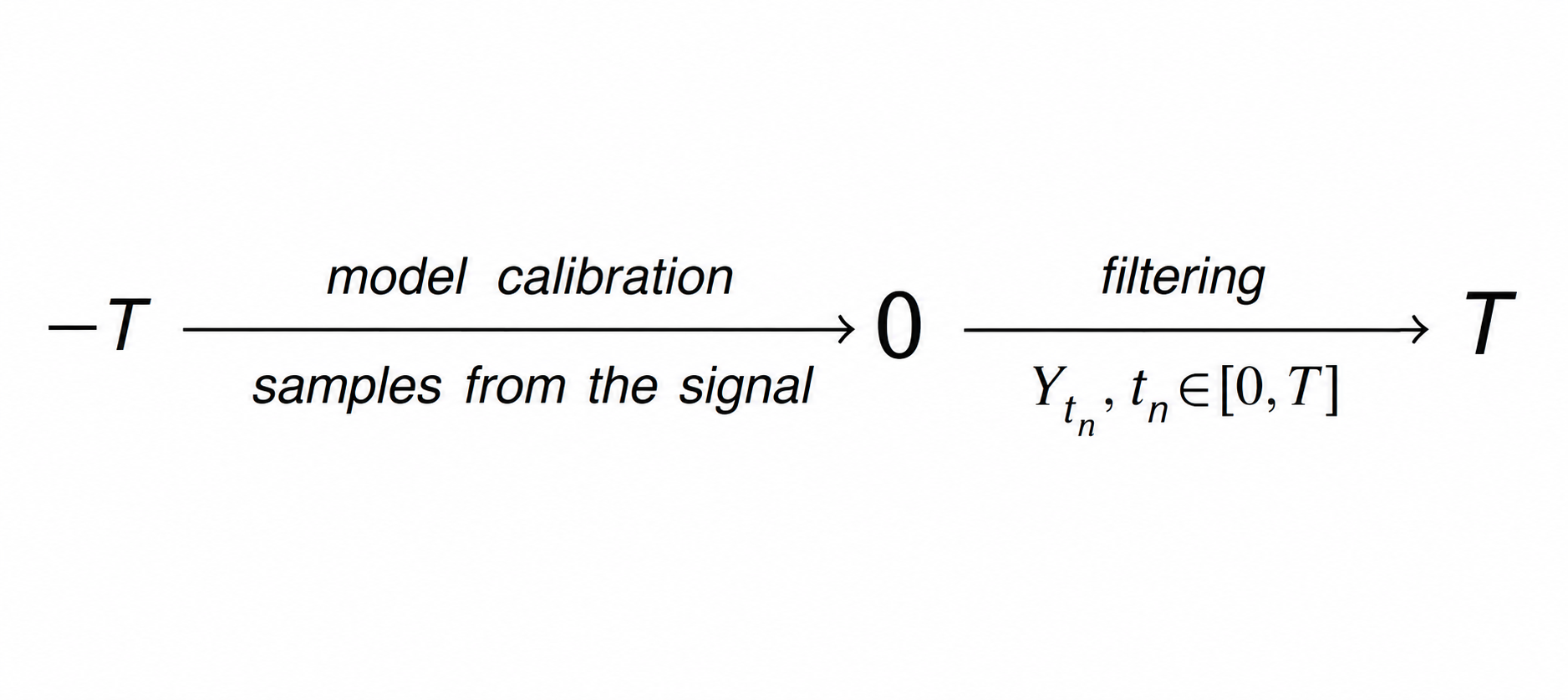}
	\caption{Overall workflow of Bayesian data assimilation. During an offline calibration phase, stochastic transport
		parameterisations are identified from historical data or high-resolution
		simulations. The calibrated stochastic model is then employed during the
		online filtering phase, where it is combined with incoming observations to
		produce sequential Bayesian estimates of the system state.}
	\label{fig:calibration_filtering}
\end{figure}

We emphasise that before Bayesian data assimilation can be carried out, the stochastic
forecast model must first be calibrated. This distinction between
\emph{model calibration} and \emph{data assimilation} is one of the central
ideas developed throughout the STUOD programme. During the calibration
stage, high-resolution numerical simulations, reanalysis products or
historical observational data are used to estimate the stochastic transport
operators that represent the effects of unresolved scales. This is an
offline procedure, performed once before the operational forecasting
period begins. Once the stochastic model has been calibrated, it is kept
fixed and used as the forecast model within the Bayesian filtering
algorithm. Data assimilation then proceeds sequentially as observations
become available, recursively updating the posterior distribution of the
system state. Figure~\ref{fig:calibration_filtering} illustrates this
workflow. The interval $[-T,0]$ corresponds to the calibration phase, in
which samples of the signal are used to estimate the stochastic model,
whereas the interval $[0,T]$ represents the operational filtering phase, in
which the calibrated model is combined with the observations
$\{Y_t:t\in[0,T]\}$ to produce sequential Bayesian estimates of the
evolving state.

\paragraph{Analysis.}

When a new observation becomes available the predictive distribution is
updated using Bayes' theorem. Let
$
g_n(z_n|x)
$
denote the likelihood of observing \(z_n\) when the system is in state
\(x\). The updated posterior distribution is then given by

\begin{equation}
	\pi_n(dx)
	=
	\frac{
		g_n(z_n|x)\,
		p_n(dx)
	}{
		\displaystyle
		\int
		g_n(z_n|y)\,
		p_n(dy)
	}.
	\label{eq:bayes_update}
\end{equation}

Equation~\eqref{eq:bayes_update} is the fundamental equation of Bayesian
data assimilation. It combines the information contained in the forecast
distribution with the newly available observation to produce the posterior
distribution.
The numerator weights each possible state according to its consistency
with the observation, while the denominator provides the normalisation
required to ensure that the posterior remains a probability measure.
In the abstract notation introduced in Figure
\ref{fig:prediction_analysis}, this nonlinear Bayesian update is written
compactly as
\[
\pi_n
=
g_n*p_n,
\]
where \(g_n\) denotes the nonlinear analysis operator acting on the
predictive distribution.


\paragraph{Recursive Bayesian filtering.}

The Bayesian filtering algorithm therefore consists of repeatedly
alternating between prediction and analysis. Beginning with an initial
distribution
$
\pi_0,
$
one constructs the sequence
\[
\pi_0
\longrightarrow
p_1
\longrightarrow
\pi_1
\longrightarrow
p_2
\longrightarrow
\pi_2
\longrightarrow
\cdots ,
\]
where each prediction step is determined by the forecast model and each
analysis step incorporates the newest observations through Bayes'
theorem.

This recursive structure makes Bayesian filtering particularly attractive
for real-time applications. New observations are incorporated
sequentially as they arrive, without requiring the complete filtering
problem to be solved again from the beginning.


\paragraph{Prior, predictive and posterior distributions.}

It is useful to distinguish three different probability distributions that
appear naturally in the Bayesian framework. The \emph{prior distribution} represents the uncertainty before any
observations have been assimilated. The \emph{predictive} (or background) distribution is obtained by
propagating the previous posterior through the forecast model.
Finally, the \emph{posterior distribution} combines the predictive
distribution with the newest observations and therefore represents the
best probabilistic estimate of the current system state.
This distinction is illustrated in
Figure~\ref{fig:prediction_analysis}. The prediction step maps the
posterior distribution to the predictive distribution through the
transition operator \(K_n\), while the analysis step maps the predictive
distribution back to the posterior distribution through Bayes' theorem.

\subsubsection*{Particle Approximations of the Posterior Distribution}

The Bayesian recursion described above is exact. Unfortunately, for
nonlinear and non-Gaussian systems the prediction and analysis operators
cannot be evaluated analytically. Even if the forecast model is known
exactly, the multidimensional integrals appearing in the prediction step
and in Bayes' formula rapidly become computationally intractable as the
dimension of the state space increases. This difficulty is particularly
severe in geophysical fluid dynamics, where operational oceanic and
atmospheric models typically contain between \(10^7\) and \(10^9\) degrees
of freedom.

A natural approach is therefore to approximate the posterior distribution
by Monte Carlo sampling. Rather than representing the posterior by a
density function, one represents it by a finite ensemble of particles,
each corresponding to a possible realisation of the hidden state. The
posterior distribution is approximated by the empirical measure

\begin{equation}
	\pi_n^N
	=
	\sum_{i=1}^{N}
	w_n^{(i)}
	\delta_{X_n^{(i)}},
	\label{eq:empirical_measure}
\end{equation}
where
$
X_n^{(i)}
$
denotes the \(i\)-th particle,
$
w_n^{(i)}\ge0,$
 $\sum_{i=1}^{N}w_n^{(i)}=1,
$
are the corresponding importance weights, and
$
\delta_x
$
denotes the Dirac probability measure concentrated at the point \(x\).
Equation~\eqref{eq:empirical_measure} replaces the infinite-dimensional
problem of evolving a probability measure by the finite-dimensional problem
of propagating an ensemble of particles. Expectations with respect to the
posterior distribution are then approximated by Monte Carlo averages,
\[
\int
\varphi(x)\,
\pi_n(dx)
\approx
\sum_{i=1}^{N}
w_n^{(i)}
\varphi(X_n^{(i)}),
\]
for suitable test functions \(\varphi\).

The Monte Carlo approximation
\eqref{eq:empirical_measure}
transforms the nonlinear filtering problem from one involving the evolution
of probability measures into the evolution of a finite set of random
samples. This observation naturally leads to the family of
\emph{Sequential Monte Carlo} (SMC) methods, more commonly known as
\emph{particle filters}. The basic idea is to propagate each particle
forward independently according to the stochastic forecast model and then
modify the ensemble whenever new observations become available so that it
remains an accurate approximation of the posterior distribution. The
Bayesian prediction and analysis operators are therefore implemented at the
level of particles rather than probability measures. Since the empirical
measure converges to the true posterior as the number of particles tends to
infinity, particle filters provide a consistent numerical approximation of
the Bayesian filtering problem while remaining entirely simulation based.
Moreover, they are applicable to highly nonlinear and non-Gaussian systems,
making them particularly attractive for stochastic transport models arising
in oceanic and atmospheric applications, where analytical representations
of the posterior distribution are generally unavailable.


\paragraph{The standard particle filter algorithm.}

The simplest member of the Sequential Monte Carlo family is the
\emph{bootstrap particle filter}. It recursively approximates the
posterior distribution by a finite ensemble of particles that are
propagated by the forecast model and periodically reweighted and resampled
using the available observations.
The approximation of the posterior distribution uses a
finite ensemble of equally weighted particles.
At the beginning of each assimilation cycle the posterior distribution is
represented by
$
\left\{
X_{n-1}^{(i)}
\right\}_{i=1}^{N},
$
where all particles have equal weights
$
w_{n-1}^{(i)}
=
\frac1N,
\qquad
i=1,\ldots,N.
$
The algorithm consists of two successive steps.\\[2mm]
\emph{Forecast.}
Each particle is propagated independently according to the stochastic
forecast model,
\[
X_n^{(i)}
\sim
K_n\!\left(X_{n-1}^{(i)},\cdot\right),
\]
producing an ensemble that approximates the predictive distribution.\\[2mm]
\begin{figure}[h]
	\centering
	\includegraphics[width=0.9\textwidth]{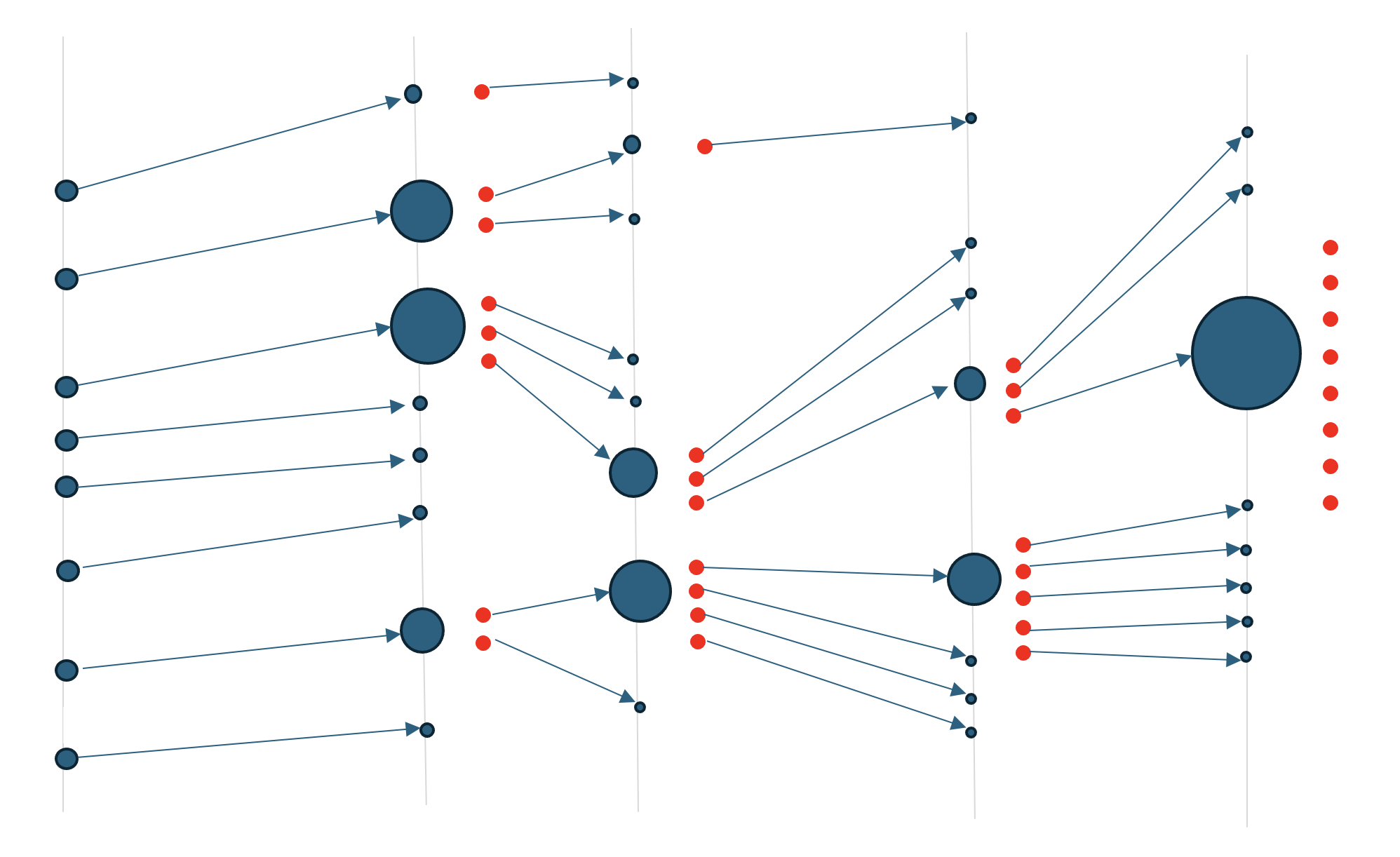}
	\caption{Illustration of the particle filtering algorithm.
		The particles evolve independently during the forecast step.
		Following the assimilation of a new observation their weights are updated
		according to the observation likelihood.
		Particles with large weights are replicated during the resampling stage,
		whereas particles with negligible weights are removed.
		The resulting ensemble provides a Monte Carlo approximation of the
		posterior distribution.}
	\label{fig:particle_filter}
\end{figure}
\emph{Analysis.}
When a new observation \(z_n\) becomes available, the particles are
weighted according to the likelihood,
$
w_n^{(i)}
\propto
g_n\!\left(z_n\,|\,X_n^{(i)}\right),
$
followed by normalisation,
\[
\sum_{i=1}^{N}w_n^{(i)}=1.
\]
Particles that are more consistent with the observations receive larger
weights, whereas particles that are inconsistent with the observations are
assigned very small weights.

A new equally weighted ensemble is then generated by sampling particles
with probabilities proportional to the importance weights. Particles with
large weights are replicated, while particles with negligible weights are
discarded. The new ensemble therefore consists of equally weighted
particles,
$
w_n^{(i)}
=
\frac1N,$
$
i=1,\ldots,N,
$
and provides the starting point for the next assimilation cycle.

The successive stages of forecasting, weighting and resampling are
illustrated schematically in Figure~\ref{fig:particle_filter}. The blue
circles represent the particles, while their sizes are proportional to
their importance weights. After the analysis step, particles that agree
well with the observations acquire large weights, whereas the remaining
particles contribute little to the approximation of the posterior
distribution. During the resampling step these highly weighted particles
are replicated, producing the offspring shown in red, while particles with
small weights disappear.

Although resampling restores equal particle weights, it does not prevent
particle collapse. In high-dimensional systems the observations are so
informative that, after only a few assimilation cycles, the offspring are
typically descended from only a very small number of ancestral particles.
As a result, the particle cloud rapidly loses diversity and becomes
concentrated in a small region of the state space. In the extreme case,
which is frequently encountered in large-scale geophysical applications,
all particles become descendants of a single ancestor, and the particle
filter effectively collapses. This phenomenon, known as the
\emph{curse of dimensionality} of particle filtering, is the principal
motivation for the advanced particle  filtering methodologies described in the following sections.

\subsubsection{Particle filtering in high-dimensional systems}

The standard bootstrap particle filter is conceptually simple and is
provably consistent as the number of particles tends to infinity.
Unfortunately, this asymptotic result is of limited practical value for
high-dimensional systems. In applications such as numerical weather
prediction, ocean modelling and climate dynamics, the state vector may
contain millions or even billions of degrees of freedom, while only a few
hundred or thousand particles can be afforded computationally.

The fundamental difficulty is that, in high dimensions, the likelihood
function becomes extremely concentrated. Following the forecast step,
different particles typically produce very different predictions at the
observation locations. Consequently, one or two particles receive almost
all of the probability mass, while the remaining particles obtain
negligible weights. Although the subsequent resampling step restores equal
weights, it simply creates many copies of these few surviving particles.
After only a small number of assimilation cycles the ensemble loses nearly
all diversity and effectively becomes descendants of a single ancestral
particle. This phenomenon is the principal manifestation of the curse of
dimensionality for particle filtering.

Modern particle filtering algorithms therefore augment the bootstrap
particle filter with additional mechanisms whose purpose is to maintain
particle diversity while preserving consistency with the target posterior
distribution. Within the applications below, we make extensive use of three
such techniques: \emph{tempering}, \emph{jittering}, and
\emph{nudging}. Although they are conceptually distinct, they are designed
to complement one another\footnote{See also \cite{VanLeeuwen2009} and \cite{VanLeeuwen2010} for further 
	developments.}.

\paragraph{Tempering.}

The principal difficulty in particle filtering for high-dimensional systems
is that the predictive and posterior distributions may differ
substantially. In this case the
importance weights become extremely concentrated, so that only a very
small number of particles receive significant weight while the remainder
effectively disappear. This phenomenon, known as particle degeneracy,
causes the effective sample size to collapse and renders the standard
bootstrap particle filter ineffective. Tempering alleviates this difficulty by replacing a single likelihood
update with a sequence of intermediate Bayesian updates. Rather than
assimilating the entire observation at once, one introduces a family of
intermediate probability distributions connecting the predictive
distribution $p(x)$ and the posterior distribution $\pi(x)$ through

\[
\pi_r(x)
\propto
p(x)L(x)^{\phi_r},
\qquad
0=\phi_0<\phi_1<\cdots<\phi_R=1,
\]
where $L(x)$ denotes the likelihood and $\phi_r$ is referred to as the
\emph{temperature}. The predictive distribution corresponds to
$\phi_0=0$, while the final posterior is recovered when $\phi_R=1$.
Equivalently, the particle weights are updated incrementally according to

\[
w_i^{(r)}
\propto
\left(
L(x_i)
\right)^{\phi_r-\phi_{r-1}},
\]
so that only a fraction of the observational information is incorporated
during each tempering step.

The sequence of temperatures is chosen adaptively rather than fixed in
advance. Starting from the current value of $\phi_{r-1}$, the next
temperature $\phi_r$ is selected so that the effective sample size (ESS)
of the intermediate weighted ensemble remains above a prescribed
threshold. In practice this can be determined efficiently using a simple
one-dimensional search procedure such as binary search. Once the
intermediate weights have been computed, the particles are resampled and, an MCMC jittering step (as described below) is applied before
continuing to the next temperature. This repeated cycle of tempering,
resampling and jittering replaces the single resampling step of the
standard bootstrap particle filter.

\begin{figure}[h]
	\centering
	\includegraphics[width=0.85\textwidth]{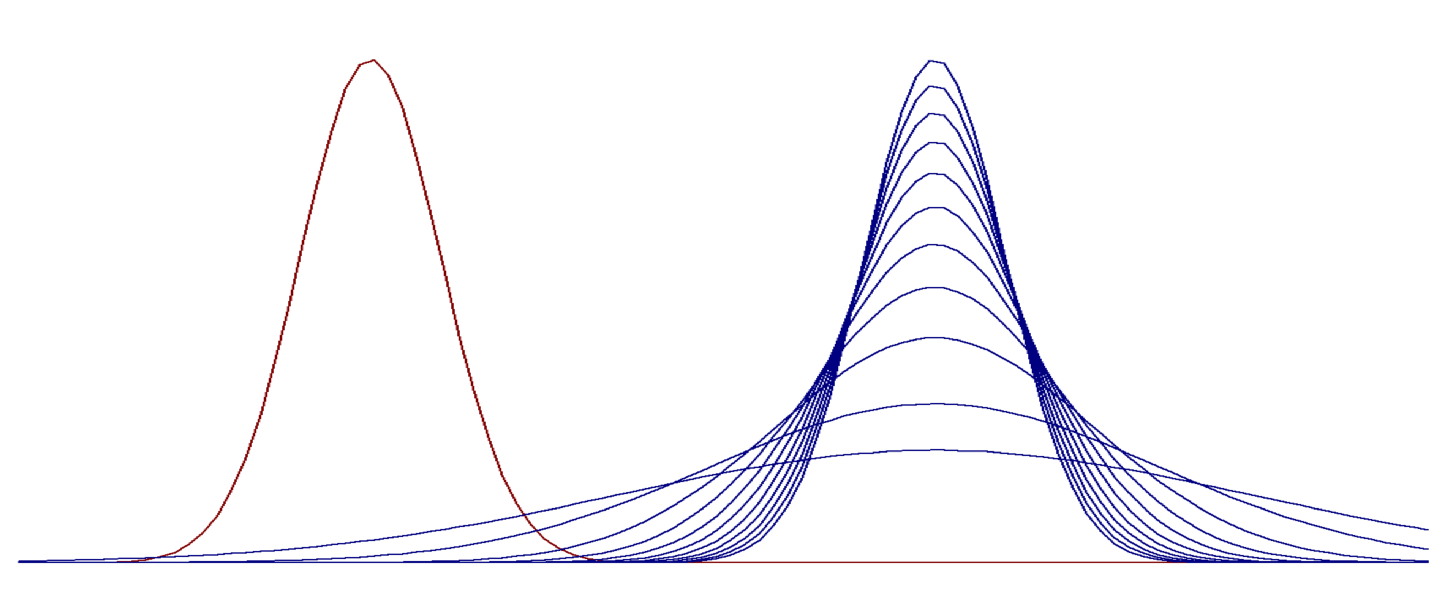}
	\caption{
		Illustration of adaptive tempering. The red curve represents the
		predictive (forecast) distribution prior to assimilating the observations,
		while the final narrow blue curve represents the posterior distribution
		obtained after the full Bayesian update. The intermediate blue curves are
		the tempered distributions,
		obtained by gradually increasing the temperature parameter.
		Instead of performing one highly concentrated likelihood update, the
		particle filter progresses through this sequence of intermediate
		distributions. At each stage the particles are reweighted, resampled, and
		optionally rejuvenated using an MCMC move, maintaining a sufficiently
		large effective sample size and preventing particle degeneracy. As the
		temperature increases, the intermediate distributions become increasingly
		concentrated around the posterior, allowing the particle ensemble to move
		smoothly from the forecast to the analysis.
	}
	\label{fig:tempering}
\end{figure}

Figure~\ref{fig:tempering} illustrates the basic idea. The red curve
represents the predictive distribution, while the narrow blue curve
corresponds to the posterior distribution obtained after assimilating the
full observation. If the posterior were imposed in a single step, only
those few particles already lying close to the blue peak would receive
appreciable weight, leading to severe particle degeneracy. Tempering
instead constructs the sequence of intermediate blue distributions shown
in the figure. Their variances decrease gradually as the temperature
increases, allowing the particle ensemble to migrate progressively from
the predictive distribution towards the posterior while maintaining
sufficient diversity at every stage. The resulting sequence of well-
conditioned updates is substantially more robust than a single highly
concentrated likelihood weighting step. Although a large number of observations may become
available simultaneously at one analysis time, they are incorporated only
gradually through the intermediate distributions. In effect, one large
Bayesian update is replaced by several smaller ones, each chosen so that
the ensemble remains informative. This adaptive procedure has proved
essential for the successful application of particle filters to the
high-dimensional models. 

\paragraph{Jittering.}

Tempering substantially reduces the severity of particle degeneracy but
cannot eliminate the loss of diversity caused by repeated  resampling.
Although resampling restores the effective sample size, it also creates many
identical copies of the same particle.  
Jittering, seeks to restore
particle diversity immediately after resampling. A principled approach is to construct the perturbation using a
Markov chain whose invariant distribution is pecisely the posterior
distribution. This is commonly achieved by performing one or more
Metropolis--Hastings (MH) updates after each resampling step. Starting
from the replicated particles, a proposal is generated, typically by
propagating the particle according to the underlying stochastic forecast
model. The proposal is then accepted or rejected according to the standard
Metropolis--Hastings acceptance probability. Since the posterior
distribution (or the intermediate distributions that arise as an effect of tempering) is invariant under the MH transition kernel, the resulting
rejuvenated ensemble preserves the target distribution exactly while
simultaneously restoring particle diversity.

In practice, tempering and jittering should be viewed as complementary
techniques. Tempering prevents severe weight collapse by replacing a
single large Bayesian update with a sequence of smaller ones, whereas
jittering restores the diversity inevitably lost during the subsequent
resampling step. Together they form one of the key ingredients enabling
particle filters to operate successfully in the high-dimensional settings. For a theoretical justification of the tempering and nudging procedure, see 
\cite{BeskosCrisanJasra2014} and  \cite{BeskosCrisanJasraKamataniZhou2017}.

\paragraph{Nudging.}

Whereas tempering and jittering act during the analysis stage, nudging
modifies the forecast dynamics themselves. The basic idea is to steer the
particle trajectories towards regions of the state space that are more
compatible with the available observations before the importance-weighting
step is performed. Instead of allowing each particle to evolve solely
according to the stochastic forecast model, an additional control term is
introduced into the dynamics that gently attracts the particles towards
the observations.

For example, if the forecast model is written abstractly as
\[
dX_t=b(X_t)\,dt+\sigma(X_t)\,dW_t,
\]
the nudged dynamics take the form
\[
d\widetilde X_t
=
\bigl(
b(\widetilde X_t)
+
f(\widetilde X_t,t)
\bigr)\,dt
+
\sigma(\widetilde X_t)\,dW_t,
\]
where the control
\(u\)
is designed using the current observations, typically through the
observation operator, so that the propagated particles remain close to the
regions of high posterior probability.

The principal advantage of nudging is that the proposal distribution is
brought closer to the posterior distribution \emph{before} the Bayesian
analysis step. Consequently, the likelihood values become much less
variable across the particle ensemble, substantially reducing the variance
of the importance weights and greatly alleviating particle degeneracy. In
high-dimensional chaotic systems this improvement can be dramatic, since
the forecast trajectories no longer drift as far from the true state
during the assimilation interval.

However, nudging fundamentally changes the proposal distribution used to
generate the particles. If the standard likelihood weights were employed,
the resulting particle approximation would converge to the wrong target
distribution. To preserve the asymptotic consistency of the particle
filter, the importance weights must therefore be corrected to account for
the discrepancy between the original transition kernel and the nudged
proposal kernel.

More precisely, if
\(K(x,\cdot)\)
denotes the original transition kernel and
\(\widetilde K(x,\cdot)\)
the kernel corresponding to the nudged dynamics, then the importance
weights become
\[
w_n^{(i)}
\propto
w_{n-1}^{(i)}
\,
L\!\left(
z_n\mid \widetilde X_n^{(i)}
\right)
\,
\frac{dK\!\left(X_{n-1}^{(i)},\cdot\right)}
{d\widetilde K\!\left(X_{n-1}^{(i)},\cdot\right)}
\!\left(
\widetilde X_n^{(i)}
\right),
\]
where the Radon--Nikodym derivative compensates exactly for the change of
measure induced by the nudging. For diffusion models this correction is
given by Girsanov's theorem, while for discrete-time models it reduces to
the ratio of the original and proposal transition densities. This
importance-weight correction guarantees that the particle approximation
remains asymptotically consistent despite the modification of the forecast
dynamics.

The nudging strength should therefore be chosen so
that the proposal distribution moves sufficiently close to the posterior
while remaining statistically close to the original forecast model.\\[2mm]

These three procedures address different aspects of particle degeneracy.
Tempering prevents abrupt weight collapse, jittering restores diversity
after resampling, and nudging constructs proposal trajectories that remain
close to regions of high posterior probability. Used together, they enable
particle filters to be applied successfully to high-dimensional
geophysical models where the standard bootstrap particle filter fails.

\subsection{Case Studies in Particle Filtering Data Assimilation}

The methodological framework developed in the preceding sections has been implemented and tested on several  geophysical models. These examples illustrate how 
particle filtering data assimilation can be applied to high dimensional filtering problems. Each case study addresses a common challenge: assimilating sparse observational data into a computationally affordable stochastic model that accurately represents the uncertainty arising from unresolved scales.

\subsubsection{Two-Dimensional Euler Equation}

The case study analysed in  \cite{CotterCrisanHolmPanShevchenko2020PF}  
is for the solution of a two-dimensional Euler equation, as introduced in  subsection~\ref{sec:05_04_calibration}, where the
Eulerian calibration methodology is described in detail. 
Each particle in
the ensemble evolves independently according to this stochastic model,
thereby generating an ensemble approximation of the predictive
distribution.

The objective is to reconstruct the evolving velocity field from sparse
noisy observations. Synthetic observations are generated from a
high-resolution deterministic reference solution, which serves as the
ground truth. These observations consist of noisy pointwise measurements
of the velocity field collected at prescribed observation locations. Two observation scenarios are
considered. In the \emph{perfect-model} setting, the observations are generated
directly from the reference solution, whereas in the \emph{imperfect-model}
setting they are obtained from a coarse-grained version of the deterministic
velocity field. The coarse graining is performed by applying an inverse Helmholtz
operator to the stream function, thereby removing scales that cannot be resolved
by the stochastic model while preserving the underlying fine-resolution dynamics.
In both cases the observations consist of noisy pointwise measurements of the
velocity field recorded at a regular array of synthetic ``weather stations''. 
The reference ("truth") is generated by a high-resolution deterministic simulation on a
$512\times512$ grid, corresponding to approximately $2.6\times10^5$ spatial degrees of
freedom. In contrast, the stochastic model used for filtering evolves on a
$64\times64$ grid, reducing the state dimension by a factor of $64$ while retaining the
large-scale flow structures. The observations consist of noisy pointwise velocity
measurements at regular arrays of synthetic weather stations comprising
$81$, $289$ or $1089$ locations. Consequently, even the densest observation network
samples only about $27\%$ of the coarse stochastic grid and less than $0.5\%$ of the
degrees of freedom of the high-resolution truth. This severe mismatch between the
dimension of the state and that of the observations is characteristic of practical
geophysical data assimilation problems.

The posterior distribution is approximated by a Sequential Monte Carlo
method that combines adaptive tempering, systematic resampling and
Metropolis--Hastings jittering.

\begin{figure}[h]
	\centering
	\includegraphics[width=\textwidth]{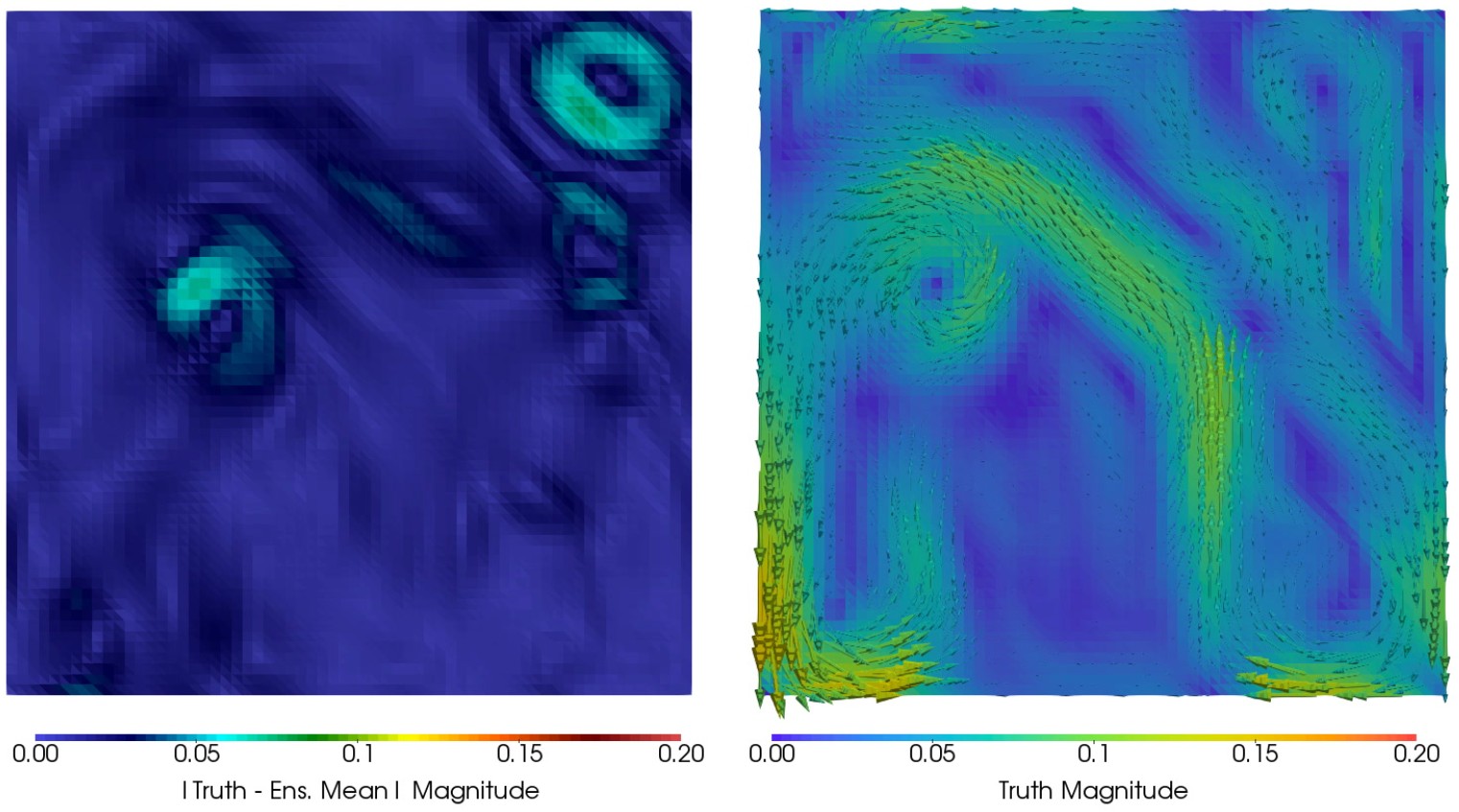}
	\caption{
		Representative data assimilation results for the stochastic
		two-dimensional Euler equations. The right panel shows the reference
		velocity magnitude together with the associated velocity vectors. The left
		panel displays the magnitude of the difference between the reference
		solution and the posterior ensemble mean obtained by the adaptive
		particle filter. The assimilation algorithm reconstructs the
		large-scale coherent vortical structures, with the largest errors
		confined to regions of strong velocity gradients.
	}
	\label{fig:euler_DA}
\end{figure}
Representative assimilation results are shown in
Figure~\ref{fig:euler_DA}. The right panel displays the reference velocity
field, while the left panel shows the pointwise error between the
posterior ensemble mean and the reference solution. The adaptive particle
filter successfully reconstructs the dominant coherent vortices, with the
remaining errors concentrated primarily in regions where the flow exhibits rapid temporal evolution.

\subsubsection{The Two-Layer Quasi-Geostrophic Equations}

This case study covered in \cite{CotterCrisanHolmPanShevchenko2020JSP} considers Bayesian data assimilation for the
two-layer quasi-geostrophic (QG) equations. This model provides a more realistic
representation of large-scale ocean circulation. The stochastic parameterisation of the unresolved transport is
obtained using the Eulerian calibration methodology described in
subsection~\ref{enough}. The resulting stochastic QG model
forms the forecast model for the particle filtering algorithm.

The governing equations and their stochastic parameterisation were
introduced in subsection~\ref{enough}. The stochastic transport
vector fields are calibrated from high-resolution deterministic
simulations and incorporated into the coarse-grid model 5through transport
noise. This enables the reduced-order model that preserve the circulation
properties of the quasi-geostrophic dynamics.
Each member
of the particle ensemble evolves independently according to the calibrated
stochastic QG equations.

Synthetic observations are generated from a high-resolution reference
simulation that serves as the truth. The observations consist of sparse,
noisy measurements of the potential vorticity field at a prescribed set of
observation locations distributed throughout the computational domain.
Independent Gaussian observational errors are assumed throughout the
numerical experiments.

The objective of the particle filter is therefore to reconstruct the full
potential-vorticity field from these sparse observations while
simultaneously quantifying the uncertainty associated with the forecast.

The reference (``truth'') is generated by a high-resolution deterministic
simulation of the two-layer quasi-geostrophic equations, while the Bayesian
filter operates on the calibrated stochastic reduced-order model described in
Section~\ref{enough}. The reference solution contains
approximately $O(10^6)$ degrees of freedom, whereas the stochastic forecast
model evolves only $O(10^4)$ degrees of freedom after model reduction. The
observations therefore represent only a very small fraction of the information
contained in the underlying flow, see 
\cite{CotterCrisanHolmPanShevchenko2020JSP} for details. 

Rather than observing the potential-vorticity field directly, the observations
consist of noisy pointwise measurements of the velocity field. These
measurements are collected at a regular array of synthetic ``weather stations''
distributed uniformly throughout the computational domain. The observation
operator projects the model velocity from the signal grid onto the observation
grid, and independent Gaussian observational errors are added. The standard
deviation of the observational noise is not prescribed arbitrarily but is chosen
to equal the local standard deviation of the velocity field over the
corresponding coarse-grid cell, providing a realistic measure of observational
uncertainty (details in \cite{CotterCrisanHolmPanShevchenko2020JSP}).

\begin{figure}[h]
	\centering
	\includegraphics[width=\textwidth]{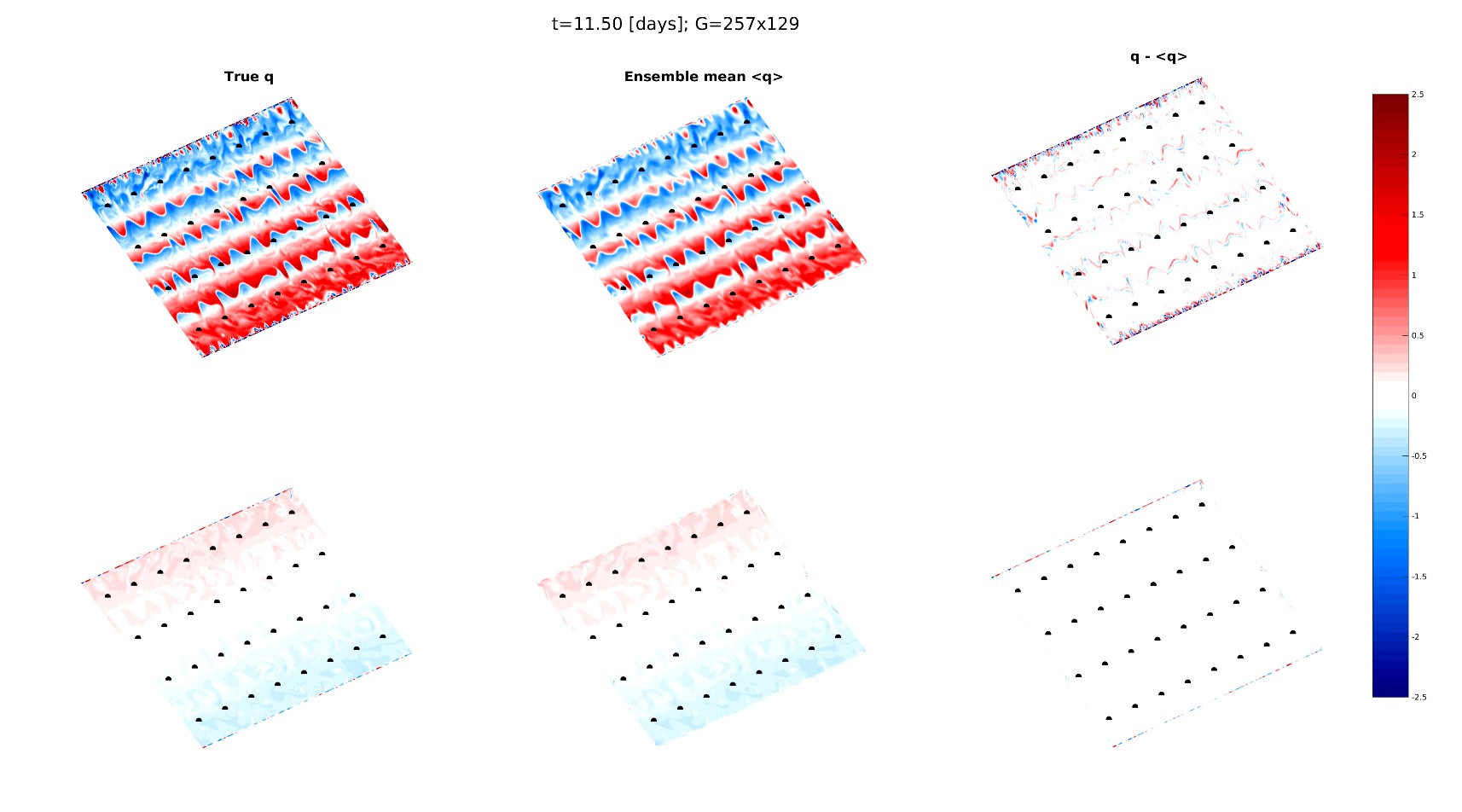}
	\caption{
		Representative Bayesian data assimilation results for the stochastic
		two-layer quasi-geostrophic equations. The left column shows the reference
		potential-vorticity field, the middle column the posterior ensemble mean
		obtained by the adaptive particle filter, and the right column the
		difference between the two. Black dots indicate the observation
		locations. The upper row corresponds to the upper layer, while the lower
		row shows the lower layer. Despite the relatively sparse observational
		network, the adaptive particle filter accurately reconstructs the
		large-scale flow structures in both layers. The remaining errors are
		primarily confined to regions of strong potential-vorticity gradients,
		illustrating the effectiveness of combining stochastic model reduction,
		adaptive nudging, tempering and Metropolis--Hastings jittering.
	}
	\label{fig:qg_DA}
\end{figure}

Several observation networks are considered in order to investigate the effect
of observational density on the performance of the particle filter. The paper \cite{CotterCrisanHolmPanShevchenko2020JSP}
reports experiments using $16$ and $32$ weather stations, while the underlying
methodology also considers denser regular observation grids. The weather
stations are placed at fixed uniformly distributed locations and remain fixed
throughout the assimilation window. Figure~\ref{fig:qg_DA} illustrates the
configuration used in the numerical experiments: the black markers indicate the
weather stations overlaid on the true potential-vorticity field in each layer.
Even with only $32$ observation locations, corresponding to a minute fraction
of the state dimension, the adaptive particle filter successfully reconstructs
the large-scale flow structures.

The stronger nonlinearities and increased sensitivity to initial
conditions exhibited by the quasi-geostrophic equations make particle
degeneracy significantly more severe than in the Euler case. Consequently,
each assimilation cycle consists of the following stages: a. 
 propagation of the particles using the stochastic QG forecast model;
b.  adaptive nudging of the forecast trajectories towards the incoming
	observations;
c. adaptive likelihood tempering;
d.  systematic resampling when the effective sample size falls below a
	prescribed threshold;
e.  Metropolis--Hastings jittering to restore ensemble diversity.

The nudging step is particularly important for this application. During
the forecast interval an observation-dependent control is incorporated
into the stochastic dynamics, steering the particles towards regions of
state space that are more consistent with the observations before the
Bayesian update is performed. Since this modifies the proposal
distribution, the particle weights are corrected using the corresponding
Radon--Nikodym derivative (computed through Girsanov's theorem), ensuring
that the particle approximation remains asymptotically consistent with the
target posterior distribution.

Combined with adaptive tempering and particle, nudging
substantially reduces the variance of the importance weights and enables
the particle filter to operate successfully despite the high-dimensional
nature of the quasi-geostrophic dynamics.

Representative assimilation results are presented in
Figure~\ref{fig:qg_DA}. The posterior ensemble mean reproduces the
large-scale circulation patterns present in both layers with remarkable
accuracy, while the error field demonstrates that small discrepancies
remain after assimilation. The observation locations are indicated by the
black markers, illustrating that the complete flow field is recovered from
relatively sparse measurements through Bayesian inference.

The numerical experiments demonstrate the importance of combining several
algorithmic enhancements. Adaptive tempering prevents weight
collapse, Metropolis--Hastings jittering restores particle diversity after
resampling, while nudging significantly improves the proposal
distribution by steering the forecast trajectories towards the
observations before the Bayesian update. Together these ingredients make
particle filtering feasible for this strongly nonlinear geophysical model.

\subsubsection{The Rotating Shallow Water Equation}

The rotating shallow water (SRSW) equation provides a
natural intermediate level of complexity between the two-dimensional
Euler equations and fully operational atmospheric models. They retain
the essential large-scale balances of geophysical fluid dynamics,
including geostrophic balance, inertia--gravity waves and nonlinear
transport, while remaining computationally tractable. The stochastic
model used in \cite{LangVanLeeuwenCrisanPotthastRSW2022} is the SALT formulation introduced in
subsection \ref{enough3}.

The truth is generated by an independent realization of the stochastic
rotating shallow water equations, while the particle filter propagates
an ensemble of stochastic forecasts. The observations consist of noisy
measurements of the pressure field taken at prescribed assimilation
times. Different observation scenarios are investigated, varying both
the number of observations available at each assimilation cycle and the
time interval between successive analyses. These experiments are
designed to quantify the influence of observation density and
assimilation frequency on the performance of the particle filter.

The Bayesian inference is performed using an adaptive particle filter. During the analysis stage, adaptive tempering introduces a
sequence of intermediate posterior distributions whose temperatures are
chosen using the effective sample size criterion. Each tempering step
is followed by multinomial resampling and a Metropolis--Hastings
jittering procedure which restores particle diversity while preserving
the target posterior distribution. No localisation or covariance
inflation is employed, making the experiments a direct test of the
high-dimensional particle filter itself.

\begin{figure}[h]
	\centering
	
	\includegraphics[width=0.8\textwidth]{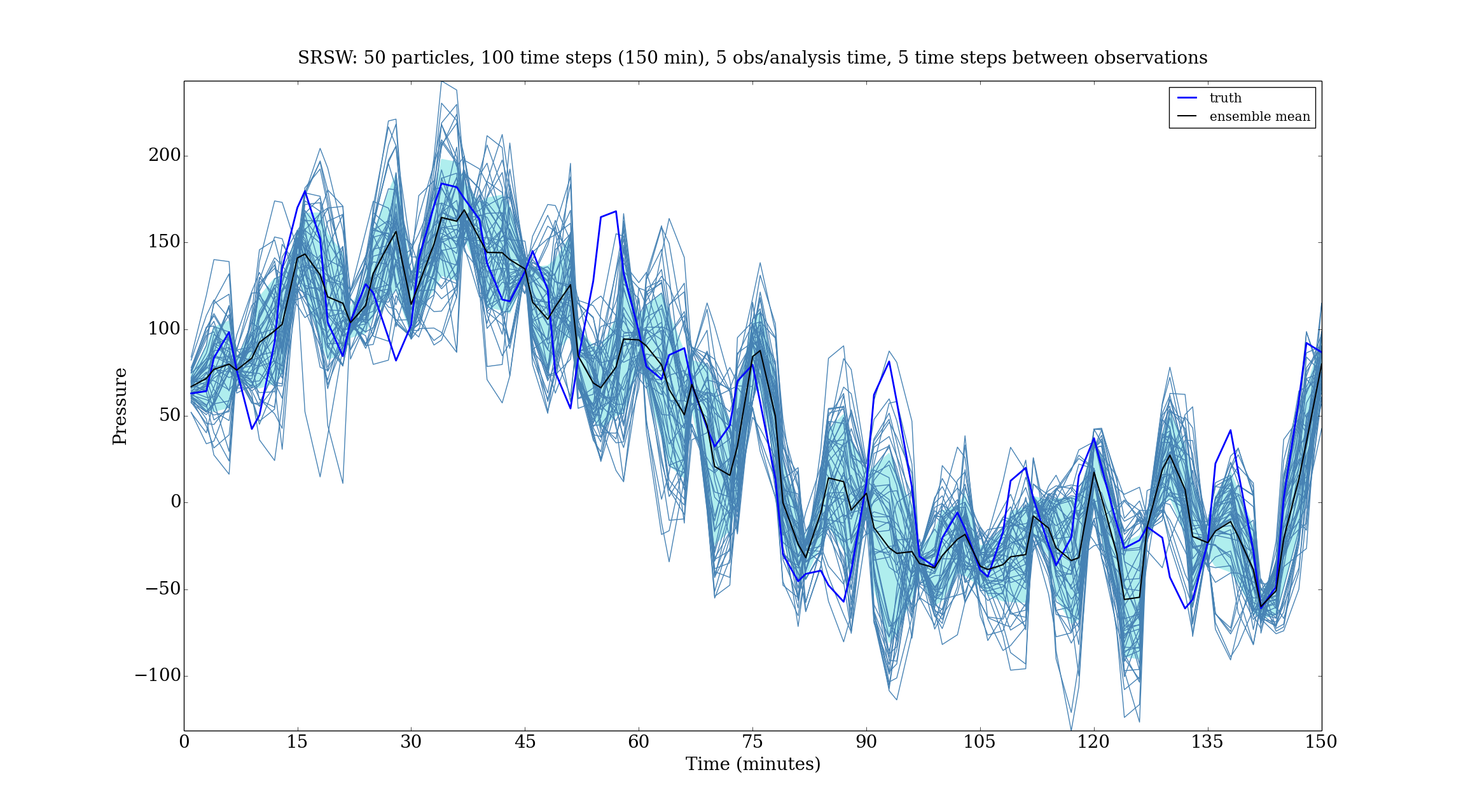}
	
	\vspace{0.8em}
	
	\includegraphics[width=0.8\textwidth]{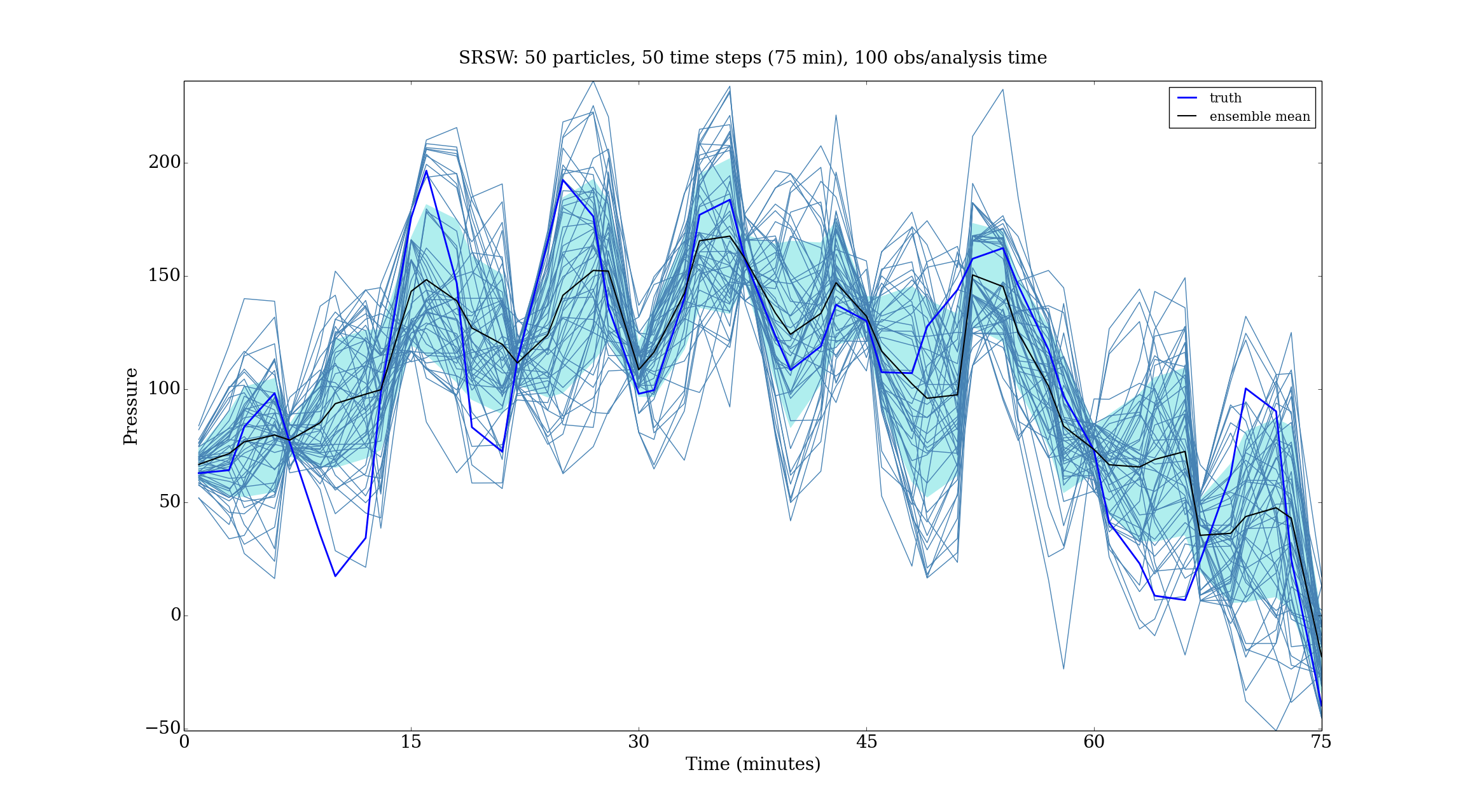}
	
	\caption{
		Bayesian data assimilation for the stochastic rotating shallow water
		equations using the adaptive particle filter with tempering and
		Metropolis--Hastings jittering.
		\textbf{Top:} assimilation over $100$ model time steps using $50$
		particles, five observations per analysis time, and observations every
		five model steps. The ensemble mean closely follows the truth while the
		ensemble spread remains well controlled.
		\textbf{Bottom:} assimilation over $50$ model time steps using $50$
		particles and $100$ observations at each analysis time. Increasing the
		number of observations produces a significantly tighter posterior
		ensemble and a more accurate reconstruction of the truth, illustrating
		the strong dependence of the filter performance on the observational
		information available.
	}
	\label{fig:srsw_da}
\end{figure}

Figure~\ref{fig:srsw_da} illustrates two representative data
assimilation experiments. In the first experiment (top panel), the
ensemble consists of $50$ particles and the system is evolved for
$100$ model time steps. Assimilation is performed every five model
steps using five pressure observations at each analysis time. The
ensemble mean tracks the truth throughout the assimilation window,
while the ensemble spread remains well controlled despite the
stochastic forcing.

The second experiment (bottom panel) investigates the effect of
increasing the observational information. The assimilation window is
reduced to $50$ model time steps, while the number of observations at
each analysis time is increased to $100$. The additional observations
produce a noticeably tighter posterior ensemble and substantially
improve the agreement between the ensemble mean and the truth. This
illustrates one of the principal conclusions of the study: increasing
either the observation frequency or the number of observations per
assimilation cycle leads to a significant improvement in the quality of
the posterior estimate.

More generally, the numerical experiments demonstrate that the adaptive
particle filter remains effective even for the high-dimensional
stochastic rotating shallow water model. Accurate posterior
distributions are obtained without employing localisation, a technique
that is usually regarded as essential for particle filtering in
geophysical applications. The experiments therefore provide strong
evidence that adaptive tempering combined with Metropolis--Hastings
jittering substantially alleviates the particle degeneracy problem in
high-dimensional Bayesian data assimilation.

\subsubsection{Bayesian inference with generative stochastic parameterisations}

The previous subsection demonstrated that adaptive particle filtering
can successfully assimilate observations for the rotating
shallow water equation when the unresolved scales are represented by a
prescribed stochastic transport model. A natural question is how such a
stochastic parameterisation should be calibrated from data. In
\cite{LobbeCrisanLang2025} this question is addressed by combining
modern generative machine learning with Bayesian data assimilation.

The principal idea is to replace the classical Gaussian calibration of
the stochastic transport noise by a data-driven calibration using a
diffusion Schrödinger Bridge (DSB). High-resolution simulations of the
rotating shallow water equations are first performed on a fine
$128\times128$ grid. A low-pass filter is then applied to construct a
coarse representation of the solution on a $32\times32$ grid, and the
difference between the two resolutions is interpreted as the unresolved
component of the dynamics. Instead of assuming that these unresolved
increments follow a Gaussian distribution, the probability law is
learnt directly from the data using a diffusion generative model. Once
trained, the DSB generates synthetic noise samples having essentially
the same statistical properties as the unresolved scales extracted from
the high-resolution simulations.

The resulting stochastic model is therefore no longer based on an
analytically prescribed covariance operator but on a learned
distribution obtained from numerical simulations. The generated samples
are subsequently incorporated into the stochastic rotating shallow
water equations, producing a hybrid model that captures the influence
of unresolved dynamics while remaining computationally inexpensive.
Bayesian inference is then carried out using the adaptive particle
filter with tempering, resampling and Metropolis--Hastings jittering
introduced in the previous subsections. 

The calibration experiments demonstrate that the diffusion
Schr\"odinger Bridge accurately reproduces the non-Gaussian statistics of
the unresolved scales. In particular, the empirical distribution of the
noise extracted from the deterministic simulations is distinctly
bimodal, and this feature is faithfully reproduced by the trained
generative model, unlike the Gaussian approximations used previously.
Forecast verification through continuous ranked probability scores
(CRPS) and rank histograms confirms that the calibrated stochastic
model provides realistic ensemble forecasts over a wide range of
forecast horizons. 

The data assimilation experiments employ ensemble sizes ranging from
50 to 200 particles together with different observation densities and
assimilation windows. The particle filter remains stable throughout the
experiments despite the considerable reduction in model dimension from
the $128\times128$ reference model to the $32\times32$ stochastic
forecast model. Mean bias and root mean square error remain below the
observation noise level in all experiments, demonstrating that the
combination of generative calibration and adaptive particle filtering
can successfully recover the evolving state of the high-resolution
system.

\begin{figure}[h]
	\centering
	\includegraphics[width=0.82\textwidth]{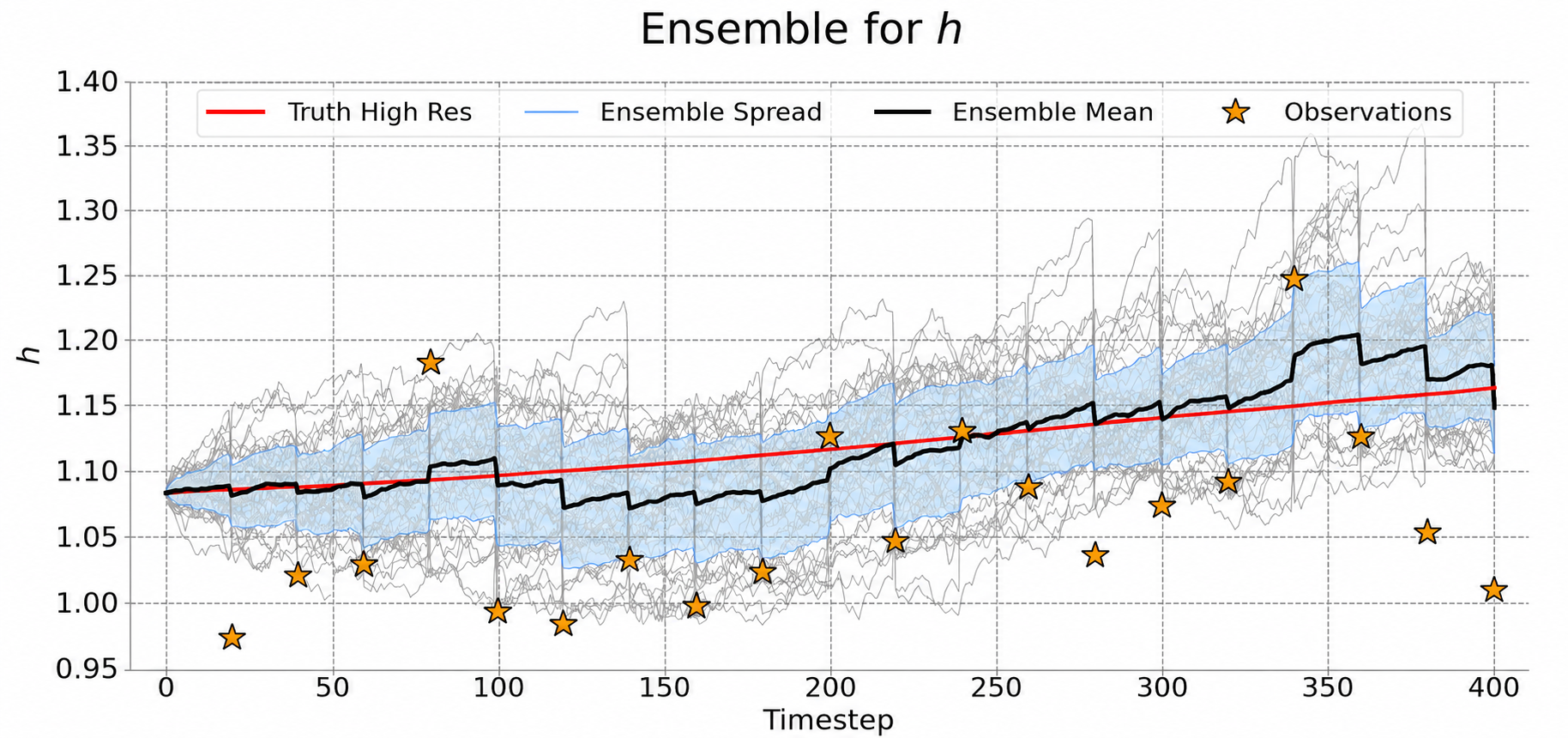}
	
	\caption{
		Bayesian data assimilation for the stochastic rotating shallow water
		equations using a stochastic parameterisation calibrated by a diffusion
		Schrödinger Bridge. The grey curves represent the ensemble members, the
		black curve is the ensemble mean, the red curve denotes the
		high-resolution reference solution, and the stars indicate noisy
		observations. Between assimilation times the ensemble spreads under the
		learned stochastic dynamics, while each analysis step contracts the
		ensemble towards the observations through adaptive tempering,
		resampling and Metropolis--Hastings jittering. The experiment
		illustrates that a reduced-order stochastic model whose unresolved
		scales are generated by a diffusion model can accurately recover the
		high-resolution solution.}
	\label{fig:gm_da}
\end{figure}

Figure~\ref{fig:gm_da} illustrates a typical filtering experiment. The
grey curves denote the ensemble members, while the black curve is the
ensemble mean and the red curve is the reference high-resolution
solution. Between assimilation times the ensemble spreads naturally
under the stochastic dynamics generated by the learned model. Whenever
new observations become available, the adaptive particle filter
contracts the ensemble towards the truth while maintaining sufficient
ensemble diversity through tempering and Metropolis--Hastings
jittering. The figure illustrates that the posterior ensemble remains
centred on the true trajectory throughout the assimilation window,
providing accurate state estimation despite the substantial model
reduction.

This work constitutes one of the first demonstrations that modern
generative models can be integrated directly into Bayesian data
assimilation for stochastic partial differential equations. Rather than
learning the posterior distribution itself, the generative model learns
the unresolved stochastic forcing required by the reduced-order model.

\subsection{Further Data Assimilation Developments }

\label{subsec:further_da_developments}

Beyond the particle-filter methodologies developed for stochastic rotating
shallow water and quasi-geostrphic models, the STUOD programme also led to
several further developments in data assimilation. These works broaden the
scope of the programme in complementary directions as detailed below:

\paragraph{Stochastic parametrisation based on Location Uncertainty.}
In the paper \cite{DufeeMeminCrisan2022}, the authors investigate whether a
stochastic parametrisation based on Location Uncertainty can replace the
ad hoc variance inflation procedures commonly used in ensemble Kalman
filters. Ensemble Kalman filters are widely used in high-dimensional data
assimilation because they can work with relatively small ensembles,
especially when combined with localisation. However, deterministic ensemble
forecasts often suffer from insufficient spread, and this is usually
corrected by artificial inflation. The paper shows that introducing
stochasticity directly into the dynamics provides a more physically
consistent alternative. The stochastic model builds uncertainty into the
forecast step itself, rather than adding variance externally. In numerical
experiments for the surface quasi-geostrophic model, the resulting
stochastic ensemble Kalman filter gives better mean-square-error
performance than the deterministic model with inflation and avoids some of
the instability associated with poorly tuned inflation parameters.

\paragraph{Comparison of stochastic parametrization schemes using data assimilation on triad models.}

The paper~\cite{ChapronCrisanHolmLangLobbeMemin2024} introduces reduced-order triad models as benchmark systems for
comparing stochastic parametrisation schemes. The work studies two
physically motivated stochastic parametrisations: Stochastic Advection by
Lie Transport (SALT) and Location Uncertainty (LU). When projected onto a
single helical Fourier triad, both parametrisations lead to the same
helicity-preserving stochastic triad model. This model is then compared
with an energy-preserving stochastic triad model. These low-dimensional
systems retain important nonlinear and conservation properties of fluid
dynamics while remaining computationally inexpensive. They therefore
provide useful testbeds for analysing data assimilation algorithms. The
paper shows that, in the absence of stochasticity, particle filters collapse
rapidly, whereas carefully calibrated stochastic transition kernels provide
sufficient ensemble spread and lead to stable filtering performance.

\paragraph{Data assimilation for energy-aware hybrid models.}

The energy-aware hybrid modelling framework has recently been extended to
Bayesian data assimilation in
\cite{ShevchenkoCrisan2025Hybrid}. The principal motivation is to combine
the computational efficiency and physical consistency of energy-aware
hybrid models with the uncertainty quantification provided by Bayesian
sequential inference. The paper considers a three-layer quasi-geostrophic
model of the Gulf Stream, where coarse-resolution simulations are enhanced
by data-driven corrections that preserve the energy distribution across
resolved and unresolved scales. Building upon the Sequential Monte Carlo
methodologies developed within the STUOD programme, the authors formulate a
particle filtering framework incorporating adaptive tempering, resampling,
nudging and Markov Chain Monte Carlo rejuvenation to assimilate sparse
noisy observations into the hybrid forecast model. Numerical experiments
demonstrate that the resulting methodology accurately tracks the evolution
of the Gulf Stream jet and associated mesoscale vortices while
significantly reducing forecast errors relative to the underlying
coarse-resolution model. Furthermore, the study shows that observations
concentrated in dynamically active regions provide nearly the same
assimilation accuracy as observations distributed throughout the entire
domain, suggesting that careful sensor placement can substantially improve
the efficiency of operational observing systems. This work illustrates how
physics-informed machine learning and Bayesian data assimilation can be
combined to produce computationally efficient forecasting methodologies
that preserve the essential physical structure of complex geophysical
flows.

\paragraph{Data assimilation for the stochastic Camassa--Holm equation.}

The Bayesian data assimilation methodologies developed within the STUOD
programme have also been applied to the stochastic Camassa--Holm equation,
a nonlinear dispersive model describing shallow water wave propagation.
The work
\cite{CotterCrisanSingh2024STUOD}
investigates the use of advanced particle filtering techniques for
assimilating noisy observations into stochastic Camassa--Holm dynamics.
The forecast model incorporates stochastic transport to represent model
uncertainty, while the Bayesian inference procedure employs Sequential
Monte Carlo methods together with adaptive tempering, resampling and
Metropolis--Hastings rejuvenation to mitigate particle degeneracy. Through
a series of numerical experiments, the authors demonstrate that the
proposed methodology is capable of accurately reconstructing the evolving
wave field from sparse noisy observations while maintaining a diverse
particle ensemble throughout the assimilation process. The study shows
that the particle filtering framework developed for stochastic transport
models extends naturally beyond large-scale geophysical circulation models
to nonlinear dispersive wave equations, illustrating the versatility of
Bayesian sequential inference for stochastic partial differential
equations.

\paragraph{Global Girsanov nudged particle filters.}

Particle filters often suffer from severe weight degeneracy when applied to
high-dimensional nonlinear systems, particularly when observations are
widely separated in time. The work
\cite{SinghHopeCollinsCotterCrisan2025} addresses this challenge by
introducing a global Girsanov nudging strategy for Sequential Monte Carlo
methods. Rather than nudging particles only locally towards the
observations, the proposed methodology constructs globally guided proposal
trajectories over the entire assimilation window, significantly increasing
the probability that particles remain close to the observed dynamics. The
resulting change of measure is accounted for exactly through Girsanov's
theorem, ensuring that the particle approximation remains consistent with
the target Bayesian posterior distribution. Numerical experiments
demonstrate that the method substantially reduces weight degeneracy,
improves effective sample sizes and produces more accurate state estimates
than conventional particle filtering algorithms, particularly for strongly
nonlinear geophysical models. The work represents a significant advance in
the design of proposal mechanisms for high-dimensional Bayesian data
assimilation and complements earlier STUOD developments on adaptive
tempering, nudging and transport-based particle filtering.

\paragraph{Sequential Markov Chain Monte Carlo for Lagrangian Data Assimilation with Applications to Unknown Data Locations.}

Most Bayesian filtering methodologies assume that the observation
operator is known precisely. In many practical applications, however,
the spatial locations at which measurements are collected are themselves
uncertain. This occurs, for example, when drifting oceanographic instruments are only approximately localized.

The work
\cite{RuzayqatBeskosCrisanJasraKantas2024}
introduces a new class of spatial filtering problems in which both the hidden
state and the observation locations are unknown random variables. The
resulting Bayesian formulation requires simultaneous inference of the
system state and the observation geometry, substantially increasing the
dimension and complexity of the filtering problem. The paper develops
Monte Carlo algorithms capable of estimating both quantities jointly and
provides a theoretical framework for filtering with uncertain
observation operators.

\subsection{Practical Considerations for Particle Filtering Data Assimilation}

The preceding sections have focused primarily on the mathematical
foundations of stochastic transport models and Bayesian data assimilation. 
During the research project, however, it became clear that the success of a
data assimilation algorithm depends just as much on a number of practical
implementation choices as on the underlying mathematical formulation.
Many of these issues are not immediately apparent from the theoretical
analysis but proved to have a significant impact on the performance of the
algorithms in numerical experiments. We conclude by summarising several of
the lessons that emerged during the project.

\paragraph{Model reduction.}

Stochastic reduced-order models provide a far
more effective framework for data assimilation than attempting to filter
the full deterministic model directly. The primary motivation for model
reduction is, of course, computational. Reducing the spatial resolution
substantially decreases the cost of propagating each particle during the
forecast step. However, the principal benefit is not simply a reduction in
computational time, but rather the opportunity to reallocate the available
computational resources to improve the quality of the Bayesian inference.
The computational savings obtained through model
reduction may be used to increase the number of particles, to introduce
extra tempering and jittering steps, or more
sophisticated nudging strategies. These improvements often have a greater impact on
filter performance than using a finer deterministic forecast model with a
small particle ensemble.

Model reduction, however, cannot be considered independently of the
stochastic parametrisation. As the spatial resolution is reduced, an
increasing proportion of the unresolved dynamics must be represented by the
stochastic transport fields
\(\{\xi_i\}\).
These fields are obtained through the calibration procedures described
earlier in this paper and are represented by a finite collection of
dominant empirical orthogonal functions. By construction, the retained
transport modes explain only a prescribed proportion of the variability of
the unresolved dynamics---typically \(90\%\), \(95\%\), or \(99\%\) of the
variance observed in the calibration data. The remaining variability is
discarded. Consequently, the stochastic parametrisation should not be
viewed as an exact representation of the unresolved scales, but rather as
their best low-dimensional approximation within the chosen reduced-order
space.

There is therefore an intrinsic balance between the degree of model
reduction and the complexity of the stochastic transport model. A more
aggressive reduction generally requires a larger number of calibrated
transport modes to recover the dominant unresolved dynamics. Conversely,
retaining an excessively large number of stochastic modes increases both
the computational cost of the forecast model and the dimensionality of the
uncertainty representation. The objective is therefore not to maximise the
number of retained modes, but to identify a reduced stochastic model that
captures the dominant transport statistics while remaining computationally
efficient.

A successful
data assimilation is achieved by balancing these competing effects. An
appropriate combination of model reduction, stochastic calibration,
ensemble size, assimilation frequency and advanced particle-filtering
techniques such as tempering, jittering and nudging often produces greater
overall accuracy than simply increasing the spatial resolution of the
forecast model. In this sense, model reduction is one of the key steps practical high-dimensional particle filtering based   data assimilation.

\paragraph*{Choice of the initial ensemble.}

The effectiveness of the data assimilation 
procedure depends crucially on the initial ensemble containing particles in
the relevant region of the state space.
This issue is particularly important for stochastic transport models.
Unlike additive stochastic forcing, transport noise preserves much of the
geometric structure of the underlying deterministic equations. In the SALT
framework, the stochastic perturbation preserves the transport structure
and constrains the dynamics to evolve on manifolds associated with the
Casimir invariants, including enstrophy. Similarly,
for the Location Uncertainty (LU) formulation, the stochastic dynamics
preserve the total energy of the system. As a
consequence, the stochastic evolution cannot rapidly transport particles
between regions of the state space having substantially different values of
these invariants.

From a practical point of view, this has an important consequence. If the
initial particle ensemble does not contain particles whose energy,
enstrophy, or other relevant conserved quantities are compatible with the
true state, then no amount of stochastic transport will recover the missing
probability mass. The observations simply identify that these particles are
inconsistent with the data, causing their importance weights to collapse.
In our numerical experiments this effect is particularly pronounced during
the first assimilation cycle: if the initial ensemble is poorly chosen,
almost all particles receive negligible weights after assimilating the
first observations and are eliminated during the subsequent resampling
stage. Once this occurs, the filter has effectively lost all information
about the true state and recovery becomes extremely unlikely.

For this reason, the construction of the initial ensemble should not be
viewed as a routine step but as an integral part of the data
assimilation algorithm.

\paragraph{Mismatch between the transition kernel and the true dynamics.}

The performance of any particle filter depends critically on the quality of
the transition kernel used during the forecast step. In principle, this
kernel should represent the true conditional distribution of the state over
one assimilation interval. In practice, however, it is generated by an
approximate numerical model and therefore inevitably differs from the true
dynamics.

The consequences of this mismatch depend strongly on the dynamical
properties of the underlying system. For systems possessing
stability or synchronisation properties, small forecast errors may decay
naturally as the dynamics evolve. In such situations, an imperfect
transition kernel may still provide a satisfactory approximation of the
true evolution, and the particle filter can recover through successive
assimilation cycles.

The situation is fundamentally different for chaotic systems, such as those
encountered in atmospheric and oceanic dynamics. In these systems, nearby
trajectories separate rapidly, so even small modelling errors introduced by
the transition kernel grow quickly during the forecast step. As a result,
the propagated particles drift away from the region of high posterior
probability before the observations are assimilated. The subsequent
importance-weighting step then assigns negligible weights to most
particles, leading to severe weight degeneracy and, ultimately, filter
collapse.

This
difficulty cannot be overcome simply by increasing the number of particles.
Instead, the proposal mechanism itself must be improved so that the
forecast particles remain close to the observations. An effective approach
is provided by \emph{nudging}, in which an additional term is
introduced into the forecast dynamics to steer each particle towards the high posterior
probability areas during the propagation step. Rather than allowing
the particles to evolve freely under an imperfect transition kernel,
nudging corrects the forecast trajectory, thereby reducing the
discrepancy between the proposal distribution and the true posterior.

\paragraph{Choice of the data assimilation interval.}

The interval between successive assimilation steps
determines how long the particles evolve under the forecast model before
they are corrected by the observations. Although the assimilation interval
is often dictated by the frequency at which observations become available,
it may also be viewed as an algorithmic parameter whose choice has a
significant impact on the overall performance of the particle filter.

If the assimilation interval is chosen too small, observations are
assimilated very frequently and the particle cloud remains close to the
true trajectory. This generally improves the statistical quality of the
posterior approximation but comes at a considerable computational cost,
since each assimilation step requires the evaluation of likelihoods,
resampling, and, in many implementations, tempering and jittering.
For large-scale applications, these operations typically dominate the computational cost of the algorithm.

Conversely, if the assimilation interval is too large, the uncertainty
introduced during the forecast step grows substantially before the next
observation is assimilated. This effect is particularly pronounced for
chaotic systems, where nearby trajectories separate rapidly. As a result,
the particle cloud spreads over increasingly large regions of the state
space and many particles drift far from the true trajectory. When the next
observation is finally assimilated, the likelihood function assigns
significant weight to only a small fraction of the particles, leading to
severe weight degeneracy and a dramatic reduction of the effective sample
size.

\begin{figure}[h]
	\centering
	\includegraphics[width=\textwidth]{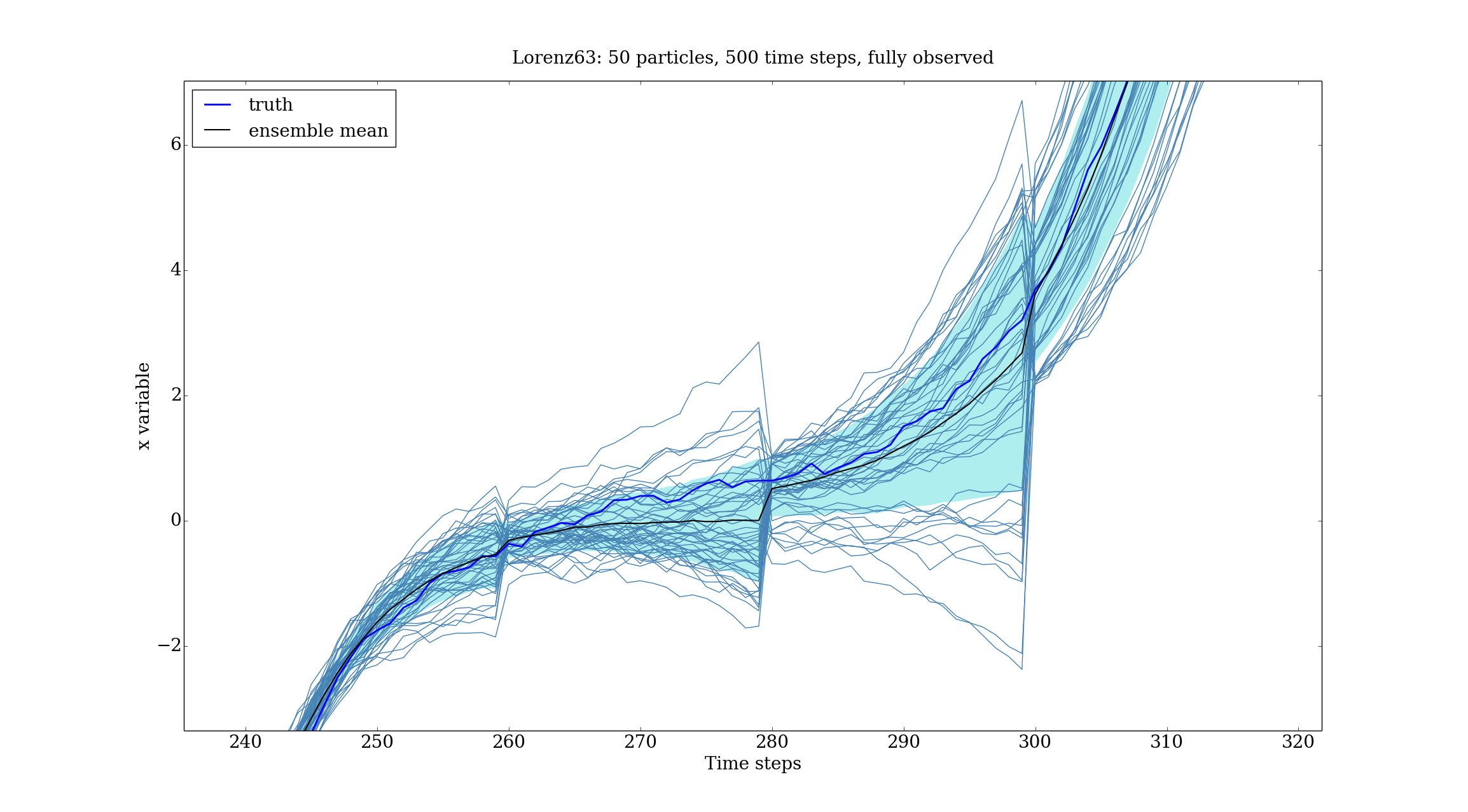}
	\caption{Illustration of the balance between forecast uncertainty and
		data assimilation. The blue curve denotes the true trajectory, the
		black curve is the ensemble mean, and the light blue curves represent
		the individual particles. Between assimilation times the forecast model
		causes the ensemble to spread as uncertainty accumulates. At each
		assimilation step the observations reduce this uncertainty by steering
		the particle cloud back towards the true trajectory. The figure
		illustrates a successful filtering cycle in which the forecast spreads
		the ensemble while the analysis step prevents uncontrolled divergence
		from the truth.}
	\label{fig:forecast_spread}
\end{figure}

Figure~\ref{fig:forecast_spread} illustrates the desirable balance between
forecast uncertainty and observational correction. Between two consecutive
assimilation times the forecast model naturally spreads the particle cloud,
reflecting the increasing uncertainty in the predicted state. At each
assimilation step the observations reduce this uncertainty by eliminating
particles inconsistent with the data and concentrating the ensemble around
the true trajectory. A successful particle filter therefore exhibits a
repeating cycle of uncertainty growth during the forecast phase followed by
uncertainty reduction during the analysis phase.

The choice of the assimilation interval cannot be considered in isolation.
It is closely coupled with several other algorithmic parameters,
particularly the number of particles and the complexity of the model. Longer assimilation intervals generally require larger
particle ensembles in order to maintain a satisfactory representation of
the posterior distribution. Similarly, increasing the number of stochastic
transport modes \(\{\xi_i\}\) introduces additional forecast variability,
which may require either more frequent observations or a larger ensemble to
avoid particle collapse. Conversely, a reduced stochastic representation
using fewer calibrated transport modes may permit longer assimilation
intervals without significantly degrading filter performance.

Successful
high-dimensional particle filtering is achieved not by optimising any one
of these parameters individually, but by balancing them simultaneously.
The assimilation interval, ensemble size, number of stochastic transport
modes, and the use of techniques such as nudging, tempering and jittering
should therefore be regarded as components of a single algorithmic design
problem rather than as independent tuning parameters.

\paragraph{Tempering and jittering}

An important practical question is the choice of the number of tempering
steps. Too few intermediate distributions provide little protection
against weight degeneracy, whereas too many increase the computational cost
without necessarily improving the quality of the posterior approximation.
In practice, the number of tempering steps should depend on the quality of
the particle ensemble immediately before the analysis step. The quantity
most commonly used to guide this choice is the \emph{effective sample size}
(ESS), which provides an indication of the concentration of the importance
weights. Adaptive tempering schemes typically introduce as many
intermediate distributions as required to maintain the ESS above a
prescribed threshold. Although the ESS has become the standard diagnostic
for adaptive tempering, it is our opinion that ESS 
is not always an ideal measure of sample quality. In particular,
the ESS reflects only the distribution of the particle weights and does not
measure how well the particle cloud represents the geometry of the
posterior distribution. Developing more informative diagnostics for
adaptive tempering remains an important open research problem.

The most widely used jittering strategy employs a
Metropolis--Hastings (MH) Markov chain whose invariant distribution is the
posterior distribution. In principle, this provides an elegant solution,
since repeated MH moves increase the diversity of the particle cloud
without introducing bias. In practice, however, this approach may be
computationally expensive. Several MH iterations are often required spread the particles, and each iteration involves repeated
evaluations of the likelihood and forecast model, which can become
prohibitively expensive for high dimensional applications.

An attractive alternative is deterministic jittering, in which the
particles are displaced using carefully designed deterministic
transformations rather than stochastic Markov chain moves. Such methods are
typically much less expensive computationally and can produce significant
improvements in ensemble diversity. However, deterministic jittering must
be employed with considerable care. Unlike Metropolis--Hastings updates, it
does not automatically preserve the posterior distribution, and an
inappropriate deterministic perturbation may introduce systematic bias into
the particle ensemble. In extreme cases, this bias can completely negate
the improvements achieved during the preceding assimilation step.

Both tempering and jittering should be regarded as adaptive rather than fixed
components of the filtering algorithm. Their successful implementation
depends critically on the quality of the particle ensemble, the complexity
of the forecast model, and the available computational budget.

\paragraph{Choice of the stochastic parametrisation.}

The choice of the stochastic parametrisation has a direct impact on both
the computational cost and the predictive skill of the reduced-order
forecast model. Several modelling decisions must be made, including the
choice between different stochastic transport formulations, such as SALT
or Location Uncertainty, the calibration methodology (Lagrangian or
Eulerian), and the representation of the unresolved transport statistics. Generative
modelling provides a promising alternative to the traditional Gaussian
representation. Rather than approximating only the
covariance structure of the unresolved transport, generative models learn
its full probability distribution directly from high-resolution
simulations. 

\paragraph{Diagnostics and monitoring.}

Assessing the performance of a particle filter requires considerably more
than monitoring the estimation error alone. This can be done by computing the effective
sample size, ensemble spread, root mean square error, continuous ranked
probability score, and rank histograms. Each of
these measures captures a different aspect of filter performance. 

\paragraph{Computational resources.}

The
available computational budget should be regarded as a resource to be
distributed optimally rather than being devoted exclusively to increasing
the spatial resolution of the forecast model. The forecast propagation of
particles is naturally parallelisable and therefore scales well on modern
high-performance computing architectures.. Determining the most effective
allocation of computational resources remains an important practical
optimisation problem.

\paragraph{Calibration procedure.}

The quality of the stochastic parametrisation depends fundamentally on the
data used during the calibration stage. An important practical question is
how much calibration data is required and over what time interval it
should be collected. Also how frequently the stochastic parametrisation should be recalibrated?
The statistics of the unresolved scales evolve over time as the underlying
flow changes. A stochastic model calibrated from one dynamical regime may
therefore become progressively less representative as the system evolves.
Rather than viewing calibration as a one-off offline procedure, it may be
preferable to update the stochastic parametrisation periodically using
more recent data. An even more attractive possibility is to perform the
calibration continuously, allowing the stochastic transport model to adapt
automatically to changes in the underlying dynamics. The development of
computationally efficient methods for adaptive or online calibration
remains an important open research problem and represents a promising
direction for future work.

\section{Conclusions and Future Directions}
\label{sec:08_conclusions_and_future_directions}

This paper has reviewed the principal theoretical and data assimilation developments achieved within Work Packages 2 and 4 of the STUOD project. Although these developments span several areas of mathematics, including stochastic analysis, geometric mechanics, uncertainty quantification, inverse problems and Bayesian inference, they can be understood as components of a single scientific programme. The central objective of this programme has been the development of stochastic transport models as mathematically rigorous and practically useful tools for representing uncertainty in geophysical fluid dynamics.
From this perspective, the various analytical, variational, calibration and data assimilation developments reviewed in the preceding sections should not be regarded as independent contributions. Rather, they form successive stages in the evolution of a coherent framework linking stochastic transport modelling with uncertainty quantification and forecasting.
A recurring theme throughout the section has been the observation that uncertainty in geophysical fluid systems should be represented through transport mechanisms rather than through externally imposed forcing terms.
This principle provided the conceptual foundation for the stochastic transport paradigm investigated throughout the programme.
The resulting programme may be summarised schematically as

\[
\text{Stochastic Transport}
\longrightarrow
\text{Analysis}
\longrightarrow
\text{Calibration}
\longrightarrow
\text{Bayesian Inference}.
\]

Viewed collectively, these developments transformed stochastic transport models from a modelling paradigm into a complete framework for uncertainty quantification in geophysical fluid dynamics. The resulting programme extends from stochastic transport theory to calibration and Bayesian inference, thereby connecting fundamental mathematical analysis with practical uncertainty quantification.

Although many challenges remain, the work reviewed in this paper establishes a foundation upon which future developments in stochastic fluid dynamics, uncertainty quantification and data assimilation can be built. In this sense, the principal legacy of this work is not a collection of individual results, but the creation of a unified mathematical framework linking stochastic transport, geometric mechanics, statistical inference and forecasting.

The developments described throughout these proceedings have established a mathematical framework for stochastic transport modelling, uncertainty quantification and Bayesian data assimilation. A natural next step is the
application of these methodologies to contemporary satellite observations.
One of the principal future objectives is
therefore the assimilation of observations from the Surface Water and Ocean
Topography (SWOT) satellite into stochastic ocean circulation models.

The SWOT mission provides sea-surface height measurements with an
unprecedented spatial resolution, making it possible to observe
mesoscale and submesoscale ocean structures that were previously
unresolved by conventional satellite altimetry. Unlike traditional nadir
altimeters, SWOT measures sea-surface height over a wide swath, producing
high-resolution observations along successive satellite ground tracks.
These observations provide an ideal testbed for the stochastic transport
models and Bayesian data assimilation methodologies developed in the
preceding sections.

The proposed data assimilation framework combines three complementary
sources of information. The first is the GLORYS12 global ocean reanalysis,
which provides dynamically consistent initial conditions and background
forecasts. The second consists of the SWOT sea-surface height observations,
which provide high-resolution measurements along the satellite swaths. The
third component is the stochastic transport model, calibrated using the
techniques described above, which supplies the probabilistic forecast model
required by Bayesian filtering.

\begin{figure}[h]
	\centering
	\includegraphics[width=\textwidth]{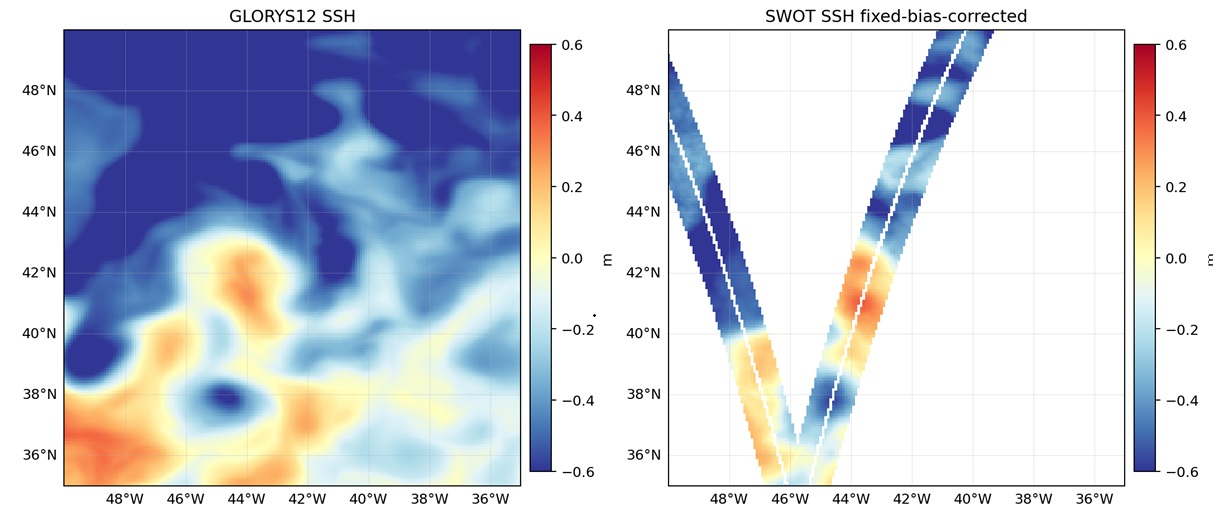}
	\caption{Comparison between the GLORYS12 sea-surface height field (left)
		and SWOT observations (right). The GLORYS12 reanalysis provides the
		background state used to initialise the stochastic forecast model, while
		the SWOT observations constitute the observational data assimilated by the
		Bayesian filtering algorithm.}
	\label{fig:future_glorys_swot}
\end{figure}

Figure~\ref{fig:future_glorys_swot} illustrates the complementary nature of
the available datasets. The left panel shows a typical sea-surface height
field obtained from the GLORYS12 reanalysis, while the right panel shows
the corresponding SWOT observations. The contrast between the two datasets
is immediately apparent. Whereas GLORYS12 provides complete spatial
coverage on a regular grid, the SWOT observations are available only along
the satellite ground tracks. The data assimilation problem therefore
consists of reconstructing the full ocean state from these sparse but
highly informative observations.

\begin{figure}[h]
	\centering
	\includegraphics[width=.8\textwidth]{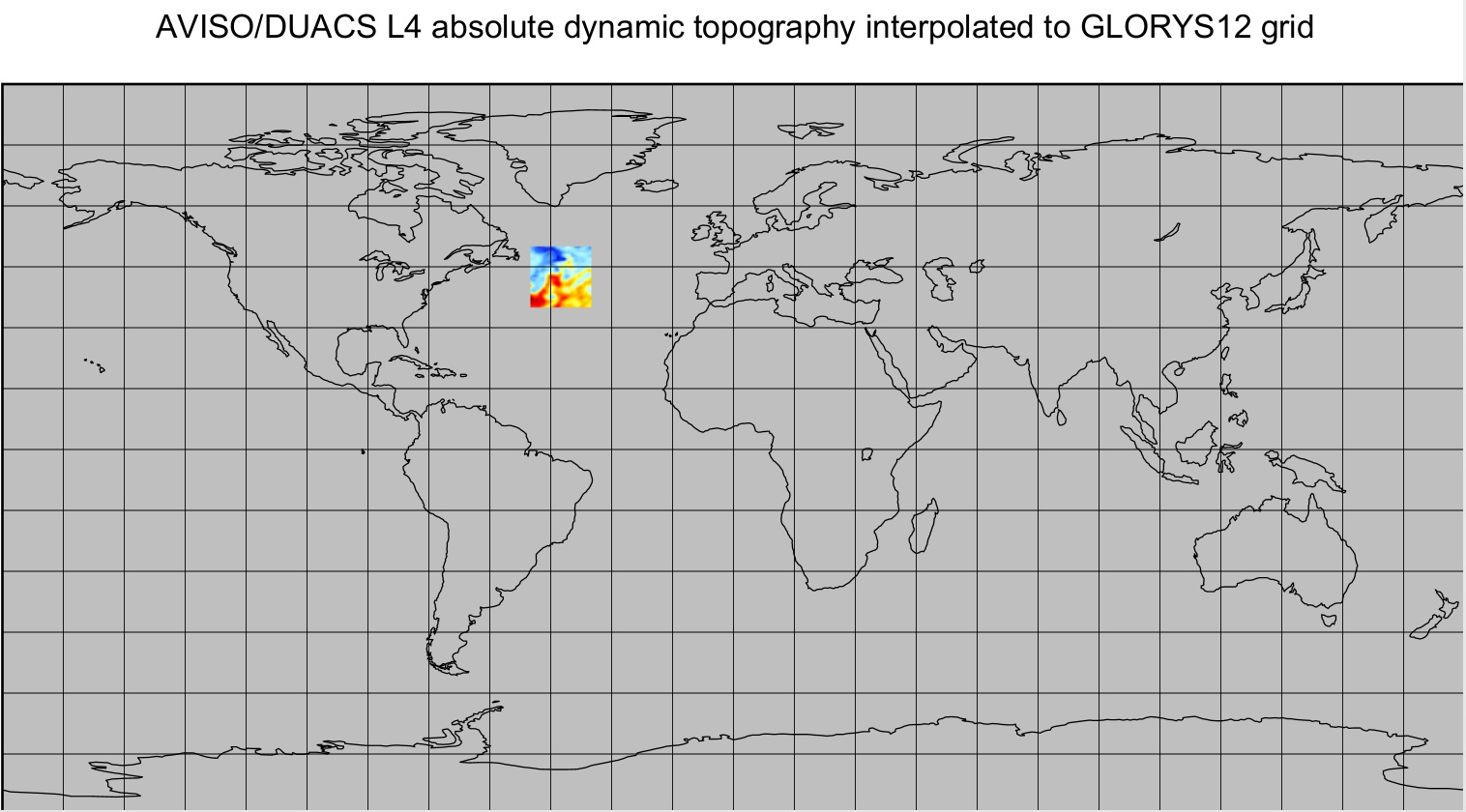}
	\caption{Location of the computational domain within the global ocean. The
		boxed region identifies the North Atlantic sector used for the initial
		SWOT data assimilation experiments.}
	\label{fig:future_domain}
\end{figure}

The geographical region that will be considered in an initial study is shown in
Figure~\ref{fig:future_domain}. The computational domain lies in the
North Atlantic, centred on the Gulf Stream extension, one of the most
energetic and dynamically complex regions of the global ocean. This region
contains strong mesoscale variability, coherent vortices and energetic
transport barriers, making it an ideal environment for evaluating
stochastic transport models and high-dimensional particle filtering
algorithms.

The selected domain covers approximately
\(35^\circ\mathrm{N}\)--\(49^\circ\mathrm{N}\) and
\(49^\circ\mathrm{W}\)--\(35^\circ\mathrm{W}\), encompassing the Gulf
Stream extension and the surrounding mesoscale eddy field. Daily GLORYS12
fields provide the initial conditions, while the SWOT observations are
assimilated sequentially as they become available. The objective is to
combine these observations with stochastic reduced-order models,
constructed using the calibration methodologies described earlier in these
proceedings, to produce dynamically consistent estimates of the evolving
sea-surface height and velocity fields.

From a methodological perspective, this programme brings together many of
the themes developed throughout these proceedings. The stochastic transport
models provide physically consistent representations of unresolved oceanic
variability; the calibration methodologies estimate the stochastic
transport coefficients directly from high-resolution data; and the Bayesian
particle filtering algorithms combine these stochastic forecasts with
satellite observations to reconstruct the evolving ocean state. At the
same time, the unprecedented spatial resolution of SWOT provides an
opportunity to evaluate these stochastic models at scales that were
previously inaccessible.

Looking further ahead, this framework opens several other promising research
directions. These include
joint assimilation of sea-surface height together with temperature and
salinity observations, adaptive calibration of stochastic transport
coefficients during the filtering process, and the direct use of
generative models for Data Assimilation.

\section*{Acknowledgements}

The research presented in these proceedings was carried out within the
framework of the ERC Synergy Grant \emph{Stochastic Transport in Upper Ocean
	Dynamics (STUOD)} (ERC Synergy Grant No.~856408). The authors gratefully
acknowledge the support of the European Research Council, whose funding has
made possible the close collaboration between participating scientists. These proceedings draw together results obtained over the course of the STUOD project by many researchers. The authors are particularly grateful to their collaborators whose work is represented throughout the volume,
including Bertrand Chapron, Darryl ~Holm, Etienne M\'emin,  Valentin Resseguier, Colin ~Cotter, Peter Korn, Franco Flandoli, Arnaud Debussche, Peter Jan van Leeuwen, Roland Potthast, Igor Shevchenko, Hamza Ruzayqat, Alex Beskos, Ajay Jasra, Nikolas Kantas
Wei Pan, Oliver Street, James Woodfield, Alexander Lobbe, Benjamin Duf\'ee, Romeo Mensah, Daniel Goodair, Erwin Luesink, Maneesh Singh.  Many of the ideas described here originate from joint research with these colleagues, and the authors are deeply indebted to them for numerous discussions and insights.

\end{document}